\RequirePackage[2024-06-01]{latexrelease}
\documentclass[twocolumn]{aastex631}

\newcommand\X{\textcolor{red}{X}}
\newcommand\B{\textbf}
\newcommand\R{\textcolor{red}}
\usepackage{array}
\usepackage{placeins}

\defcitealias{Madhusudhan2023K2-18b}{M23}

\shorttitle{Transmission spectrum of TOI-1231~b}

\shortauthors{Sarkar et al.}

\graphicspath{{./}{figures/}}

\begin{document}

\title{Transmission Spectrum of the Benchmark Temperate Exo-Neptune TOI-1231~b}

\author[0000-0002-2705-5402]{Subhajit Sarkar} \affiliation{School of Physics and Astronomy, Cardiff University, The Parade, Cardiff CF24 3AA, UK}

\author[0000-0002-4869-000X]{Nikku Madhusudhan} \affiliation{Institute of Astronomy, University of Cambridge, Madingley Road, Cambridge CB3 0HA, UK}

\author[0009-0008-9026-5534]{Lorenzo Pica-Ciamarra} \affiliation{Institute of Astronomy, University of Cambridge, Madingley Road, Cambridge CB3 0HA, UK}

\author[0000-0002-0931-735X]{M\aa ns Holmberg} \affiliation{Space Telescope Science Institute, 3700 San Martin Drive, Baltimore, MD 21218, USA}

\author[0009-0001-9140-3299]{Frances E. Rigby} \affiliation{Institute of Astronomy, University of Cambridge, Madingley Road, Cambridge CB3 0HA, UK}

\author[0000-0002-8837-0035]{Julianne I. Moses}
\affiliation{Space Science Institute, Boulder, CO 80301, USA}

\author[0009-0008-6843-037X]{Megan Mealing} \affiliation{School of Physics and Astronomy, Cardiff University, The Parade, Cardiff CF24 3AA, UK}

\correspondingauthor{Nikku Madhusudhan, Subhajit Sarkar}
\email{nmadhu@ast.cam.ac.uk, SarkarS3@cardiff.ac.uk} 

\begin{abstract}
The JWST is revolutionizing our understanding of the temperate sub-Neptune population through atmospheric spectroscopy. The nature of these planets remains debated, as their bulk properties are compatible with a range of interior scenarios, including mini-Neptunes, hycean worlds, and gas dwarfs, with different predicted atmospheric compositions. While theoretical studies have predicted compositional diagnostics for shallow- versus deep-atmosphere scenarios, there is a critical need for empirical constraints for a temperate planet that is a priori known to possess a deep H$_2$-rich atmosphere. The temperate exo-Neptune TOI-1231 b provides one such benchmark target. In this work, we present the JWST near-infrared (0.65--5.2 $\mu$m) transmission spectrum of TOI-1231 b, observed with NIRISS single-object slitless spectroscopy and NIRSpec G395H, representing the first for a temperate exo-Neptune. The density of TOI-1231 b requires a thick H$_2$-rich atmosphere, making the planet a keystone reference case for testing mini-Neptune scenarios for sub-Neptunes. We report a strong detection of CH$_4$ ($\ln B = 54.5$--$69.6$) and moderate to strong evidence for CO$_2$ ($\ln B = 2.9$--$6.6$). We do not find significant evidence for any other prominent molecule, although we find high 95\% upper limits on the mixing ratios of NH$_3$ and CO, both of which are expected in deep H$_2$-rich atmospheres. We also do not find any significant evidence for sulfur-bearing species that have been inferred for some temperate sub-Neptunes. This composition is consistent with expectations for a temperate Neptune possessing a deep H$_2$-rich atmosphere with no distinct surface. We discuss the implications of our results for the characterization of temperate sub-Neptunes.

\end{abstract}

\keywords{Exoplanets(498)  --- Exoplanet atmospheres(487) -- Exoplanet atmospheric composition (2021) --- JWST (2291) --- Infrared spectroscopy(2285)}

\section{Introduction} \label{sec:intro}

With the advent of JWST, the frontier of small temperate exoplanets and their atmospheres is now being explored. JWST offers the capability to probe in detail the atmospheric compositions and the resulting implications for the interior structures of this population through transmission spectroscopy \citep[e.g.][]{Barstow2016,greene2016, Morely2017, Lustig-Yaeger2019, madhusudhan_habitability_2021}. Sub-Neptunes are the most abundant class of currently known exoplanets. We use the term `sub-Neptune' for planets with radii between $\sim$$1.5-4 R_{\oplus}$ with densities suggestive of significant volatile content or hydrogen-rich atmospheres \citep{Lopez2014, Rogers2015}.  We also use the term `exo-Neptune' for planets at the upper size limit of this range, $\sim4R_{\oplus}$, with radii similar to that of Neptune. We define `temperate' planets as those with equilibrium temperatures of $T_{\mathrm{eq}}\lesssim400$ K.

A number of JWST transmission spectra have already been published for temperate sub-Neptune planets \citep{Madhusudhan2023K2-18b, Madhusudhan2025K2-18b, madhusudhan_exploring_2025, holmberg_possible_2024, Benneke2024, Cadieux2024, rigby_jwst_2025}. Additionally, JWST transmission spectra have been obtained for sub-Neptune-sized planets at warmer temperatures ($T_{\mathrm{eq}} \sim 400-700$ K) \citep{Schlawin2024, Banerjee2024, Piaulet-Ghorayeb2024, Ahrer2025, Wallack2024, Wallack2026, Teske2025, Alderson2025, Banerjee2024, Beatty2024}, as well as hotter planets with $ T_{\rm eq}> 700$ K \citep{Alderson2024, Davenport2025, Bello-Arufe2025, Alam2025, Fisher2026, Radica2024_LTT9779}. 

Sub-Neptunes are more amenable to  atmospheric characterization by transmission spectroscopy than small rocky planets, due to the fact that they can have H$_2$-rich envelopes with large scale heights, which enhance their atmospheric observability. These planets, however, remain a fundamental open question in exoplanet science. Indeed, their bulk densities are degenerate and consistent with a wide array of possible internal structures, spanning gas dwarfs (rocky planets with a thick H$_2$/He envelope), mini-Neptunes (planets with a rocky core, icy mantle and a thick H$_2$/He envelope), and water worlds (planets with a large water fraction) \citep[e.g.][]{rogers2010,  madhu2020, Nixon21_ocean, Luque2022, Rigby2024_gasdwarf}.  The latter include hycean worlds, which are water-rich planets with a shallow H$_2$/He envelope \citep{madhusudhan_habitability_2021,rigby_ocean_2024}. Determining the nature of sub-Neptune exoplanets has important implications for the origins of the radius valley and for the potential habitability of this population.

The degeneracy in interior composition may, however, be resolved through atmospheric observations. Baseline expectations for a temperate, deep H$_2$-rich atmosphere include the presence of prominent CNO molecules, such as methane (CH$_4$), ammonia (NH$_3$), and water (H$_2$O), which are expected in thermochemical equilibrium under those conditions \citep{Lodders2002,Visscher2006,Moses2013}. The photochemical breakdown of these molecules can be replenished through vertical mixing from the deep atmosphere. 
Breakdown products flow down into deeper layers where the parent molecules are regenerated under conditions where thermochemical equilibrium can dominate. A key diagnostic of a deep atmosphere in a temperate planet is therefore the presence of NH$_3$ and CH$_4$, which often indicate deep atmospheric recycling. CO and CO$_2$ may also be present in a high-metallicity deep atmosphere, with a CO$_2$/CO ratio of $\lesssim$1 \citep{Hu2021, Hu2021b, Tsai2021, Yu2021}.

In contrast, a shallow atmosphere atop a solid or liquid surface would present a different chemical signature.  Deep atmospheric recycling is prevented if the atmosphere is sufficiently shallow that temperatures at its base do not become high enough for thermochemical reactions to recycle photochemical products back into their parent species  \citep{Hu2021, Tsai2021, Yu2021, Madhusudhan2023K2-18b}. As a result, a fraction of the initial NH$_3$ and CH$_4$ is progressively replaced by photochemical products. The absence of NH$_3$ in a temperate H$_2$-rich atmosphere is therefore generally considered diagnostic of a shallow H$_2$-rich envelope with a solid or liquid surface. NH$_3$ is not readily recycled from strongly bonded photochemical products such as N$_2$, HCN, HC$_3$N or other nitriles at the low temperatures and pressures expected above a shallow solid surface. Furthermore, NH$_3$ is highly soluble, and may dissolve in a liquid water ocean surface if present. By contrast, H$_2$O and CH$_4$ are more stable in a low-temperature H$_2$-dominated atmosphere, allowing a larger fraction of these species to persist over time \citep[but see][]{Madhusudhan_chem_2023,  Cooke2024, wogan_jwst_2024}.  The presence of an ocean in equilibrium with the atmosphere can lead to CO$_2$ becoming the dominant carbon carrier \citep{Hu2021}, resulting in a high CO$_2$/CO ratio, potential photochemical production of CH$_4$, and an absence of NH$_3$.  Soluble photochemical products such as SO$_2$ and CH$_3$OH might also may also be absent in the presence of an ocean, while potentially remaining detectable in a solid-surface scenario \citep{Hu2021, Tsai2021}. While NH$_3$ depletion may be also due to a reduced magma ocean present beneath a deep H$_2$-rich atmosphere \citep[e.g.,][]{daviau_experimental_2021, Rigby2024_gasdwarf, shorttle_distinguishing_2024}, such cases are also expected to exhibit a high CO/CO$_2$
ratio, enabling them to be distinguished from shallow-atmosphere scenarios \citep{Rigby2024_gasdwarf}.

With the advent of JWST, it has become possible to test these diagnostics observationally. In particular, non-detections of NH$_3$ and CO have been used in the case of the habitable-zone sub-Neptune exoplanet K2-18 b to interpret its observations as indicative of a shallow atmosphere overlying a water layer, possibly indicating hycean conditions \citep{Madhusudhan2023K2-18b, Hu2025}. Similarly, non-detections of the
same two species in the temperate sub-Neptune TOI-270 d following JWST observations have been used to suggest either a hycean scenario  \citep{holmberg_possible_2024, constantinou_atmospheric_2026} or a miscible-envelope scenario \citep{Benneke2024}.

However, uncertainties in theoretical considerations \citep[e.g.,][]{Cooke2024} and a lack of empirical observations of temperate and unambiguously deep atmospheres have resulted in challenges to the above interpretations. Hence, there is a critical need for atmospheric characterization of a `benchmark' temperate Neptune, i.e. a temperate planet at the larger-radius edge of the sub-Neptune regime whose atmosphere is known a priori to be deep based on its density. Characterizing such an atmosphere would provide a crucial test of predictions for deep-atmosphere scenarios in the temperate regime. 

The temperate exo-Neptune planet TOI-1231~b provides exactly such a benchmark. TOI-1231 b was discovered by the Transiting Exoplanet Survey Satellite in 2021 \citep{Burt2021}. It is close to Neptune in size (3.65$^{+0.16}_{-0.15}$ R$_{\oplus}$) and mass (15.4$\pm3.3$ M$_{\oplus}$), with those of Neptune being 3.9 R$_{\oplus}$ and 17.2 M$_{\oplus}$. This results in a bulk density of 1.74 g/cm$^3$, which requires a large H$_2$-rich envelope. TOI-1231~b orbits an M3 dwarf (0.476$^{+0.015}_{-0.014}$ R$_{\odot}$, 3553$^{+51}_{-52}$ K). Its $T_{\mathrm{eq}}$, assuming a typical Bond albedo of $A_B = 0.3$, is 301.5$^{+3.8}_{-3.7}$ K, intermediate between the two sub-Neptunes recently characterized by JWST, K2-18 b (254.9$\pm$3.9 K, \citealp{Benneke2019}) and TOI-270 d (354$\pm$8 K, \citealp{VanEylen2021}).

TOI-1231 b also serves as a valuable reference for Neptune-sized planets at higher temperatures. A few JWST spectra from warmer Neptune-sized planets have been published. GJ 3470 b is a warm Neptune (4.57 R$_{\oplus}$, $T_\mathrm{eq}$ $\sim 600-700$ K) \citep{Awiphan2016, Kosiarek2019} for which detections of H$_2$O, CO$_2$, CH$_4$ and SO$_2$ have been reported, along with a low CH$_4$ abundance of $\sim$10$^{-4}$ and an atmosphere with  $\sim$100$\times$ solar metallicity \citep{Beatty2024}. 
In addition, the ultra-hot Neptune LTT 9779 b (4.72 R$_{\oplus}$, $T_\mathrm{eq}$ = 1978 K, \citealp{Jenkins2020}) has been recently studied in both transmission and emission with JWST \citep{Coulombe2025, Radica2025}, with the transmission spectrum showing muted spectral features indicative of H$_2$O and/or CH$_4$, high metallicity and clouds. 
A JWST emission spectrum has also been published for the warm Neptune GJ 436 b (4.19 R$_{\oplus}$, $T_\mathrm{eq}$ = 686 K) \citep{Turner2016}, showing weak spectral features and only tentative evidence for CO$_2$ at $2\sigma$ \citep{Mukherjee2025}.

In this work, we present the first JWST transmission spectrum of a temperate exo-Neptune, TOI-1231 b, using NIRISS single-object slitless spectroscopy (SOSS) and NIRSpec G395H in the near-infrared (0.65–5.2 $\mu$m). The present observation is a keystone for our understanding of deep H$_2$-rich atmospheres in the temperate regime, as the first ever transmission spectrum of a temperate Neptune-sized exoplanet. This will also serve as a reference point for the characterization of temperate sub-Neptunes and help break possible degeneracies between shallow- and deep-atmosphere scenarios. 

This work is structured as follows. In Section \ref{sec:obs}, we describe the observations, data reduction and lightcurve analyses used to generate the transmission spectra. We then describe the atmospheric retrievals and the resulting constraints on the atmospheric properties in Section \ref{sec:Retrievals}. In Section \ref{sec:summary}, we summarize our findings and discuss their implications and future directions.

\section{Observations and Data Reduction} 
\label{sec:obs}

We report the transmission spectrum of TOI-1231 b in the near-infrared (0.65-5.2 $\mu$m) obtained using the JWST NIRISS \citep{Doyon2012}  and  NIRSpec \citep{Ferruit2014} instruments.
We observed two primary transits of the planet, one with each instrument, as part of the JWST GO Program 3557 (PI: N. Madhusudhan). 
The first transit was observed
using NIRSpec G395H on 2024 April 30 between
03:49:43.892 UTC and 09:17:34.708 UTC, for a total exposure time of 5.48 hr, which is $\sim 1.7 \times$ the transit duration. The observation was made in the bright object time series (or BOTS) mode with the F290LP filter, the SUB2048 subarray, and the NRSRAPID readout pattern, with the spectra dispersed over two detectors: NRS1 and NRS2. The two detectors  span wavelength ranges of 2.73-3.72 $\mu$m and 3.82-5.17 $\mu$m, respectively, with a gap between them at 3.72-3.82 $\mu$m. The G395H grating offers the highest resolution mode of NIRSpec, with R$\sim$2700. The spectroscopic time series observation consisted of 2726 integrations, with seven groups per integration.   The host star was too bright for direct target acquisition, so another nearby target, 2MASS 10265697-5227497, within the splitting distance of the science target was used for this purpose.

The second observation was conducted using the NIRISS SOSS instrument mode.  The main science exposure lasted from 2025 January 21 20:38:57.231 UTC to 2025 January 22 07:13.221 UTC, totaling an exposure time of 5.48 hr. The observation used the GR700XD grism (R$\sim$700), the CLEAR filter, the SUBSTRIP256 subarray, and the NISRAPID readout pattern, providing wavelength coverage of $\sim$ 0.8–2.8 $\mu$m for the first spectral order (R$\sim$700) and $\sim$ 0.6-1.4 $\mu$m for the second order (R$\sim$1400) spectrum. The exposure consisted of 1195 integrations, with two groups per integration. An additional
short exposure was obtained with the F277W filter, consisting of 10 integrations and 12 groups per integration. Target acquisition was achieved on the main source.

\begin{figure*}
\begin{center}  
        \includegraphics[width=0.49\textwidth]{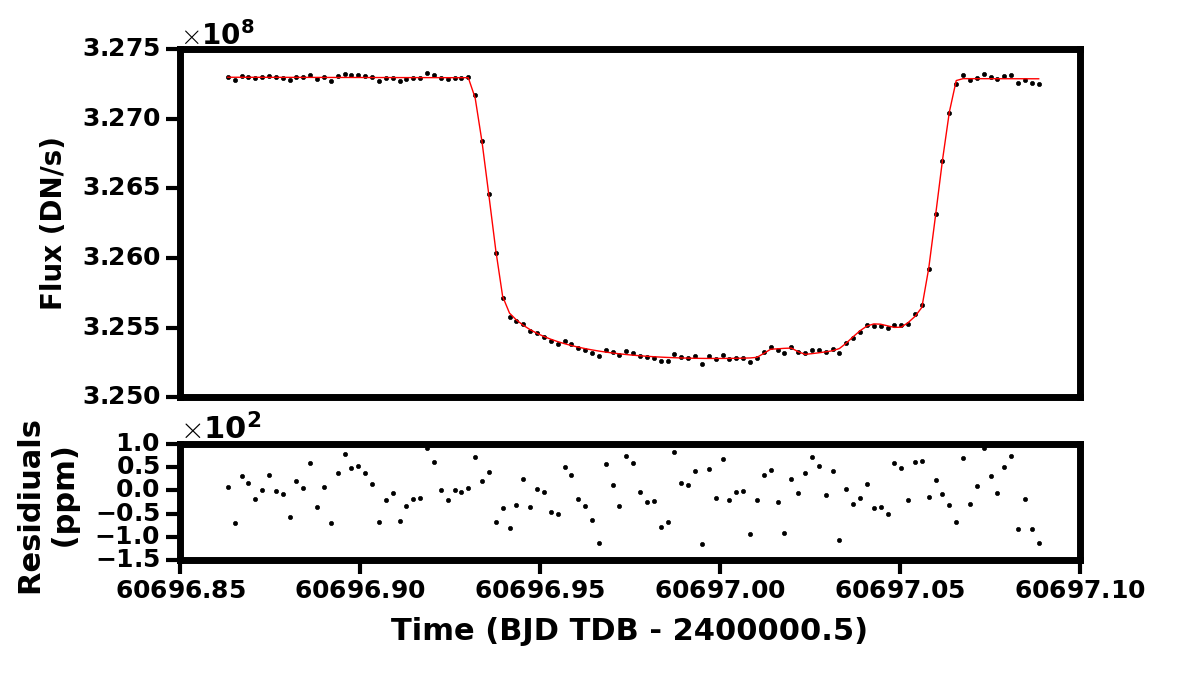}
    \includegraphics[width=0.49\textwidth]{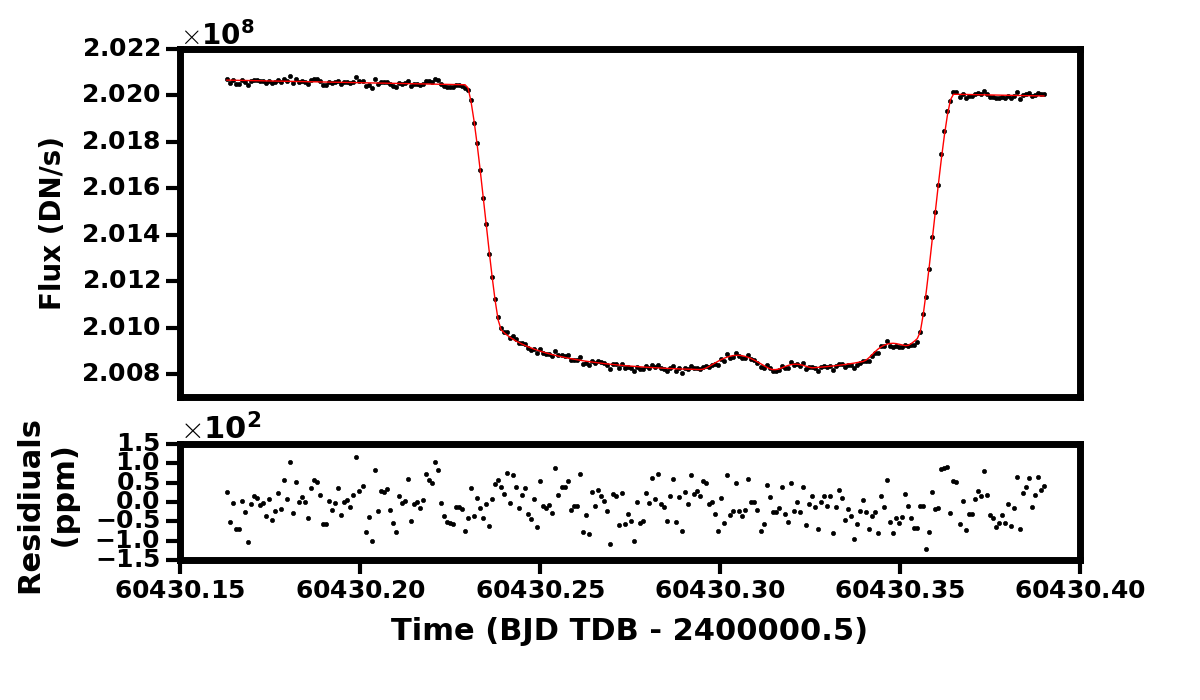}   
\end{center}
    \caption{White light curves for transits of TOI-1231 b observed with two JWST instruments at different epochs. Left: NIRISS.  Right: NIRSpec.  In each case top panel shows light curve with best fit model (red line), while the bottom panel shows the residuals.  Spot crossing events are evident in both the light curves.  The white light curves shown here are from the \texttt{JexoPipe} reduction and are binned to one point every 10 integrations.
    } 
    \label{fig:wlc} 
\end{figure*}

\subsection{NIRISS Data Reduction}

We reduce the NIRISS data using two independent pipelines: a NIRISS version of \texttt{JexoPipe} \citep{Sarkar2024} and the latest version of the \texttt{JExoRES} pipeline \citep{holmberg2023, Madhusudhan2023K2-18b}.
In \texttt{JexoPipe}, we start with the .uncal files and apply the following Stage 1 steps from official JWST calibration pipeline: data-quality array initialization, saturation flagging, superbias subtraction, reference pixel correction, linearity correction, ramp fitting, and gain scaling. Between the
reference pixel correction and linearity correction steps, we apply a 1/f noise subtraction step. For each group image, we subtract a group-scaled background model (from commissioning program 1541) and then obtain a difference image using a background-subtracted group-wise median (modulated by the white light curve).  This reveals the 1/f noise.  We then `deband' the science images by subtracting the column-wise medians of the corresponding difference image. The sky background is added back before proceeding to linearity correction.

In Stage 2, we combine all segments, apply the assign World Coordinate System (WCS) and flat-field steps from the official pipeline, and then apply custom bad pixel flagging, background subtraction and bad-pixel correction steps.  The latter consists of a combination of temporal and spatial interpolation to fill in bad-pixel values \citep{Sarkar2024}. 
We then further correct 5$\sigma$ outliers using a rolling median image.  The wavelength solution is obtained using PASTASOSS \citep{Baines2023b}. In Stage 3, we perform a custom zeroth order subtraction, utilizing the F277W image, followed by order separation.  An aperture of 37 pixels is centered on each trace, and box extraction of the 1-D stellar spectra performed for each order.  The pipeline is run twice, with the white light curve from the first run used to modulate the background-subtracted group-wise medians in the second run.

The \texttt{JExoRES} pipeline proceeds as follows.
Starting with the raw .uncal files, Stage 1 proceeds using a combination of official JWST calibration pipeline steps, including saturation flagging, superbias subtraction, linearity correction, and fitting of the group-level ramp, and customized steps. We omit the jump step due to there being only two groups per integration. We apply a custom 1/f noise subtraction step before linearity correction.  Briefly, this involves subtraction of a model of the star (modulated by the white light curve) and sky background from all group-level images to reveal the column-wise banding due to 1/f noise, which is subsequently subtracted from the original science images.  Stage 2 takes the integration-level ramp images from Stage 1 and applies the flat-field correction.  

We then proceed to Stage 3,  where we first search for outliers in the time series of each pixel, using a rolling median (with a window size of 7 integrations) and sigma clipping (5$\sigma$). Next, we perform order tracing and wavelength calibration using PASTASOSS. We then refine the order tracing according to \cite{holmberg2023}. We perform a custom background subtraction to remove the sky background in a two-step process, as described in \citet{Madhusudhan2023K2-18b}.  Moreover, there was significant contamination from field stars, particularly above 2.4 $\mu$m. For this reason, we did not consider the spectrum past 2.4 $\mu$m (this is also the case for the \texttt{JexoPipe} spectrum). We use box extraction with an aperture of 35 pixels to obtain the spectrum. We consider both order 1 and order 2 spectra. We then searched for additional outliers in the light curves of each spectral channel, which we removed from further analysis. The pipeline is iterated updating the white light curve and background model for the 1/f step until convergence is achieved.

\begin{figure*}
    \includegraphics[width=\textwidth]{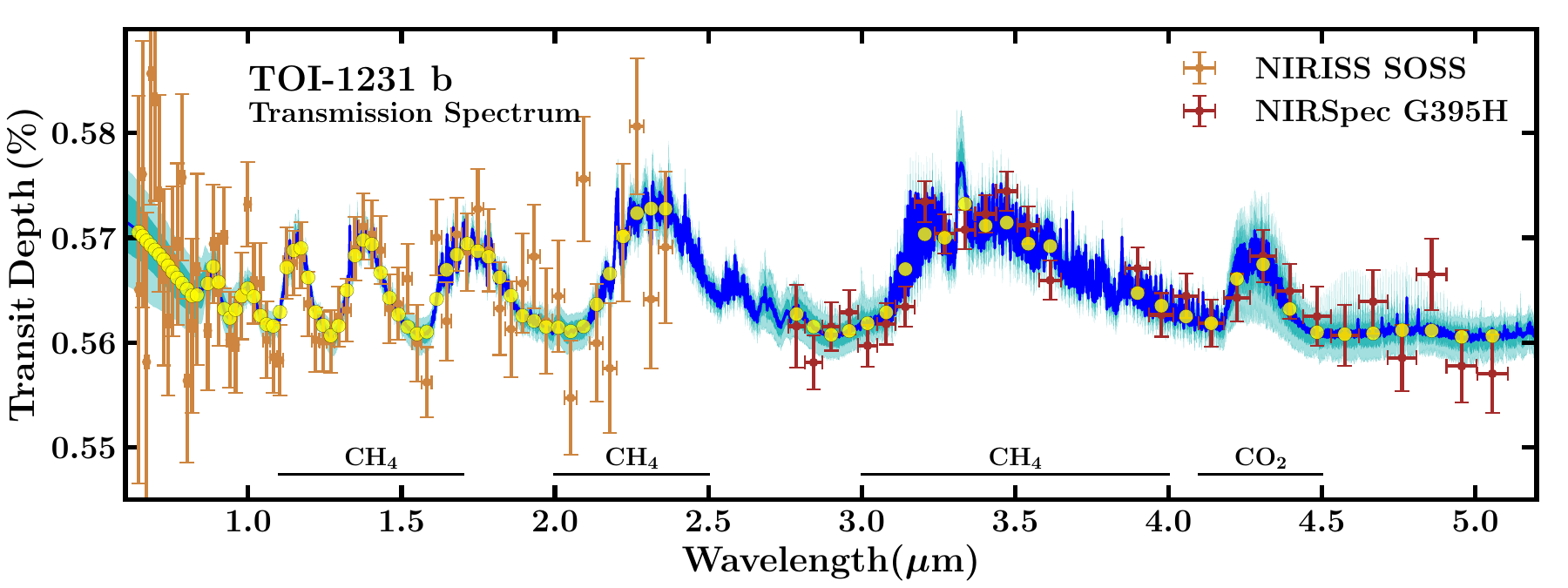}
    \caption{The transmission spectrum of TOI-1231~b. The observed JWST spectrum and retrieved model fits are shown for the fiducial retrieval using the \texttt{JexoPipe} data, considering the six prominent CNO molecules, with a single offset and no stellar heterogeneity, as discussed in Section~\ref{sec:Retrievals}. For visual clarity, the data shown here have been binned to R$\sim$50. A retrieved offset of 79.38 ppm has been added to the NIRSpec data. The dark blue line indicates the median retrieved spectrum, with the shaded regions indicating the 1$\sigma$ and 2$\sigma$ uncertainties. The yellow points show the median retrieved spectrum binned at the same resolution as the plotted data.
    }
    \label{fig:spectrum} 
\end{figure*}

\subsection{NIRSpec Data Reduction}

We again use \texttt{JexoPipe} and \texttt{JExoRES}  to independently reduce the NIRSpec data.
In \texttt{JexoPipe},  Stage 1 begins with the raw group-level .uncal files.  We apply the standard pipeline steps: data-quality array initialization, superbias subtraction, saturation flagging, linearity correction, and dark current subtraction.  We then apply a custom group-level background subtraction step to remove 1/f noise and sky background. We then apply the standard jump-detection, ramp fitting and gain-scaling steps. In Stage 2, we first combine all the data segments and then apply the assign WCS step. As for NIRISS, we apply the custom bad-pixel flagging, bad-pixel correction, and rolling-median outlier detection step.
We complete Stage 2 by applying the official 2-D extraction and wavelength-correction steps.  In Stage 3, we extract the 1-D stellar spectra.  We apply an aperture of 9 pixels centered on the spectral trace.   We then apply an optimal extraction algorithm \citep{horne_optimal_1986} to obtain the 1-D stellar spectrum per integration. 

For \texttt{JExoRES}, we perform Stage 1, beginning with the .uncal files and applying 
superbias subtraction, saturation flagging, reference pixel correction, linearity correction, dark-current subtraction and jump-detection (5$\sigma$ threshold) steps.  Next, a group-level background subtraction is performed by masking the spectral trace ($\pm10$ pixels from the center), bad pixels, and outliers, and then obtaining the mean in each pixel column.  This mean value is then subtracted from all the pixels in that column.  Finally, ramp fitting is performed.  Stage 2 assigns the wavelength solution.  Flat-fielding is omitted.  In stage 3, there is another outlier detection step followed by optimal extraction  of the 1-D stellar spectra.  We discard channels that have $>20\%$
of their flux masked.
 
\begin{table*}
\centering
\begin{tabular}{lcccccc}
\hline \hline
Parameter & NIRISS & NIRSpec\\ \hline
Planet-to-star radius ratio, $R_\mathrm{p} / R_\mathrm{s}$  & $0.07509^{+0.00011}_{-0.00017}$ & $0.07464^{+0.00014}_{-0.00011}$\\ 
Mid-transit time, $t_0$ (BJD - 2400000.5)  & $60696.998167_{-0.000034}^{+0.000024} $   & $60430.297108_{-0.000015}^{+0.000013}$ 
\\
Inclination, $i$ ($^{\circ}$) &$89.769^{+0.034}_{-0.019} $&  $89.955^{+0.010}_{-0.014}$\\ 
Normalised semi-major axis, $a / R_\mathrm{s}$ & $59.65^{+0.46}_{-0.26}$  & $61.371^{+0.041}_{-0.057}$\\ 
First LDC, $u_1$&   $0.147^{+0.018}_{-0.022}$&  $0.111^{+0.011}_{-0.013}$\\ 
Second LDC, $u_2$ & $0.214^{+0.040}_{-0.030}$&  $0.145^{+0.022}_{-0.019}$ \\

\hline\hline
\end{tabular}
\caption{Parameter estimation from the \texttt{JexoPipe} white light curve analyses of our NIRISS SOSS and NIRSpec G395H observations of TOI-1231 b.}
\label{tab:wlc_params}
\end{table*}

\subsection{White Light Curves}
\label{WLC_fitting}

For both the NIRSpec and NIRISS observations, we see signs of spot occultations in the white light curves (Figure \ref{fig:wlc}). 
We process white light curves from the \texttt{JexoPipe} reduction as follows. 
We obtain the NIRISS order 1 white light curve between 0.8-2.0 $\mu$m.  We fit the light curve using a model consisting of a transit model generated with \texttt{pylightcurve} \citep{tsiaras_2016_pylightcurve}, a linear trend and a spot-correction factor.  The latter is defined as the time-dependent ratio between the semi-analytical \texttt{SPOTROD} model \citep{Beky2014} with spots and the corresponding spotless model, and is
multiplied into the analytic transit model.  We use a custom-modified, fully Python-based implementation of the \texttt{SPOTROD} algorithm.  Based on an analysis of the residuals from spotless transit fits, we model the light curve using two spots and a single spot contrast.   We identify outliers on the white light curve as points lying beyond $\pm 5\sigma$ from a rolling median. For those integrations, we replace the 1-D stellar spectra using a
linear interpolation of spectra from adjacent integrations. We use \texttt{DYNESTY} \citep{Speagle2020} with nested sampling to fit for mid-transit time ($t_0$), planet-to-star radius ratio ($R_\mathrm{p}/R_\mathrm{s}$), normalized semi-major axis ($a/R_\mathrm{s}$), inclination angle ($i$), two quadratic limb-darkening coefficients (LDCs) ($u_1$ and $u_2$) (using the \citealp{Kipping2013} parameterization), and two parameters describing the linear trend.  We additionally fit for two position coordinates and size of each spot, along with a single spot contrast.  To facilitate the computation, we bin the light curve
to one point every 10 integrations.
Prior to fitting, we inflate the light curve errors such that the average error-bar size matches the standard deviation of the out-of-transit residuals. We perform this fit 20 times and adopt result with the lowest chi-squared value for the final analysis.

For NIRSpec, we construct the white light curve by combining data from the NRS1 and NRS2 detectors.  We use the same model set-up and fitting routine as described for the NIRISS white light curve. Based on analysis of residuals from spotless transit fits, we fit for three spots with a single spot contrast.  In all these cases, we keep the orbital period fixed to 24.245586 days \citep{Burt2021} and assume an eccentricity of zero. 
The final fitted parameter values are given in Table \ref{tab:wlc_params}. 

For \texttt{JExoRes}, we model the NIRISS order 1 and NIRSpec white light curves using \texttt{batman} \citep{kreidberg_batman_2015}. We fit for the same parameters as above, except for the spot parameters, which are not included because the spot-crossing events are instead masked. The masked integrations are listed in Section~\ref{SLC_fitting}. We additionally include an error-inflation parameter, which is fit simultaneously with the other model parameters. Posterior distributions are sampled using \texttt{MultiNest} \citep{Feroz2009}. The resulting parameters are consistent with those obtained from the \texttt{JexoPipe} analysis.

\subsection{Spectral Light Curves}
\label{SLC_fitting}

For the \texttt{JexoPipe} NIRISS data, we obtain a low-resolution transmission spectrum by binning the pixel-level light curves to R$\sim$50. In our model, we fix $t_0$, $a/R_s$ and $i$ to the  white light values.  We also fix all spot parameters, except for the spot contrast, to the white light values.  The spectral light curves are not binned in time.  We use \texttt{DYNESTY} to fit for $R_\mathrm{p}/R_\mathrm{s}$, $u_1$, $u_2$, spot contrast, two linear-trend parameters, and an error-inflation parameter.  We then derive a high-resolution transmission spectrum at the 2-pixel level as follows.  We perform a non-linear least-squares fit using the Levenberg–Marquardt algorithm implemented via  \texttt{LMFit} \citep{Newville2025}.  We fit for $R_\mathrm{p}/R_\mathrm{s}$ and two linear-trend parameters, while fixing $u_1$, $u_2$ and the spot contrast to the values from the corresponding low-resolution bin.

For the \texttt{JExoRES} NIRISS data, the spectra are binned to R $=$ 50. We fix $t_0$, $a/R_\mathrm{s}$ and $i$ to the values from the \texttt{JExoRES} white light curve fit.
We fit each light curve using \texttt{MultiNest} \citep{Feroz2009}, fixing the system parameters and masking the two spot crossings. Specifically, we mask integrations 765-850 and 895-1045. We fit for $R_\mathrm{p}/R_\mathrm{s}$, the two LDCs, the two linear-trend parameters, and an error inflation parameter. For both \texttt{JexoPipe} and \texttt{JExoRES},  we consider wavelengths between 0.65-0.85 $\mu$m from the second order spectrum and 0.85-2.4 $\mu$m from the first order spectrum.

For the \texttt{JexoPipe} NIRSpec data, we first bin the light curves to R$\sim$100 to obtain an initial low-resolution spectrum. We apply the same model used for the white light curve, fixing $t_0$, $a/R_s$ and $i$ to the \texttt{JexoPipe} NIRSpec white light values.  We also fix all spot parameters, except for the spot contrast, to the white light values. The spectral light curves are not binned in time.  Using \texttt{DYNESTY}, we fit for $R_\mathrm{p}/R_\mathrm{s}$, $u_1$, $u_2$, two linear-trend parameters and the spot contrast.
Finally, for the spectral retrievals, we obtain high-resolution transmission spectra (at the 4-pixel and 2-pixel levels), using \texttt{LMFit}.  In these fits, $u_1$, $u_2$ and spot contrast are fixed to the values for the corresponding low-resolution bin.  Each light curve is fit for $R_\mathrm{p}/R_\mathrm{s}$ and two linear-trend parameters.  

For the \texttt{JExoRES} NIRSpec data, we first
bin the light curves to R$\sim$20 to fit for the LDCs at low resolution, while fixing the system parameters $t_0$, $a/Rs$ and $i$ to the values from  the corresponding white light curve fit.
As for the \texttt{JExoRES} NIRISS data reduction, we mask the spot-crossing events. Specifically,     this involved masking integrations 1580-1950 and 2130-2370. We then bin the light curves at the 2-pixel level and fit them using the Levenberg–Marquardt algorithm, while keeping the LDCs fixed to the values from the corresponding R$\sim$20 bin. This procedure yields the high-resolution \texttt{JExoRES} spectrum used for the retrieval analysis.

\begin{figure*}
    \includegraphics[width=\textwidth]{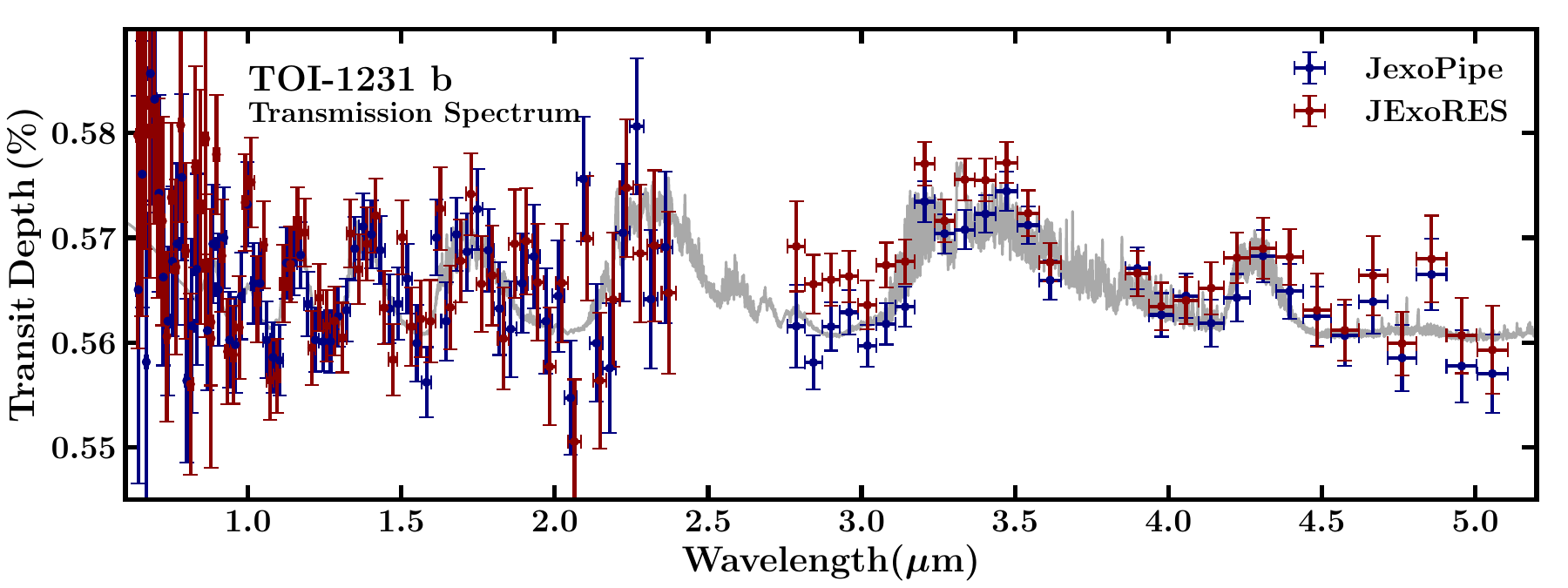}
    \caption{ Transmission spectra for NIRISS and NIRSpec from the two pipelines used in this study: \texttt{JexoPipe} and \texttt{JExoRES}.  The spectra are binned to R$\sim$50 for visual clarity. The grey line indicates the median retrieved spectrum from Figure~\ref{fig:spectrum}. The corresponding offset of 79.38 ppm has been added to NIRSpec data from both reduction pipelines for ease of comparison.}
    \label{fig: JexoPipe_JExoRES_spectra}
\end{figure*}

The combined transmission spectrum from NIRISS and NIRSpec using \texttt{JexoPipe} is shown in Figure \ref{fig:spectrum} together with a retrieved model (as described further in Section \ref{sec:Fiducial cases}).
Transmission spectra from each reduction are shown for comparison in Figure \ref{fig: JexoPipe_JExoRES_spectra}.  For NIRISS, the final transmission spectrum uncertainties at R$\sim$50 and $\sim$1.3 $\mu$m are $\sim$1.5-1.6 $\times$ larger than the predictions from \texttt{Pandexo} \citep{Batalha2017}.  For NIRSpec at R$\sim$50 and $\sim$4 $\mu$m, the uncertainties are $\sim$ 1.2-1.3 $\times$ the \texttt{Pandexo} estimate. We attribute this excess to uncertainties arising from the fitting process and residual noise and systematics which are not accounted for in \texttt{Pandexo}.  These include the processing of 1/f noise and contaminants in NIRISS, as well as uncertainties associated with spot modeling and decorrelation. 

As an additional consistency check, we also run light curve fits on the \texttt{JexoPipe} data, masking out the same integrations as used in the  \texttt{JExoRES} fits.  For these, we apply a similar process to the spotted-model fits, except that we do not apply a spot model and use Markov Chain Monte Carlo sampling with \texttt{emcee} \citep{foreman-mackey_emcee_2013} for the white light curve and low-resolution spectral light curve fits.

\section{Atmospheric Retrievals} \label{sec:Retrievals}

We retrieve the atmospheric properties at the day-night terminator of TOI-1231~b using the near-infrared transmission spectrum in the 0.65-5.2 $\mu$m range derived above. The spectrum spans NIRISS order 1 and order 2, as well as NIRSpec NRS1 and NRS2, as discussed above. We begin with a fiducial retrieval scenario, including six prominent CNO chemical species that may be expected in temperate H$_2$-rich atmospheres, along with a full treatment of clouds/hazes, pressure-temperature ($P$-$T$) profile, stellar heterogeneity, and offsets between detectors. We then systematically explore the potential presence of additional species in the atmosphere of TOI-1231~b.  We do so  by first applying an expanded set of 11 molecules used in previous studies of temperate sub-Neptunes \citep{Madhusudhan2023K2-18b, holmberg_possible_2024}.
We then extend the search to a suite of 201 species, listed in Appendix \ref{sec:app_list}, including both simple molecules and complex organic compounds, considered individually. 
Finally, we present the atmospheric constraints derived for TOI-1231~b.

\subsection{Retrieval Set-up}
\label{sec:retrievals/setup}
We conduct atmospheric retrievals using the \texttt{POSEIDON} retrieval code \citep{macdonald2017, macdonald_poseidon_2024},
treating the H$_2$-dominated atmosphere as plane-parallel and in hydrostatic equilibrium. \texttt{POSEIDON} uses \texttt{PyMultiNest} \citep{Buchner2014} to implement \texttt{MultiNest} \citep{Feroz2009} for Bayesian inference and parameter estimation, given the priors for each model parameter. We adopt 500-1000 live points and a sampling efficiency of 0.3 for all our retrieval runs. 
We include as free parameters the abundances of select chemical species (with the remaining gas assumed to be H$_2$/He in solar proportions) as well as parameters describing the $P$-$T$ profile and cloud/haze properties. For the fiducial cases, we parameterize the $P$-$T$ profile following \citet{Madhusudhan2009}, which includes six free parameters. Clouds and hazes are treated as inhomogeneous, including a gray cloud deck and hazes following the \citet{pinhas2019} parameterization. This leads to the following cloud/haze free parameters: a cloud-top pressure ($P_{\rm c}$), a Rayleigh enhancement factor for the hazes ($a$), a scattering slope ($\gamma$) such that the scattering cross-section $\sigma \sim a \lambda^\gamma$, and a terminator coverage fraction for both clouds and hazes ($\phi$). The reference pressure ($P_{\rm ref}$), corresponding to the white light radius inferred from the NIRISS observations (as discussed in Section \ref{WLC_fitting}), is also retrieved. 

To allow for possible offsets between different detectors \citep[e.g.][]{Madhusudhan2023K2-18b, moran_high_2023}, we consider three cases of different combinations of offsets: no offsets, one offset for the whole NIRSpec spectrum relative to NIRISS ($\delta_{\rm NIRSpec}$), and separate offsets between NIRSpec NRS1 and NRS2 with respect to the NIRISS ($\delta_{\rm NRS1}$ and $\delta_{\rm NRS2}$). This approach follows previous analyses of JWST transmission spectra from NIRISS and NIRSpec G395H for sub-Neptunes \citep{Madhusudhan2023K2-18b, rigby_jwst_2025}. Furthermore, we also consider the possibility of unocculted stellar spots/faculae being present by adding three free parameters to the model: the photospheric temperature ($T_{\rm phot}$), the temperature of the heterogeneity ($T_{\rm het}$), and the fraction of the stellar disk covered by the heterogeneity ($f_{\rm het})$. This results in a total of 11-16 free parameters excluding the molecular mixing ratios, depending on the treatment of offsets and stellar heterogeneity treatment. We report our priors in Table \ref{tab:priors} in Appendix \ref{sec:app_priors}.

We use the cross-sections included in the native \texttt{POSEIDON} v1.2 opacity database when available. In particular, these cross-sections are obtained from the following line lists: \citet{Polyansky2018} for H$_2$O, \citet{yurchenko_exomol_2024} for CH$_4$, \citet{Li2015} for CO,  \citet{yurchenko_exomol_2020} for CO$_2$, \citet{Coles2019} for NH$_3$,  \citet{Barber2014} for HCN, \citet{HITRAN2020} for CH$_3$Cl and CS$_2$, \citet{owens_exomol_2024} for OCS, and \citet{hargreaves_spectroscopic_2019} for N$_2$O. Cross-sections based on the \citet{HITRAN2020} line list were also used for HC$_3$N. For all other species, cross-sections at $P = 1$ bar and $T = 300$ K (or the closest available) are obtained from the \texttt{HITRAN} database  \citep{HITRAN2020}. We use \texttt{ExoMol} cross-sections for HCl \citep{li_direct_2011}, KCl \citep{barton_exomol_2014}, HF \citep{hill_temperature-dependent_2013, li_reference_2013, coxon_improved_2015}, H$_2$CS \citep{mellor_exomol_2023}, NaOH and KOH \citep{owens_exomol_2021},  as these are not available in the \texttt{HITRAN} or original \texttt{POSEIDON} database. We also include collision-induced absorption for H$_2$-H$_2$, H$_2$-He, H$_2$-CH$_4$, CO$_2$-H$_2$ and CO$_2$-CH$_4$ \citep{karman_update_2019}. Our forward models are computed on a grid including 25000 equally spaced wavelength points between 0.5 $\mu$m - 5.2 $\mu$m, before being binned to the resolution of the observed spectrum. We consider this set-up for each of the three offset scenarios described above.

\begin{figure*}
	\includegraphics[width=\textwidth]{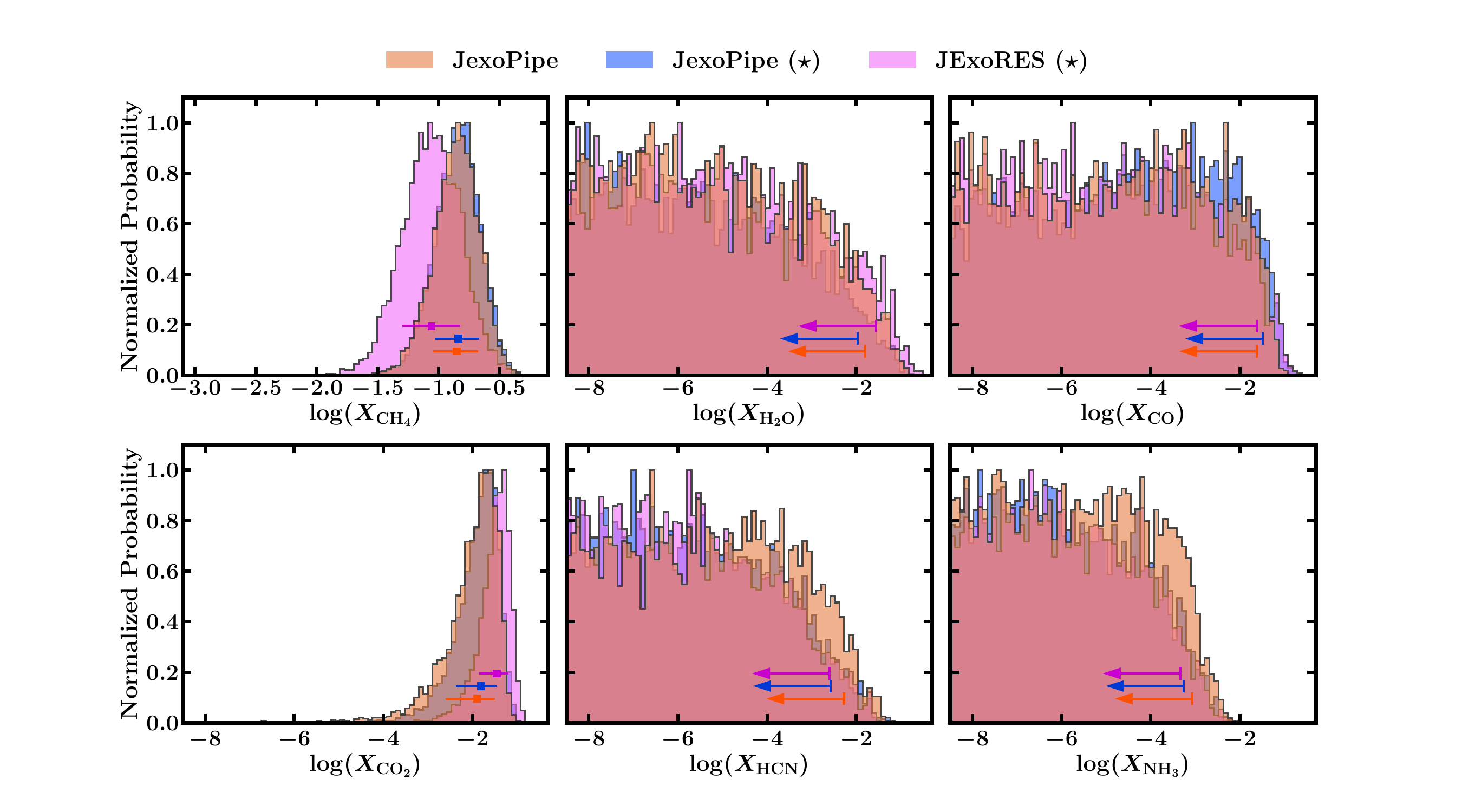}
 \caption{Posterior distributions for the abundances of the six molecular species included in our fiducial retrievals with a single offset, both with (indicated by a $\star$ symbol) and without stellar heterogeneity for \texttt{JexoPipe}, and with stellar heterogeneity included for \texttt{JExoRES}. 
 The median retrieved values and 1$\sigma$ error bars are shown for CH$_4$ and CO$_2$, while arrows indicate the 95\% upper limits for all other species.}
\label{fig:6mol_posteriors}
\end{figure*}

\subsection{Fiducial Cases}
\label{sec:Fiducial cases}

We begin with our fiducial case which includes six prominent CNO species, which may be expected in a temperate Neptune-like atmosphere: CH$_4$, CO$_2$, CO, H$_2$O, NH$_3$, and HCN \citep{madhusudhan_exoplanetary_2016, Hu2021b, Hu2021, Madhusudhan_chem_2023}.  For robustness, we first use the  \texttt{JexoPipe} spectrum and then validate the results using the \texttt{JExoRES} spectrum.

\subsubsection{JexoPipe Data}
\label{subsec:fiducial case}

 We start by adopting the  \texttt{JexoPipe} 2-pixel data for NIRSpec and the R$\sim$50 NIRISS data. Here, we assess the molecular abundances for each molecule using retrievals under different assumptions for offsets (i.e. zero, one or two offsets) and stellar heterogeneity. We summarize our results in Tables \ref{tab:abundances} and \ref{tab:physical_properties}. We report the logarithmic Bayes factor for the model preference as $\ln B$. We estimate a typical uncertainty on the Bayes factor of $\Delta \ln B \sim \pm 0.5$.

We find strong evidence for at least one offset being present. In particular, the one-offset scenario is preferred at $\ln B = 12.0$ over the no-offset case and at $\ln B = 1.8$ over the two-offset case. This is consistent with several previous analyses of NIRISS and NIRSpec data \citep[e.g.][]{Madhusudhan2023K2-18b, holmberg_possible_2024, rigby_jwst_2025,constantinou_atmospheric_2026}. In none of the offset cases considered do we find substantial evidence for stellar heterogeneity. In particular, stellar heterogeneity is disfavored in both the zero-offset ($\ln B = -2.6$) and one-offset ($\ln B = -2.0$) cases, and only weakly preferred ($\ln B =1.2$) when two offsets are adopted. We note that this is likely  the result of the star spot correction process applied during the data reduction stage (Section \ref{sec:obs}). As a result of the above, we adopt the one-offset case without stellar heterogeneity as the primary case for the \texttt{JexoPipe} dataset. 

We obtain strong evidence for abundant CH$_4$ across all cases considered. In particular, in our primary case,
we obtain a model preference at $\ln B = 69.6$ for CH$_4$ being present with a median abundance $\log (X_{\rm CH_4}) =-0.85^{+0.18}_{-0.19}$. We also find strong evidence for CO$_2$, at $\ln B = 6.1$, in the primary one-offset case, at $\sim 1 \%$ abundance,  $\log (X_{\rm CO_2})=-1.90^{+0.40}_{-0.71}$. However, the evidence for CO$_2$ decreases into the moderate category, to $\ln B = 2.9$, when two offsets are allowed. The model preference values for these two species do not change significantly when stellar heterogeneity is included.
No other species is detected, although several result in comparatively high upper limits. In our primary case, up to $\log X_{\rm CO}<-1.92$ is allowed by the 95\% upper limit, with a comparable 95\% upper limit being found for water ($\log X_{\rm H_2O} < -2.24$) and HCN ($\log X_{\rm HCN} < -2.60$). We show posterior distributions for the abundances of the six species in Figure \ref{fig:6mol_posteriors}.
Relatively high abundances of NH$_3$ are also allowed, with an upper limit  $\log (X_{\rm NH_3}) <-3.34$.   
The retrieved abundances or 95\% upper limits for all six molecules in the fiducial model, across all offset cases (with and without stellar heterogeneity), are listed in Table \ref{tab:abundances}.

We discuss other constraints on the atmospheric properties in Section~\ref{subsec:AtmosphericConstraints}. In particular, we find moderate-to-strong preference for the inclusion of clouds/hazes in the model, depending on the treatment of stellar heterogeneity. We find no preference for the consideration of a non-isothermal $P$-$T$ profile ($\ln B = 0.1$) over an isothermal photosphere. 

\subsubsection{JExoRES Data}
\label{sec:jres}
As a robustness check on the above results, we verify whether they are sensitive to the choice of data reduction pipeline. To do so, we use our second dataset, obtained with \texttt{JExoRES} (again with 2-pixel resolution in NIRSpec and R$\sim$50 binning in NIRISS). We note that a core difference between the \texttt{JexoPipe} data reduction and that considered here is the treatment of occulted star spots. In particular, while in the \texttt{JexoPipe} data (adopted for Section \ref{subsec:fiducial case}) these were modelled and corrected for at the light curve fitting stage, in the \texttt{JExoRES} dataset the spots are masked out of the light curves (Section \ref{sec:obs}). It is thus unsurprising that, in the one-offset case we consider, the \texttt{JExoRES} dataset results in moderate ($\ln B = 3.3$) evidence for stellar heterogeneity, in contrast to the result obtained with \texttt{JexoPipe}. As a further consistency check, we also analyse the \texttt{JexoPipe} dataset using the same masking prescription as \texttt{JExoRES}, which yields consistent results, with stellar heterogeneity preferred at $\ln B = 2.2$. 

When considering the \texttt{JExoRES} reduction, our main results, namely evidence for CO$_2$ at percent-level abundance and CH$_4$ at $\sim 10\%$, are confirmed regardless of the treatment of stellar heterogeneity, as shown in Table \ref{tab:abundances}. Their significances are also comparable to those found in Section \ref{subsec:fiducial case}, at $\ln B = 4.4$ to $6.0$ for CO$_2$ and $\ln B = 54.5$ to $55.4$ for CH$_4$, with the stronger significances obtained when stellar heterogeneity is included. While weak peaks in the posteriors of NH$_3$ and H$_2$O are visible when stellar heterogeneity is not included, they are not apparent when it is considered. The upper limits when stellar heterogeneity is included are also similar to those in Section \ref{subsec:fiducial case}, as shown in Table \ref{tab:abundances}. Overall, the retrieved properties using data from both pipelines are consistent.

\vspace{1cm}
\subsection{Exploration of Additional Trace Species}

\label{subsec:additional_species}

While our fiducial six-molecule, one-offset model already provides a good fit to the combined NIRISS and NIRSpec data (Figure \ref{fig:spectrum}) with no statistically significant residuals, we perform retrievals including an expanded range of molecules to maintain methodological consistency with previous analyses \citep[e.g.][]{Madhusudhan2023K2-18b, holmberg_possible_2024, pica-ciamarra_systematic_2025, rigby_jwst_2025, constantinou_atmospheric_2026}.  In addition, these serve as further robustness tests,  confirming that the fiducial primary model remains preferred even under a more extended molecular hypothesis, as no additional absorbers achieve significant increases in the Bayesian evidence obtained.  

First, we apply an expanded fiducial model following the set-up described in \cite{Madhusudhan2023K2-18b} (hereafter \citetalias{Madhusudhan2023K2-18b}), again using the priors in Table \ref{tab:priors} in Appendix \ref{sec:app_priors}, and starting with the \texttt{JexoPipe} data binned to 2 pixels for NIRSpec and R$\sim$50 for NIRISS. This set-up includes the six prominent CNO molecules discussed in Section \ref{subsec:fiducial case}, along with five additional species: DMS, CS$_2$, CH$_3$Cl, OCS and N$_2$O. Again, we adopt a one-offset scenario for this analysis. Consistent with the findings in Sections \ref{subsec:fiducial case} and  \ref{sec:jres}, stellar heterogeneity is included only when analysing the \texttt{JExoRES} data.

While TOI-1231 b possesses a deep H$_2$-rich atmosphere and is not generally expected to be habitable \citep[but see][]{Seager2021}, we include these five trace species  that have been proposed as potential biosignature gases \citep{seager2016}, with the consideration that abiotic production may also be possible in certain contexts \citep[e.g.][]{Hu2025}. \cite{Seager2021} discuss the possibility of aerial biospheres  and include TOI-1231 b as a potential candidate.
However, the main motivation for including these species here is to testpotential abiotic formation pathways for any of these species, such as DMS production mechanisms proposed in \citealp{Hu2025} and  expected photochemical formation of CS$_2$ in temperate H$_2$-rich atmospheres \citep{Zahnle2009soot,Moses2024}, and to ensure methodological consistency with previous retrieval studied of temperate sub-Neptunes, such as K2-18 b \citep{Madhusudhan2023K2-18b}. 

We confirm, using the \texttt{JexoPipe} data, the strong evidence for CH$_4$ found with our fiducial primary case, albeit at somewhat lower levels, $\ln B = 22.4$, while the model preference for CO$_2$ decreases to $\ln B = 3.1$. Comparable significances of $\ln B = 19.3$ and $\ln B =5.1$, respectively, are obtained when performing the same retrieval on the \texttt{JExoRES} data. Including the five species does not result in a significant difference in Bayesian evidence relative to our primary six-molecule case, with Bayes factors $\ln B = -0.1$ and $\ln B = -0.5$ for the expanded model with \texttt{JexoPipe} and \texttt{JExoRES}, respectively, compared to the fiducial model. For the six CNO species already considered above, we find abundance estimates and upper limits consistent with the fiducial primary case. We show the posterior distributions for the abundances of all 11 species considered, as well as the temperature at 1 $\mu$bar, in Figure \ref{fig:compareK218}.

Next, we explore whether there is any significant evidence for additional species that we have not considered thus far. We proceed with an agnostic exploration of the \texttt{JexoPipe} data, following the canonical-model approach described in \citet{Madhusudhan2025K2-18b} and \citet{pica-ciamarra_systematic_2025}. The above retrievals resulted in robust inferences for CH$_4$ and CO$_2$, and led to comparatively high upper limits for NH$_3$ and CO, which are also expected from theoretical predictions \citep[e.g.][]{Hu_photo21,  Rigby2024_gasdwarf, wogan_jwst_2024}. Therefore, we proceed with retrievals including these four molecules in the model, along with an additional species tested one at a time. Furthermore, we assume an isothermal temperature profile, given that we find no meaningful preference for the full $P$-$T$ profile over an isothermal profile in the fiducial case, and no stellar heterogeneity, as no evidence for it is found in the \texttt{JexoPipe} dataset (Section \ref{subsec:fiducial case}).

This 4+X set-up for a canonical model in the present case is a higher-dimensional variant of the 2+X set-up previously considered in recent studies \citep{Madhusudhan2025K2-18b,welbanks_challenges_2025, pica-ciamarra_systematic_2025}, and was adopted -- albeit with a different baseline set -- in \citet{rigby_jwst_2025}. As those studies highlight, this set-up is appropriate for identifying signs of potential excess absorption consistent with a given species. However, it is not sufficient to claim definitive evidence of a species' presence without complementary assessment of physical plausibility and potential degeneracies \citep{welbanks_challenges_2025, pica-ciamarra_systematic_2025}. 

In this work, `X' denotes any one of 201 species added individually, spanning hydrocarbons, sulfur-bearing species, nitriles, amines, and other nitrogenated hydrocarbons, and a suite of simple molecules whose presence may not be physically implausible in the atmosphere of this planet. Following the above arguments, we construct our canonical models to include a single offset, applied to the full NIRSpec dataset. 

To minimize the computational cost of this exploration, we carry out these retrievals with 500 live points rather than 1000, and we use a version of the \texttt{JexoPipe} data in which the NIRSpec spectrum is been binned to 4 pixels, while for NIRISS we continue using the R$\sim$50 binning.  No species considered reaches a strong evidence level ($\ln B \geq 5.0$, \citealp{Trotta2008}). Two species match the threshold for moderate preference ($\ln B = 2.5$, \citealp{Trotta2008}): 3-carene ($\ln B = 3.4$) and allylamine ($\ln B = 2.8$). Furthermore, three more species reach the lower threshold of $\ln B = 2.0$, which some previous works \citep{pica-ciamarra_systematic_2025, rigby_jwst_2025} adopted as a means to account for up to $\pm 0.5$ uncertainty in the estimation of $\ln B $:  butylamine ($\ln B = 2.3$), cadaverine ($\ln B = 2.3$), and limonene ($\ln B = 2.1$). 

We further assess the reliability of these inferences by considering a 6+X set-up, in which we expand our baseline model to include all six standard CNO species (CH$_4$, CO$_2$, CO, H$_2$O, NH$_3$, and HCN). For these retrievals, we return to the higher-resolution 2-pixel binning for NIRSpec and we use 1000 live points. Of the five species listed above, only one remains above the $\ln B = 2.5$ threshold, 3-carene at $\ln B = 2.8$, with two more crossing the lower $\ln B = 2.0$ threshold (allylamine, at $\ln B = 2.3$ and limonene, at $\ln B = 2.1$). When stellar heterogeneity is also included, the significance for 3-carene increases to $\ln B = 3.2$, as does that for limonene ($\ln B = 2.6$), while the remaining species do not cross the $\ln B = 2.0$ threshold. 

Lastly, we verify whether the preference for these species may be the result of a specific data reduction pipeline. To do so, we carry out the same 6+X retrievals with stellar heterogeneity for the five species identified above on the \texttt{JExoRES} data as well. We find that none results in $\ln B \geq 2.0$ preference, with the most preferred species, now limonene, resulting in $\ln B = 1.8$.  Given both the low physical plausibility of these complex organic species in a Neptune-like atmosphere, and their evidence being dependent on the data reduction pipeline used, we conclude that there is no significant evidence, at present, for excess absorption in the atmosphere of TOI-1231~b. Future observations could further assess the potential contributions of any of the species considered in this work.

\subsection{Atmospheric Constraints}

\label{subsec:AtmosphericConstraints}

\begin{table*}[]
\centering
\begin{tabular}{l|cccccc}
\hline \hline
Offsets     & CH$_4$                       & CO$_2$                       &CO                 &H$_2$O               & NH$_3$            & HCN \\ \hline
0 (\texttt{JP}) & $-3.78^{+0.40}_{-0.33}$ (63.3)  & $-6.05^{+0.70}_{-0.66}$ (5.8) & $< -4.99$ & $<-5.99$ & $<-6.49$  & $<-6.62$  \\ \hline                          
0 ($\star$\texttt{JP}) & $-0.87^{+0.18}_{-0.25}$ (64.2)  & $-2.55^{+0.58}_{-0.80}$ (5.6) & $< -2.07$ & $<-4.30$ & $<-5.42$  & $<-5.04$  \\ \hline
1 (\texttt{JR}) &  $-1.19^{+0.25}_{-0.27}$ (54.5) & $-1.84^{+0.40}_{-0.62}$ (4.4)  & $<-2.53$   & $<-1.08$ &  $<-2.62$  & $<-2.06$\\ \hline
1 ($\star$\texttt{JR}) & $-1.06^{+0.24}_{-0.24}$ (55.4) & $-1.45^{+0.26}_{-0.39}$ (6.0) & $< -1.98$ & $<-1.97$ & $<-3.70$ & $<-3.15$ \\ \hline
1 (\texttt{JP}) &  $-0.85^{+0.18}_{-0.19}$ (69.6)        & $-1.90^{+0.40}_{-0.71}$ (6.1)  & $<-1.92$   & $<-2.24$ &  $<-3.34$  & $<-2.60$ \\ \hline
1 ($\star$\texttt{JP}) & $-0.84^{+0.17}_{-0.19}$ (68.8) & $-1.81^{+0.36}_{-0.56}$  (6.6) & $<-1.77$ & $<-2.46$ & $<-3.66$ & $<-3.03$ \\ \hline
2 (\texttt{JP}) &  $-0.88^{+0.17}_{-0.19}$  (61.0) & $-2.19^{+0.51}_{-0.93}$  (2.9)   & $<-2.14$  & $<-1.94$  & $<-2.67$  & $<-2.20$ \\ \hline
2 ($\star$\texttt{JP}) &  $-0.91^{+0.15}_{-0.15}$ (61.7) & $-2.32^{+0.51}_{-0.89}$ (2.9) & $<-2.09$   & $<-2.21$  & $<-3.18$  & $<-2.33$ \\ \hline  \hline

\end{tabular}
\caption{Retrieved median molecular abundances and 95\% upper limits in the fiducial six-molecule retrievals. The symbol $\star$ denotes that unocculted stellar heterogeneities were included in the corresponding model, while \texttt{JP} and \texttt{JR} refer to retrievals carried out on the \texttt{JexoPipe} and \texttt{JExoRES} data, respectively. The molecular abundances are given as $\log_{10}$ of the volume mixing ratios, shown for the different offset combinations described in Section \ref{sec:retrievals/setup}. When evidence is found for the presence of a given species in the atmosphere of TOI-1231~b, the median retrieved abundances are reported; otherwise, the 95\% upper limits are indicated. The model preferences in favor of including CH$_4$ and CO$_2$ are also reported (in brackets) as the Bayes factor ($\ln B$) of the full model over an identical case without the relevant species.}
\label{tab:abundances}
\end{table*}

As no robust evidence was found for additional species, we adopt the one-offset six-molecule model without stellar heterogeneity (primary case) described in Section \ref{sec:retrievals/setup} as the representative case to determine atmospheric constraints for TOI-1231~b using the \texttt{JexoPipe} spectrum. We report in Table \ref{tab:abundances} the  constraints on the chemical composition of the atmosphere using each of the \texttt{JexoPipe} and \texttt{JExoRES} cases. Here, we quote results from the retrievals carried out on the \texttt{JexoPipe} data, but we note that these are generally consistent with those obtained when using the \texttt{JExoRES} data. Most importantly, two carbon-bearing species are inferred. Strong evidence ($\ln B = 69.6$) is found for the presence of CH$_4$ in the atmosphere, detected at a log-mixing ratio of $\log( X_{\rm CH_4}) = -0.85^{+0.18}_{-0.19}$. Evidence at $\ln B = 6.1$ is also found for CO$_2$, at a lower abundance of $\log (X_{\rm CO_2}) = -1.90^{+0.40}_{-0.71}$. We find 95\% upper limits on the remaining species: $\log(X_{\rm CO}) < -1.92$, $\log( X_{\rm H_2O}) <-2.24$, $\log (X_{\rm NH_3}) < -3.34$, and $\log (X_{\rm HCN} )<-2.60$. 

Strong evidence is found for a combination of clouds and hazes, at $\ln B = 7.9$. Removing either clouds or hazes individually from the model results in a decrease of $\sim 3$ in the Bayesian evidence ($\ln Z$), indicating moderate evidence for each component independently. However, when stellar heterogeneity is included, the evidence for clouds and hazes together decreases to $\ln B = 4.1$, and the evidence for each component independently becomes weak to moderate.

When adopting a non-isothermal $P$-$T$ profile we infer a 1 $\mu$bar temperature of $333.4^{+37.3}_{-35.2}$~K. This is consistent with the result obtained when assuming an isothermal atmosphere, the temperature of which is found to be $332.6^{+39.8}_{-35.6}$ K. As mentioned in Section \ref{subsec:additional_species}, no meaningful preference for or against an isothermal $P$-$T$ profile was found. 
The priors for all cases described above are reported in Table \ref{tab:priors} in Appendix \ref{sec:app_priors}. We also summarize the above results in Table \ref{tab:retrieval_summary} in Appendix \ref{sec:app_list}.

\begin{table*}
\centering
\begin{tabular}{l|ccccccccc}
\hline \hline
Offsets      & $T_{1 \mathrm{\mu bar}}$ (K) & $\phi$                 & $\log a$               & $\gamma$                 & $\log (\frac{P_{\mathrm{c}}}{\mathrm{bar}})$ & $\log (\frac{P_{\mathrm{ref}}}{\mathrm{bar}})$ & OS1/ppm & OS2/ppm & $\ln Z$ \\ \hline

0 (\texttt{JP}) & $356.0_{-49.0}^{+51.1}$      & $0.61_{-0.06}^{+0.06}$ & $9.33_{-0.91}^{+0.48}$ & $-7.29_{-0.64}^{+1.09}$  & $-4.74_{-0.21}^{+0.31}$              & $-3.61_{-0.35}^{+0.40}$                & -       & -       & 13038.27 \\ \hline
0 ($\star$\texttt{JP}) & $273.2_{-51.1}^{+45.7}$      & $0.49_{-0.17}^{+0.11}$ & $8.73_{-4.85}^{+0.97}$ & $-7.71_{-1.68}^{+1.85}$  & $-3.64_{-0.94}^{+1.69}$              & $-4.99_{-0.49}^{+0.39}$                & -       & -       & 13035.66 \\ \hline
1 (\texttt{JR}) & $374.2_{-44.8}^{+49.5}$      & $0.82_{-0.08}^{+0.08}$ & $8.36_{-1.46}^{+1.10}$ & $-13.66_{-2.98}^{+3.68}$ & $-3.89_{-0.55}^{+0.52}$              & $-3.04_{-0.50}^{+0.47}$                & $68.17_{-11.67}^{+11.26}$   & -                         & 12595.11 \\ \hline
1 ($\star$\texttt{JR}) & $356.1^{+42.7}_{-40.2}$ & $0.77^{+0.14}_{-0.16}$ & $2.55^{+4.42}_{-4.18}$ & $-11.69^{+7.04}_{-5.40}$ & $-3.54^{+0.50}_{-0.53}$ & $-4.60^{+0.51}_{-0.52}$ & $36.63^{+12.38}_{-12.79} $ & - & 12598.43 \\ \hline
1 (\texttt{JP}) & $333.4_{-35.2}^{+37.3}$      & $0.84_{-0.11}^{+0.10}$ & $8.13_{-1.67}^{+1.19}$ & $-14.72_{-3.10}^{+4.11}$ & $-3.45_{-0.44}^{+0.46}$              & $-2.72_{-0.35}^{+0.38}$               & $79.38_{-10.18}^{+10.30}$   & -                         & 13050.26 \\ \hline
1 ($\star$\texttt{JP}) & $331.5^{+35.8}_{-33.6}$ & $0.83^{+0.11}_{-0.12}$ & $7.48^{+1.54}_{-3.22}$ & $-14.19^{+4.92}_{-3.54}$ & $-3.40^{+0.48}_{-0.42}$ & $-3.07^{+0.45}_{-0.51}$ & $73.83^{+10.82}_{-11.74}$ & - & 13048.23 \\ \hline
2 (\texttt{JP}) & $349.4_{-36.8}^{+39.7}$      & $0.84_{-0.10}^{+0.09}$ & $8.10_{-1.66}^{+1.22}$ & $-14.30_{-3.21}^{+3.99}$ & $-3.64_{-0.40}^{+0.44}$              & $-2.96_{-0.34}^{+0.36}$                & $88.07_{-11.99}^{+12.38}$ & $73.25_{-11.34}^{+11.09}$ & 13048.46 \\ \hline 
2  ($\star$\texttt{JP}) & $377.0_{-36.0}^{+35.8}$      & $0.84_{-0.11}^{+0.09}$ & $1.18_{-3.33}^{+4.19}$ & $-11.10_{-5.96}^{+7.42}$ & $-3.78_{-0.34}^{+0.35}$              & $-4.61_{-0.42}^{+0.46}$                & $79.47_{-12.30}^{+12.14}$ & $34.12_{-16.11}^{+16.32}$ & 13049.69 \\ \hline \hline
\end{tabular}
\caption{Retrieved atmospheric properties for TOI-1231~b for the fiducial six-molecule retrieval cases. The symbol $\star$ denotes that unocculted stellar heterogeneities were included in the corresponding model, while \texttt{JP} and \texttt{JR} refer to retrievals carried out on the\texttt{JexoPipe} and \texttt{JExoRES} data, respectively.  Properties include the photospheric temperature at 1~$\mu$bar, cloud/haze parameters, and the reference pressure. We also show the retrieved offsets for the one offset cases (OS1) and two offset  cases (OS1 and OS2).}
\label{tab:physical_properties}
\end{table*}

\section{Summary and Discussion} \label{sec:summary}

We report the first transmission spectrum of a temperate Neptune-sized planet, TOI-1231 b, observed with JWST NIRISS SOSS and NIRSpec G395H between 0.65 and5.2 $\mu$m. When using the \texttt{JexoPipe} data reduction pipeline, we find strong evidence for CH$_4$, at $\ln B = 69.6$, with a mixing ratio of $\log (X_\mathrm{CH_4})= -0.85^{+0.18}_{-0.19}$, and for CO$_2$ at $\ln B = 6.1$ with mixing ratio $\log (X_\mathrm{CO_2})=-1.90^{+0.40}_{-0.71}$, considering our fiducial model with one offset. Consistent results, albeit with somewhat lower significances ($\ln B = 55.4$ for CH$_4$ and $\ln B = 6.0$ for CO$_2$), are found when adopting data reduced with the \texttt{JExoRES} pipeline. We find moderate evidence for stellar heterogeneity in the \texttt{JExoRES} analysis, while this is disfavored in the \texttt{JexoPipe} data, for which possible contributions from star spots are accounted for during the data reduction. We do not find evidence for additional species, although approximately percent-level abundances of HCN, CO and H$_2$O, and up to $\sim 10^{-3}$ for NH$_3$, remain permitted within the 95\% upper limits.   
A wide-ranging search including over 200 possible species using the \texttt{JexoPipe} data shows that a few complex species, such as 3-carene and limonene, can result in moderate preference in the \texttt{JexoPipe} analysis. However, this is found to be strongly pipeline-dependent, with no species among the 201 considered resulting in $\ln B \geq 2.0$ in both the \texttt{JexoPipe} and \texttt{JExoRES} data.

\subsection{Expectations for a Temperate Deep Atmosphere}

These results are consistent with the expectations for a temperate Neptune-like planet with a deep H$_2$-rich atmosphere.  CH$_4$ is detected at high abundance, as predicted for high atmospheric metallicity (e.g. $\gtrsim$100$\times$ solar) and deep-atmosphere recycling of photochemical products. While NH$_3$ is not detected, the comparatively high 95\% upper limit is consistent with its possible presence in the atmosphere. Further observations are needed to improve sensitivity to this molecule, especially in order to interpret its robust non-detections and lower upper limits in the H$_2$-rich atmospheres of several temperate sub-Neptunes \citep{Madhusudhan2023K2-18b, holmberg_possible_2024, Benneke2024, Hu2025, rigby_jwst_2025, constantinou_atmospheric_2026}. 

In particular, although we find a high upper limit on NH$_3$, it is important to consider possible reasons for its non-detection. A natural explanation is that CH$_4$ has stronger bands in the relevant wavelength region, while our observations do not have a high enough signal-to-noise ratio (SNR) for unambiguous detection of NH$_3$. However, there may be processes that lead to NH$_3$ depletion in the atmosphere. For instance, a low NH$_3$ mixing ratio could result from low elemental nitrogen abundance during planet formation. Alternatively, a moderately high internal heat flux could potentially result in a temperature profile in the deep atmosphere that shifts the disequilibrium quench point for CO-CH$_4$ to conditions where CH$_4$ is dominant, while the quench point for NH$_3$-N$_2$ occurs where N$_2$ is dominant. Another possibility is that NH$_3$ may be more readily destroyed by photochemistry than CH$_4$, although we note that photochemical models for temperate H$_2$-rich planets predict that NH$_3$ should survive into the near-infrared photosphere \citep{Moses2013,Hu2021b,Tsai2021,Ohno2022,wogan_jwst_2024}.  Finally, it is also possible that some amount of NH$_3$ could dissolve in water droplets condensed in layers below those probed by transmission spectroscopy (a ``cold trap''), or be sequestered in upper-tropospheric NH$_4$SH clouds if the tropopause temperature is cold enough, reducing the NH$_3$ abundances in the upper layers.  

The lack of detection of sulfur species in our TOI-1231 b spectra is also interesting.  Although H$_2$S is readily destroyed by photochemical processes in the upper atmosphere, it is expected to survive to at least the $\sim$1-10 mbar level in temperate Neptune-like atmospheres, where it may be detectable, with subsequent sulfur photochemical products such as CS$_2$ also being potentially detectable \citep{Zahnle2009soot,Moses2024,Mukherjee2025_CS2,Veillet2025}. CS$_2$ has been inferred in the slightly warmer sub-Neptune TOI-270 d \citep{holmberg_possible_2024,Benneke2024}, but not in K2-18 b \citep{Madhusudhan2023K2-18b}. The tropopause temperatures are unlikely to be low enough for H$_2$S condensation, however, they may possibly be sufficiently cold for NH$_4$SH formation and condensation in the upper troposphere, which could deplete both sulfur and nitrogen from the observable upper atmosphere.  This possibility depends on both the planet's albedo (and thus its atmospheric temperatures) and the tropospheric abundances of H$_2$S and NH$_3$.  Another photochemical product, S$_8$, could also potentially condense and contribute to highly scattering stratospheric hazes \citep{Zahnle2009soot}. However, it is important to be cautious in interpreting these results as evidence for a depletion in sulfur species, as the inferred 95\% upper limits still allow high abundances of H$_2$S ($\log X_{ \rm H_2S} \leq -2.03$, in the 4+X setup) and CS$_2$ ($\log X_{\rm CS_2} \leq -2.43$ in the 4+X setup, and $\log X_{\rm CS_2} \leq -2.38$ in the M23-like set-up) to be present in the atmosphere.

\subsection{Comparison to Warm and Hot Neptunes}

Spectral features in TOI-1231 b are strong, in contrast to muted spectral amplitudes observed in the ultra-hot Neptune LTT-9779 b \citep{Radica2024_LTT9779}.  The latter planet is over 1500 K higher in equilibrium temperature, such that chemical equilibrium would favor CO or CO$_2$ as the main carbon-carrier over CH$_4$. However, the LTT-9779 b spectrum cannot distinguish H$_2$O and CH$_4$.  The muted features in comparison to TOI-1231 b may also result from the high inferred metallicity and/or cloud coverage in LTT-9779 b.  

The warm Neptune GJ 3470 b \citep{Beatty2024} has a lower CH$_4$ abundance of $\log(X_\mathrm{CH_4})=-4.05^{+0.25}_{-0.27}$, and a slightly lower CO$_2$ abundance of  $\log(X_\mathrm{CO_2})=-2.47^{+0.61}_{-0.43}$.  This difference seems consistent with the higher $T_\mathrm{eq}$ and inferred high internal temperature $T_\mathrm{int}$ for GJ~3470 b, which lead to increased carbon carriage by CO over CH$_4$, resulting from thermochemical equilibrium and transport-induced quenching. The GJ 3470 b spectra also show evidence for SO$_2$ and H$_2$O, which are notably not detected in TOI-1231 b, likely due to its lower atmospheric temperatures, although substantial H$_2$O remains allowed by the upper limits. The detection of H$_2$O vapor on GJ~3470 b is consistent with the absence of a water cold trap at higher $T_\mathrm{eq}$. 

Finally, the emission spectrum of the warm Neptune GJ 436 b \citep{Mukherjee2025} shows weak evidence for CO$_2$ at $2\sigma$ 
with atmospheric models implying metallicities of $\geq80\times$ or $\geq300\times$ solar, depending on the modeling assumptions.  The possible presence of CO$_2$ along with an absence of CH$_4$ contrasts with the findings in the present work, where CH$_4$ dominates as the main carbon carrier.  This is consistent with expectations due to the higher temperature of GJ 436 b.  A previous Hubble WFC3 transmission spectrum  showed a featureless spectrum, suggestive of cloudy and/or high metallicity atmosphere \citep{Knutson2014}.  This may again indicate a propensity for increased cloud or haze formation at temperatures higher than seen for TOI-1231 b.

\begin{figure*}
    \centering
    \includegraphics[width=\textwidth]{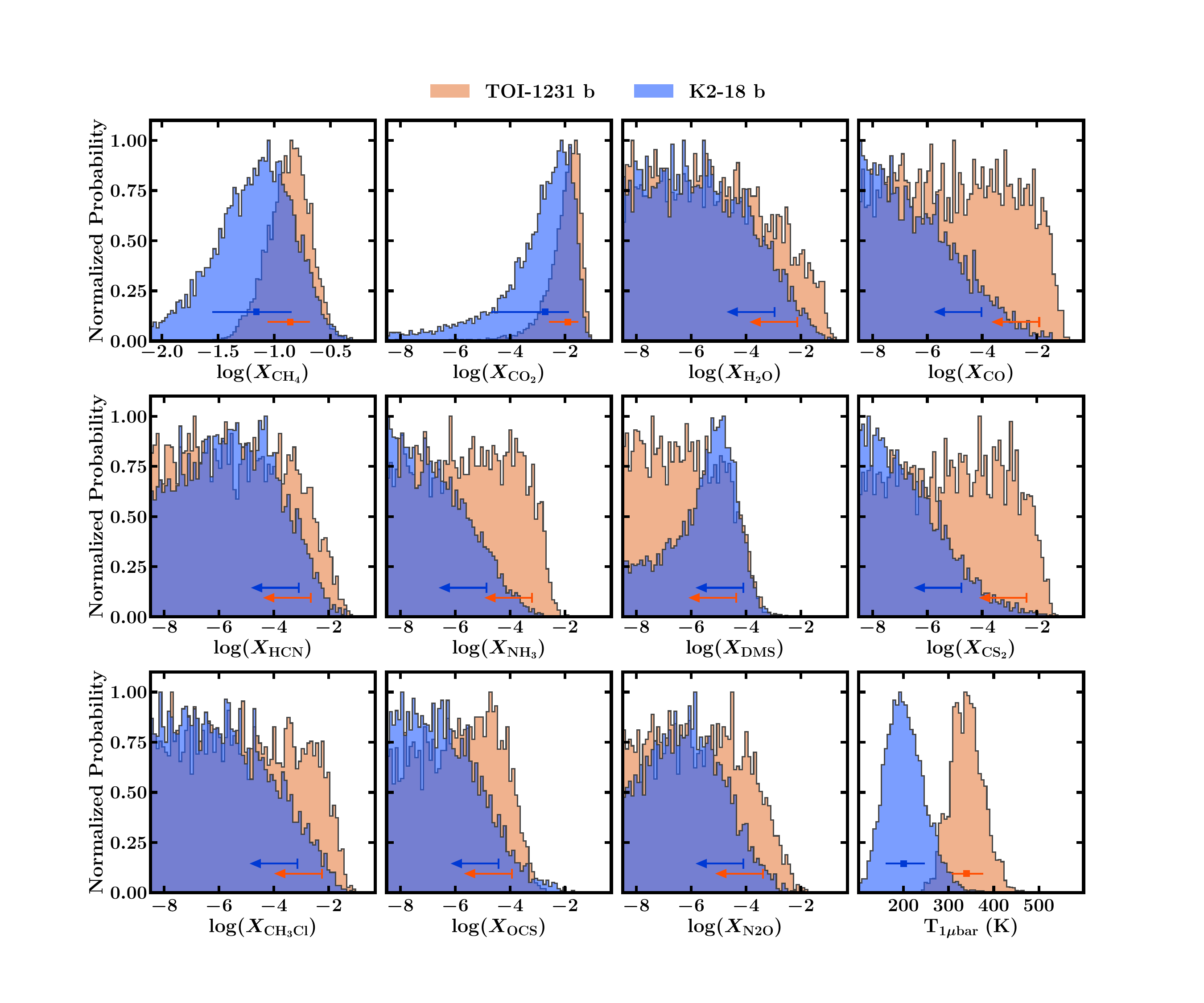}
    \caption{Comparison of the posterior distributions for the abundances of all 11 molecules in the \citetalias{Madhusudhan2023K2-18b}-informed set-up discussed in Section \ref{subsec:additional_species}, as well as the temperature at 1$\mu$bar.  We compare results for TOI-1231~b (blue) and K2-18~b (orange). The K2-18 b posteriors correspond to a new retrieval conducted in this work, using the same model and assumptions applied to TOI-1231 b. 
    The median retrieved values and 1$\sigma$ error bars are shown for $T_{\rm 1 \mu bar}$ and for the abundances of CH$_4$ and CO$_2$, while arrows indicate the 95\% upper limits for the abundances of all other species. Due to differences in priors and molecular opacities used between this work and \citetalias{Madhusudhan2023K2-18b}, minor discrepancies are present in the K2-18~b results compared to \citetalias{Madhusudhan2023K2-18b}.     
    }
    \label{fig:compareK218}
\end{figure*}

\subsection{Comparison to K2-18~b}

Our observations provide an opportunity for comparative assessment between a bona fide deep Neptune-like atmosphere and sub-Neptune atmospheres such as those of K2-18 b \citep{Madhusudhan2023K2-18b}, TOI-270~d \citep{Benneke2024, holmberg_possible_2024, constantinou_atmospheric_2026}, and TOI-732~c \citep{rigby_jwst_2025}. 
TOI-1231 b is considerably more massive than other temperate planets for which molecular detections have been reported from JWST observations, the closest in mass being K2-18 b at $M_{\rm p} = 8.63 \pm 1.35$ $\mathrm{M}_\oplus$, compared to $M_{\rm p} = 15.4 \pm 3.3$ $\mathrm{M}_\oplus$ for TOI-1231 b. 
As discussed earlier, K2-18 b has been suggested to be a candidate hycean planet, with a shallow atmosphere above a water layer \citep{madhusudhan_habitability_2021}. This was supported by atmospheric abundance constraints from JWST near-infrared transmission spectra obtained using the same instrument configurations as those used in this study (\citetalias{Madhusudhan2023K2-18b}). However, as discussed above, the bulk parameters of K2-18~b remain consistent with a degenerate set of internal structure solutions, including both shallow- and deep- atmosphere scenarios. The chemical constraints obtained for TOI-1231~b  therefore provide a benchmark for a temperate planet with a known deep H$_2$-rich atmosphere.

In order to carry out a like-to-like comparison of K2-18 b and TOI-1231 b, we perform a new retrieval for the former, using the same priors and opacities used for TOI-1231 b in this work. We use the \texttt{JExoRES} transmission spectrum from \cite{Madhusudhan2023K2-18b} for this retrieval. Using Bayesian model comparison, we find that for K2-18 b a one-offset model with no stellar heterogeneity is favored, consistent with findings in \citet{Madhusudhan2023K2-18b}. We use the 11-molecule \citetalias{Madhusudhan2023K2-18b}-like set-up with a single offset, as described in Section \ref{subsec:additional_species} for both planets.

The results of this new K2-18 b retrieval are summarized in Figure \ref{fig:compareK218}, together with the corresponding results for TOI-1231 b, where we show the posterior distributions for all 11 molecules included in the model.  We note that the results obtained for K2-18~b present some minor differences with those reported in \citetalias{Madhusudhan2023K2-18b}. These differences arise mostly from the slightly different choice of priors (for consistency with the TOI-1231~b retrievals), and from the different molecular opacities used by the retrieval codes. Overall, however, our results remain consistent with those reported in the literature \citep{Madhusudhan2023K2-18b, Hu2025}.

Consistent with the rest of this work, for TOI-1231~b we primarily consider the results obtained using the \texttt{JexoPipe} data, although the following considerations also hold true when the \texttt{JExoRES} data with stellar heterogeneity included (Section \ref{subsec:additional_species}). While the median retrieved abundances for CH$_4$ and CO$_2$ are somewhat higher for TOI-1231~b than for K2-18~b, the inferred abundances of these two prominent species are consistent between the two planets within the 1-$\sigma$ uncertainties. Similarly, while higher upper limits are found for several species on TOI-1231~b, the corresponding posterior distributions exhibit similar shapes for the two planets. However, some other species show notably different posterior distributions: CO, NH$_3$, DMS, and CS$_2$. In particular, the significantly higher ($\sim 2$~dex) upper limits found for CO and NH$_3$ on TOI-1231~b relative to K2-18~b  result in significantly greater probability mass being present at high abundances for these two species in TOI-1231 b. Similarly, while CS$_2$ is also not inferred with any statistical significance, it presents a high upper limit at $\log X_{\rm CS_2} \leq -2.38$ on TOI-1231~b, compared to $\log X_{\rm CS_2} \leq -4.75$ on K2-18~b, which is also reflected in the posterior distribution shape. DMS, however, appears to behave differently.  When considering the \texttt{JexoPipe} data, DMS is the only one of the 11 molecules considered in this set up for which the retrieved 95\% upper limit in TOI-1231~b is lower than that in K2-18~b, albeit marginally at $-4.35$ compared to $-4.10$. This is also the case when using the \texttt{JExoRES} data, where we additionally find that the HCN upper limit is lower on TOI-1231 b than on K2-18 b. The DMS posterior distribution also appears to have a significantly different shape, with a defined peak at $\log (X_\mathrm{DMS})\sim - 5$ for K2-18~b and no such peak for TOI-1231~b.

Finally, the temperatures at 1 $\mu$bar differ by $\sim 140$ K ($T_{\rm 1\mu bar} = 339.5^{+36.8}_{-33.3}$ K for TOI-1231~b, and $T_{\rm 1\mu bar} = 200.6^{+46.9}_{-40.1}$ K for K2-18~b). This difference is consistent with expectations based on the equilibrium temperatures of the two planets. It is also consistent with the higher upper limit inferred for H$_2$O on TOI-1231~b, since higher atmospheric temperatures reduce the effectiveness of cold-trapping, which is expected to be the primary mechanism responsible for H$_2$O depletion in the observable atmospheres of these temperate planets.

\subsection{Future Directions} \label{sec:future}

TOI-1231~b is the first temperate Neptune-sized exoplanet observed with JWST. We robustly detect CH$_4$ at $\ln B = 54.5$-$69.6$, and find evidence for CO$_2$ at $\ln B = 2.9$-$6.6$ across the study. In our nominal case, we find abundances of $\log (X_\mathrm{CH_4}) = -0.85^{+0.18}_{-0.19}$ and $\log (X_\mathrm{CO_2}) = -1.90^{+0.40}_{-0.71}$. We do not detect any other species, but the 95\% upper limits allow up to $\sim 0.1\%$ NH$_3$ and HCN and upto $\sim 1\%$ of CO and H$_2$O, with even higher upper limits inferred depending on the case considered (Table \ref{tab:abundances}). When considering each of 201 organic and/or sulfur-bearing molecules individually in the retrievals, none were found to be statistically significant across both pipelines.

The retrieved CH$_4$ and CO$_2$ mixing ratios, together with the high upper limits for other species, are relatively straightforward to explain for TOI-1231 b in the context of a deep, high-metallicity, H$_2$-rich atmosphere. In contrast, the relative abundances of CO$_2$ and CH$_4$ and much lower upper limits for CO and NH$_3$ on K2-18~b \citep{madhusudhan_carbon-bearing_2023}, as well as for NH$_3$ in TOI-270~d \citep{holmberg_possible_2024, benneke_jwst_2024, constantinou_atmospheric_2026} and TOI-732~c \citep{rigby_jwst_2025}, are more difficult to explain in the context of a deep-atmosphere, mini-Neptune scenario \citep[see][]{Cooke2024}.

While further observations are necessary, our results are consistent with theoretical predictions for deep versus shallow H$_2$-rich atmospheres, and thus support the interpretation that the observations of K2-18 b and TOI-270 d may indicate hycean worlds with shallow H$_2$-rich envelopes.  Additional  observations with NIRISS and NIRSpec are required to increase the SNR of the final transmission spectrum.  Complementary observations using ground-based high-resolution spectroscopy, as recently conducted for the temperate sub-Neptune TOI-732~c \citep{Cheverall2024, Cabot2024}, could also provide further evidence for CH$_4$ and NH$_3$.

Our findings provide an important benchmark against which observations of sub-Neptunes can be compared, helping to disentangle their possibly degenerate compositions and interior structures. This may aid in distinguishing between deep and shallow H$_2$-rich atmospheres, e.g.  mini-Neptunes versus hycean worlds, among the increasing numbers of sub-Neptunes with JWST transmission spectra. 

\vspace{0.695cm}
{\it Acknowledgements:}  This work is based on observations made with the NASA/ESA/CSA James Webb Space Telescope as part of Cycle 2 GO Program 3557 (PI: N. Madhusudhan). Support was
provided from JWST-GO-03557, via NASA through a grant from the Space Telescope Science Institute, which is operated by the Association of Universities for Research in Astronomy, Inc., under NASA contract NAS 5-03127. We thank NASA,
ESA, CSA, STScI, everyone whose efforts have contributed to the JWST, and the exoplanet science community for the thriving current state of the field. This work is supported by research grants to N.M. from the UK Research and Innovation (UKRI) Frontier Grant (EP/X025179/1) and the UK Science
and Technology Facilities Council (STFC). N.M., F.E.R., and L.P.C. acknowledge support from UKRI STFC toward the doctoral studies of F.E.R. (UKRI grant 2605554) and L.P.C. (UKRI grant 2886925). J.M. acknowledges support from NASA XRP grant No. 80NSSC23K0281. This research has made use of the NASA Exoplanet Archive, which is operated by the California Institute of Technology, under contract with the National Aeronautics and Space Administration under the Exoplanet Exploration Program. M.M. acknowledges support from the UK Science and Technology Facilities Council (STFC) through the Cardiff Doctoral Training Partnership grant ST/Y509152/1. S.S. acknowledges seedcorn funding from the DiRAC HPC Facility (project dp417). This work used the DiRAC Data Intensive service (CSD3) at the University of Cambridge, managed by the University of Cambridge University Information Services on behalf of the STFC DiRAC HPC Facility (www.dirac.ac.uk). The DiRAC component of CSD3 at Cambridge was funded by BEIS, UKRI, and STFC capital funding and STFC operations grants. DiRAC is part of the UKRI Digital Research Infrastructure.

{\it Author Contributions:} N.M. conceived, planned and led the project. N.M. led the JWST proposal with contributions from S.S., F.E.R and J.M. S.S. led the writing of the manuscript with contributions from all authors. S.S., M.H., N.M. and M.M. conducted the data reduction and analyses. N.M. and L.P.C. conducted the atmospheric retrievals. N.M., J.M., L.P.C., F.E.R. and S.S. conducted the theoretical interpretation.

{\it Data Availability:} Some/all of the data presented in this paper were obtained from the Mikulski Archive for Space Telescopes (MAST) at the Space Telescope Science Institute. The specific observations analyzed can be accessed via \dataset[DOI:10.17909/djm8-jm79]{https://www.doi.org/10.17909/djm8-jm79}. The transmission spectra reported in this work are available on the Open Science Framework at https://osf.io/mryd6.

\vspace{5mm}
\newpage
\appendix

\section{Priors for Retrievals}
\label{sec:app_priors}

We report in Table \ref{tab:priors} the Bayesian priors used for all retrievals presented in this work.  
\begin{table}[h]
\centering
\begin{tabular}{lll}
\hline \hline
Parameter             & Bayesian Prior         & Description                           \\[0.5mm] \hline
$\log(X_{\rm X})$     & $\mathcal{U}$(-12, -0.3) & Mixing ratio of each chemical species \\[0.5mm]
$T_{\rm 1 \mu bar}$/K & $\mathcal{U}$(100, 800)  & Reference temperature at 1$\mu$bar  \\[0.5mm]
$\alpha_1 / \mathrm{K}^{-\frac{1}{2}}$  & $\mathcal{U}$(0.02, 2.00) & P-T profile curvature  \\[0.5mm]
$\alpha_2/ \mathrm{K}^{-\frac{1}{2}}$& $\mathcal{U}$(0.02, 2.00)  & P-T profile curvature  \\[0.5mm]
$\mathrm{log}(P_1/\mathrm{bar})$   & $\mathcal{U}$(-6, 0) & P-T profile region limit  \\[0.5mm]
$\mathrm{log}(P_2/\mathrm{bar})$  & $\mathcal{U}$(-6, 0) & P-T profile region limit   \\[0.5mm]
$\mathrm{log}(P_3/\mathrm{bar})$   & $\mathcal{U}$(-2, 0)& P-T profile region limit  \\[0.5mm]
$\mathrm{log}(P_\mathrm{ref}/\mathrm{bar})$   & $\mathcal{U}$(-6, 0) & Reference pressure at the retrieved NIRISS white-light radius $R_\mathrm{p} = 3.9$ R$_\oplus$\\[0.5mm]
$\mathrm{log}(a)$  & $\mathcal{U}$(-4, 10)& Rayleigh enhancement factor \\[0.5mm]
$\gamma$   & $\mathcal{U}$(-20, 2)& Scattering slope \\[0.5mm]
$\mathrm{log}(P_\mathrm{c}/\mathrm{bar})$  & $\mathcal{U}$(-6, 0)& Cloud top pressure \\[0.5mm]
$\phi$  & $\mathcal{U}$(0, 1)& Cloud/haze coverage fraction\\[0.5mm]
$\delta_\mathrm{NIRSpec} / \mathrm{ppm}$  & $\mathcal{U}$(-300, 300)& NIRSpec dataset offset \\[0.5mm]
$\delta_\mathrm{NRS1} / \mathrm{ppm}$  & $\mathcal{U}$(-300, 300)& NIRSpec NRS1 dataset offset \\[0.5mm]
$\delta_\mathrm{NRS2} / \mathrm{ppm}$ & $\mathcal{U}$(-300, 300) & NIRSpec NRS2 dataset offset \\[0.5mm]
$T$/K & $\mathcal{U}$(100, 800)  & Temperature (isothermal)$^\dagger$\\[0.5mm] 
$T_\mathrm{phot} /$  K (TOI-1231)  & $\mathcal{N}(3353, 50^2)$& Stellar photosphere temperature for TOI-1231 \citep{Burt2021}\\[0.5mm]
$T_\mathrm{het} /$  K (TOI-1231) & $\mathcal{U}(2100, 4264)$ & Stellar heterogeneity temperature for TOI-1231 \\[0.5mm]
$T_\mathrm{phot} /$  K (K2-18)  & $\mathcal{N}(3457, 39^2)$& Stellar photosphere temperature for K2-18 \citep{Benneke2019}\\[0.5mm]
$T_\mathrm{het} /$  K (K2-18) & $\mathcal{U}(2100, 4148)$ & Stellar heterogeneity temperature for K2-18 \\[0.5mm]
$f_\mathrm{het} $ & $\mathcal{U}$(0, 0.5)& Stellar Heterogeneity coverage fraction\\ \hline
\end{tabular}
\caption{Priors used in all retrievals presented in this work. \\ 
$^\dagger$: Used only in the canonical retrievals with 201 species.}

\label{tab:priors}
\end{table}

\section{Retrieval set-up}
\label{sec:app_list}

\begin{figure*}[h]
    \centering
    \includegraphics[width=\linewidth]{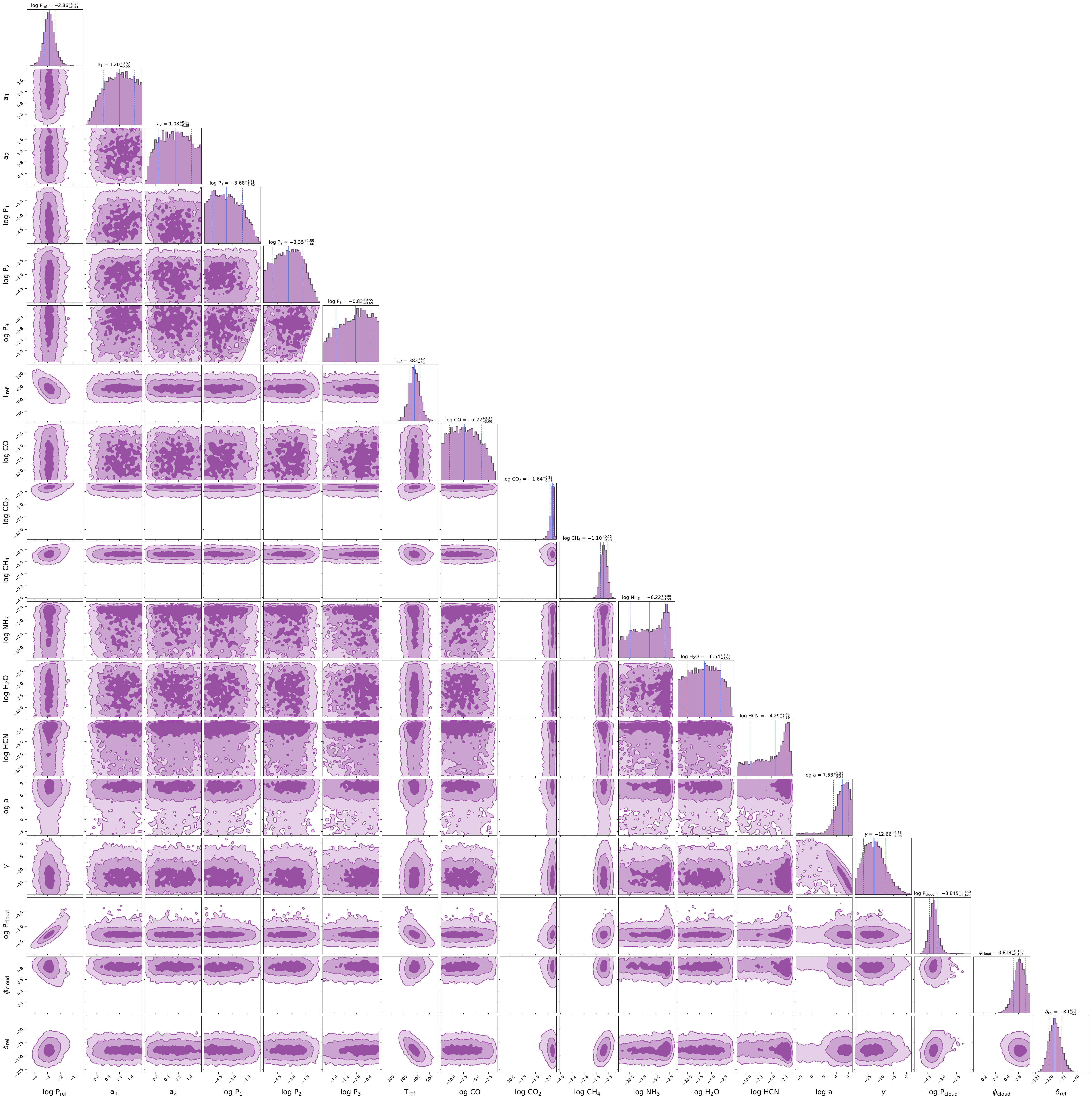}
    \caption{Corner plot for the six-molecule model including one offset and no stellar heterogeneity, using the \texttt{JexoPipe} data.}
    \label{fig:jpipe_corner_plot}
\end{figure*}

\begin{figure*}[h]
    \centering
    \includegraphics[width=\linewidth]{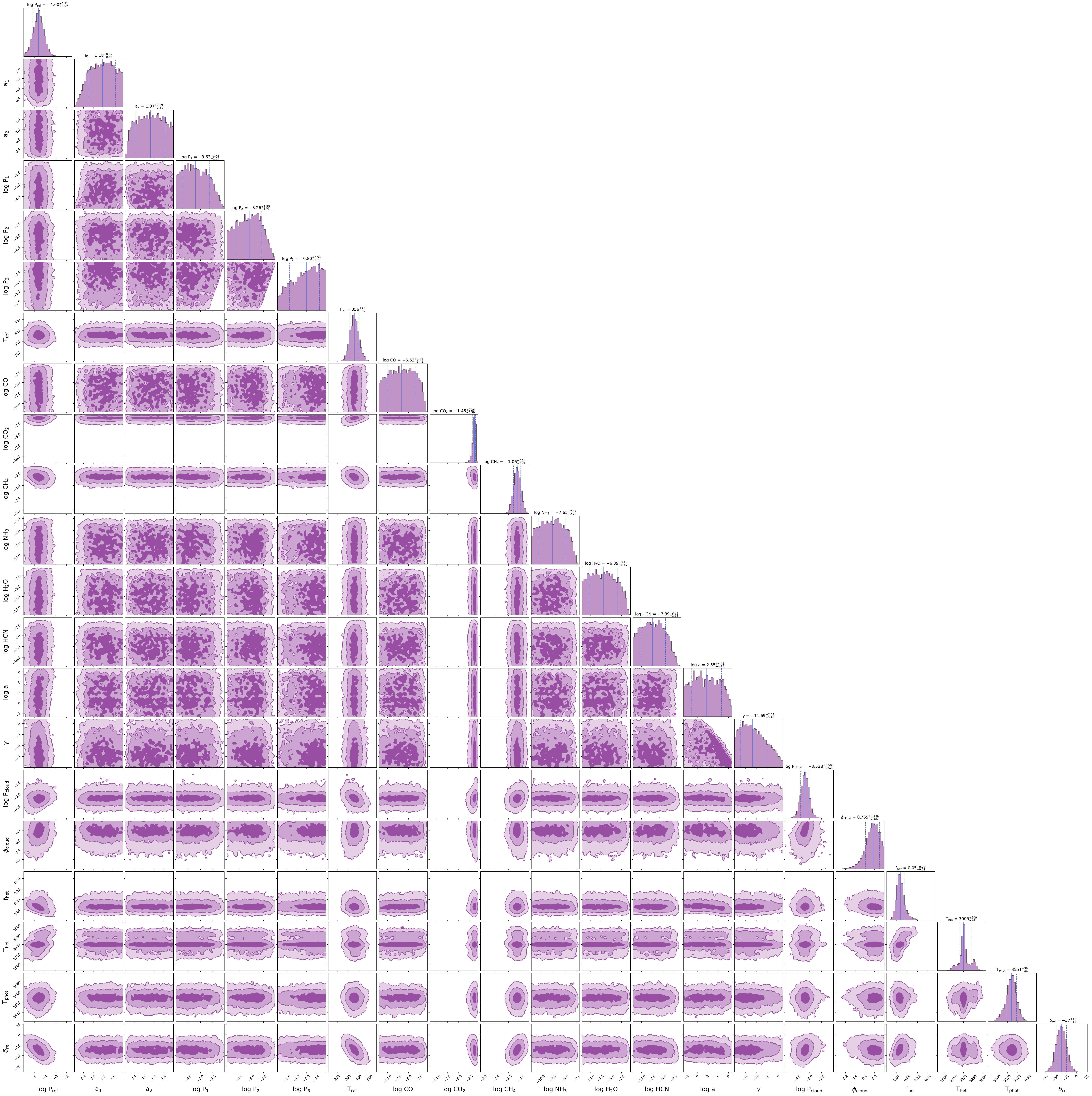}
    \caption{Corner plot for the six-molecule model including one offset and also stellar heterogeneity, using the \texttt{JExoRES} data.}
    \label{fig:jres_corner_plot}
\end{figure*}

The range of retrieval models considered, and their results, are listed in Table \ref{tab:retrieval_summary}. The full corner plots from our main cases with both \texttt{JexoPipe} and \texttt{JExoRES} data are shown in Figures \ref{fig:jpipe_corner_plot} and \ref{fig:jres_corner_plot}, respectively. We list the molecules considered in our 4+X exploration discussed in Section \ref{subsec:additional_species} below in Appendix \ref{subsec:list_species}.

\subsection{List of Species Considered}
\label{subsec:list_species}
Allylamine, P-xylene, (-)-beta-pinene, Limonene, Methylamine, Propylene, 3-carene, Methanethial, Methanesulfonyl chloride, Hydrochloric acid, 2-methylstyrene, Hexamethylphosphoramide, Quinoline, Pentane-1,5-diamine, 1-propanethiol, Isobutylene, 1,2,3,4-tetrahydronaphthalene, Toluene, 2-carene, Propylbenzene, Ethylamine, Morpholine, 1,2,3,4-tetramethylbenzene, Ethylenediamine, Ethylbenzene, Butylamine, Thiophosphoryl chloride, Cumene, Methanethiol, Trimethylbenzene, Sec-butylbenzene, 2-methyl-1-pentene, 4-ethyltoluene, 1-decene, 2-mercaptoethanol, 2,2-dimethylbutane, 1-nonene, N,n-dimethylformamide, (-)-alpha-pinene, 2-methyl-2-propanethiol, 1-butene, 1-pentene, M-xylene, Butane, 2,4,4-trimethyl-2-pentene, Cyclooctane, Diisopropylamine, 63493-28-7, 1-ethyl-2-methylbenzene, 1-octene, 1,2,3,5-tetramethylbenzene, 4-methylpyridine, Diethyl sulfate, Perchloromethyl mercaptan, Nitrogen trifluoride, Benzene, Diethylamine, Potassium chloride, Hydrogen cyanide, Sodium hydroxide, Tert-butylbenzene, 3-methylpentane, Pentyl nitrate, Propylene sulfide, Dimethyl sulfoxide, 1-nitropropane, 2-methyl-2-butene, N,n-diethylaniline, Thionyl fluoride, Nitrous acid, Cyanoacetylene, 1-heptene, Butyl isocyanate, Decane, Nitroethane, Heptane, 2-vinylpyridine, (+)-limonene, Propionitrile, 2-methylpyridine, Methyl nitrite, Methacrylonitrile, Trifluoronitrosomethane, 2-nitropropane, Methyl isocyanate, Cis-4-methyl-2-pentene, Thiophene, 2-methylaziridine, 2-methyl-1-butene, Potassium hydroxide, Hydrofluoric acid, Acrylonitrile, Sulfur hexafluoride, Isobutyronitrile, Tetradecane, O-toluidine, Sulfuryl fluoride, Trans-2-pentene, 1-hexene, Mesitylene, 1-undecene, Isopropylamine, Nitromethane, Dimethyl sulfate, 2,4-diisocyanato-1-methylbenzene, 1,1-dimethylhydrazine, 3-methylhexane, 1,3-butadiene, Nitrobenzene, Methyl isothiocyanate, Myrcene, Tert-amylamine, Isoprene, Thiirane, Naphthalene, 4-vinylcyclohexene, Trimethylamine, Peroxyacetyl nitrate, Dimethylamine, Sulfuryl chloride, Benzonitrile, Chloromethane, Tetrahydrothiophene, 3-ethyltoluene, Hydrazine, Hydroxide, Pyridine, 4-methyl-1-pentene, 2-propanethiol, Trans-2-butene, Cyclodecane, 2,4-dimethylpentane, Pentadecane, Allyl isothiocyanate, Methylhydrazine, Octane, Propyne, Neopentane, Isocyanic acid, But-2-ynenitrile, 2,2,4-trimethylpentane, Dimethyl disulfide, Piperidine, Styrene, Propane, Cyclohexanethiol, Acetone cyanohydrin, Isobutane, 3-methyl-1-butene, Nicotine, Cyclopentane, 2-methyl-2-pentene, Allene, Nonane, 1-butyne, Diethyl sulfide, Ethyl nitrite, Cis-2-pentene, Isopentane, Triethylamine, Undecane, Cycloheptene, Ethanethiol, Cyclopentene, Dimethyl sulfide, N,n-diethylformamide, 2-methylpentane, Cyclohexene, 2,6-diethylaniline, Cycloheptane, Diisobutylene, 2,3-dimethylbutane, Tridecane, Hexadecane, Cyclopropane, Aniline, Isobutyl mercaptan, Hexane, Pentane, Cyclohexane, Benzenethiol, Thiophosgene, NaH, 3picoline, CS$_2$, OCS, oxylene, O$_3$, CH$_3$OH, CH$_3$, C$_2$H$_6$, OH+, N$_2$O, CH$_3$CN, H$_2$S, NO$_2$, C$_2$H$_4$, SO$_2$, PH$_3$, H$_2$O, C$_2$H$_2$.

\newpage 

\begin{table*}[ht]
\centering
\renewcommand{\arraystretch}{1.1}
\begin{tabular}{lll}
\hline\hline
\parbox[t]{3cm}{Retrieval\vspace{1mm}} &
\parbox[t]{8cm}{Description\vspace{1mm}} &
\parbox[t]{6cm}{Results\vspace{1mm}} \\[1mm]
\hline

\parbox[t]{3cm}{Fiducial six-molecule\vspace{1mm}} &
\parbox[t]{8cm}{CH$_4$, CO$_2$, CO, H$_2$O, NH$_3$, HCN, full P--T profile, inhomogeneous clouds/hazes, stellar heterogeneity, fixed planet mass, and detector offsets (zero, one, or two)} &
\parbox[t]{6cm}{Strong preference for at least one offset; only weak preference for a second offset} \\[8mm]
\hline

\parbox[t]{3cm}{Stellar heterogeneity test} &
\parbox[t]{8cm}{Fiducial model with stellar heterogeneity versus without} &
\parbox[t]{6cm}{With a single offset, in \texttt{JexoPipe} $\ln B = 2.0$ against stellar heterogeneity; in \texttt{JExoRES}, moderate preference in favor ($\ln B = 3.3$) is found} \\[12mm]
\hline

\parbox[t]{3cm}{P-T profile test} &
\parbox[t]{8cm}{Fiducial model with one offset with full (non-isothermal) P-T profile versus with isothermal P-T} &
\parbox[t]{6cm}{No preference for non-isothermal model ($\ln B = 0.1$ in favor of full P-T)} \\[5mm]
\hline

\parbox[t]{3cm}{Cloud/haze test} &
\parbox[t]{8cm}{Fiducial model with one offset with clouds+hazes versus only removing clouds, versus only removing hazes, and versus removing both (clear atmosphere)} &
\parbox[t]{6cm}{Clouds+hazes strongly preferred ($\ln B = 7.9$) over clear case; moderate preference for clouds+hazes over only clouds ($\ln B = 3.5$) and over only hazes ($\ln B = 3.2$)} \\[12mm]
\hline

\parbox[t]{3cm}{CH$_4$ detection} &
\parbox[t]{8cm}{Fiducial model with one offset with CH$_4$ included versus excluded} &
\parbox[t]{6cm}{Strong preference for CH$_4$ ($\ln B = 69.6$)} \\[4mm]
\hline

\parbox[t]{3cm}{CO$_2$ detection} &
\parbox[t]{8cm}{Fiducial model with one offset with CO$_2$ included versus excluded} &
\parbox[t]{6cm}{Strong preference for CO$_2$ ($\ln B = 6.1$)}\\[4mm]
\hline\hline

\end{tabular}
\caption{Summary of atmospheric retrievals for TOI-1231~b. The first row describes the fiducial model. Unless otherwise specified, the results quoted here are as obtained from the one-offset, six-molecule, no stellar heterogeneity retrieval carried out on the \texttt{JexoPipe} data.}
\label{tab:retrieval_summary}
\end{table*}

\FloatBarrier

\bibliography{ref2,references, refs1}{}

@MISC{Baines2023b,
       author = {{Baines}, Tyler and {Espinoza}, N{\'e}stor and {Filippazzo}, Joseph and {Volk}, Kevin},
        title = "{Characterization of the visit-to-visit Stability of the GR700XD Wavelength Calibration for NIRISS/SOSS Observations}",
 howpublished = {Technical Report JWST-STScI-008571, 12 pages},
         year = 2023,
        month = nov,
        pages = {8571},
       adsurl = {https://ui.adsabs.harvard.edu/abs/2023jwst.rept.8571B}
}

@ARTICLE{Burt2021,
       author = {{Burt}, Jennifer A. and {Dragomir}, Diana and {Molli{\`e}re}, Paul and {Youngblood}, Allison and {Garc{\'\i}a Mu{\~n}oz}, Antonio and {McCann}, John and {Kreidberg}, Laura and {Huang}, Chelsea X. and {Collins}, Karen A. and {Eastman}, Jason D. and {Abe}, Lyu and {Almenara}, Jose M. and {Crossfield}, Ian J.~M. and {Ziegler}, Carl and {Rodriguez}, Joseph E. and {Mamajek}, Eric E. and {Stassun}, Keivan G. and {Halverson}, Samuel P. and {Villanueva}, Steven and {Butler}, R. Paul and {Wang}, Sharon Xuesong and {Schwarz}, Richard P. and {Ricker}, George R. and {Vanderspek}, Roland and {Latham}, David W. and {Seager}, S. and {Winn}, Joshua N. and {Jenkins}, Jon M. and {Agabi}, Abdelkrim and {Bonfils}, Xavier and {Ciardi}, David and {Cointepas}, Marion and {Crane}, Jeffrey D. and {Crouzet}, Nicolas and {Dransfield}, Georgina and {Feng}, Fabo and {Furlan}, Elise and {Guillot}, Tristan and {Gupta}, Arvind F. and {Howell}, Steve B. and {Jensen}, Eric L.~N. and {Law}, Nicholas and {Mann}, Andrew W. and {Marie-Sainte}, Wenceslas and {Matson}, Rachel A. and {Matthews}, Elisabeth C. and {M{\'e}karnia}, Djamel and {Pepper}, Joshua and {Scott}, Nic and {Shectman}, Stephen A. and {Schlieder}, Joshua E. and {Schmider}, Fran{\c{c}}ois-Xavier and {Stevens}, Daniel J. and {Teske}, Johanna K. and {Triaud}, Amaury H.~M.~J. and {Charbonneau}, David and {Berta-Thompson}, Zachory K. and {Burke}, Christopher J. and {Daylan}, Tansu and {Barclay}, Thomas and {Wohler}, Bill and {Brasseur}, C.~E.},
        title = "{TOI-1231 b: A Temperate, Neptune-sized Planet Transiting the Nearby M3 Dwarf NLTT 24399}",
      journal = {\aj},
         year = 2021,
        month = sep,
       volume = {162},
       number = {3},
          eid = {87},
        pages = {87},
          doi = {10.3847/1538-3881/ac0432},
archivePrefix = {arXiv},
       eprint = {2105.08077},
 primaryClass = {astro-ph.EP},
       adsurl = {https://ui.adsabs.harvard.edu/abs/2021AJ....162...87B}
}

@ARTICLE{Madhusudhan2009,
       author = {{Madhusudhan}, N. and {Seager}, S.},
        title = "{A Temperature and Abundance Retrieval Method for Exoplanet Atmospheres}",
      journal = {\apj},
         year = 2009,
        month = dec,
       volume = {707},
       number = {1},
        pages = {24-39},
          doi = {10.1088/0004-637X/707/1/24},
archivePrefix = {arXiv},
       eprint = {0910.1347},
 primaryClass = {astro-ph.EP},
       adsurl = {https://ui.adsabs.harvard.edu/abs/2009ApJ...707...24M}
}

@ARTICLE{Polyansky2018,
       author = {{Polyansky}, Oleg L. and {Kyuberis}, Aleksandra A. and {Zobov}, Nikolai F. and {Tennyson}, Jonathan and {Yurchenko}, Sergei N. and {Lodi}, Lorenzo},
        title = "{ExoMol molecular line lists XXX: a complete high-accuracy line list for water}",
      journal = {\mnras},
         year = 2018,
        month = oct,
       volume = {480},
       number = {2},
        pages = {2597-2608},
          doi = {10.1093/mnras/sty1877},
archivePrefix = {arXiv},
       eprint = {1807.04529},
 primaryClass = {astro-ph.EP},
       adsurl = {https://ui.adsabs.harvard.edu/abs/2018MNRAS.480.2597P}
}

@ARTICLE{Coles2019,
       author = {{Coles}, Phillip A. and {Yurchenko}, Sergei N. and {Tennyson}, Jonathan},
        title = "{ExoMol molecular line lists - XXXV. A rotation-vibration line list for hot ammonia}",
      journal = {\mnras},
         year = 2019,
        month = dec,
       volume = {490},
       number = {4},
        pages = {4638-4647},
          doi = {10.1093/mnras/stz2778},
archivePrefix = {arXiv},
       eprint = {1911.10369},
 primaryClass = {astro-ph.SR},
       adsurl = {https://ui.adsabs.harvard.edu/abs/2019MNRAS.490.4638C}
}

@ARTICLE{Barber2014,
   author = {{Barber}, R.~J. and {Strange}, J.~K. and {Hill}, C. and {Polyansky}, O.~L. and 
	{Mellau}, G.~C. and {Yurchenko}, S.~N. and {Tennyson}, J.},
    title = "{ExoMol line lists - III. An improved hot rotation-vibration line list for HCN and HNC}",
  journal = {\mnras},
archivePrefix = "arXiv",
   eprint = {1311.1328},
 primaryClass = "astro-ph.SR",
     year = 2014,
    month = jan,
   volume = 437,
    pages = {1828-1835},
      doi = {10.1093/mnras/stt2011},
   adsurl = {http://adsabs.harvard.edu/abs/2014MNRAS.437.1828B}
}

@ARTICLE{Buchner2014,
       author = {{Buchner}, J. and {Georgakakis}, A. and {Nandra}, K. and {Hsu}, L. and {Rangel}, C. and {Brightman}, M. and {Merloni}, A. and {Salvato}, M. and {Donley}, J. and {Kocevski}, D.},
        title = "{X-ray spectral modelling of the AGN obscuring region in the CDFS: Bayesian model selection and catalogue}",
      journal = {\aap},
         year = 2014,
        month = apr,
       volume = {564},
          eid = {A125},
        pages = {A125},
          doi = {10.1051/0004-6361/201322971},
archivePrefix = {arXiv},
       eprint = {1402.0004},
 primaryClass = {astro-ph.HE},
       adsurl = {https://ui.adsabs.harvard.edu/abs/2014A&A...564A.125B}
}

@article{Feroz2009,
    author = {Feroz, F. and Hobson, M. P. and Bridges, M.},
    title = "{MultiNest: an efficient and robust Bayesian inference tool for cosmology and particle physics}",
    journal = {Monthly Notices of the Royal Astronomical Society},
    volume = {398},
    number = {4},
    pages = {1601-1614},
    year = {2009},
    month = {09},
    issn = {0035-8711},
    doi = {10.1111/j.1365-2966.2009.14548.x},
    url = {https://doi.org/10.1111/j.1365-2966.2009.14548.x},
    eprint = {https://academic.oup.com/mnras/article-pdf/398/4/1601/3039078/mnras0398-1601.pdf},
}

@ARTICLE{Trotta2008,
       author = {{Trotta}, Roberto},
        title = "{Bayes in the sky: Bayesian inference and model selection in cosmology}",
      journal = {Contemporary Physics},
         year = 2008,
        month = mar,
       volume = {49},
       number = {2},
        pages = {71-104},
          doi = {10.1080/00107510802066753},
archivePrefix = {arXiv},
       eprint = {0803.4089},
 primaryClass = {astro-ph},
       adsurl = {https://ui.adsabs.harvard.edu/abs/2008ConPh..49...71T}
}

@ARTICLE{Teske2025,
       author = {{Teske}, Johanna and {Batalha}, Natasha E. and {Wallack}, Nicole L. and {Kirk}, James and {Wogan}, Nicholas F. and {Gordon}, Tyler A. and {Alam}, Munazza K. and {Aguichine}, Artyom and {Wolfgang}, Angie and {Wakeford}, Hannah R. and {Scarsdale}, Nicholas and {Adams Redai}, Jea and {Moran}, Sarah E. and {L{\'o}pez-Morales}, Mercedes and {Meech}, Annabella and {Gao}, Peter and {Batalha}, Natalie M. and {Alderson}, Lili and {Gagnebin}, Anna},
        title = "{JWST COMPASS: NIRSpec/G395H Transmission Observations of TOI-776 c, a 2 R$_{{\ensuremath{\oplus}}}$ M Dwarf Planet}",
      journal = {\aj},
         year = 2025,
        month = may,
       volume = {169},
       number = {5},
          eid = {249},
        pages = {249},
          doi = {10.3847/1538-3881/adb975},
archivePrefix = {arXiv},
       eprint = {2502.20501},
 primaryClass = {astro-ph.EP},
       adsurl = {https://ui.adsabs.harvard.edu/abs/2025AJ....169..249T}
}

@ARTICLE{Cadieux2024,
       author = {{Cadieux}, Charles and {Doyon}, Ren{\'e} and {MacDonald}, Ryan J. and {Turbet}, Martin and {Artigau}, {\'E}tienne and {Lim}, Olivia and {Radica}, Michael and {Fauchez}, Thomas J. and {Salhi}, Salma and {Dang}, Lisa and {Albert}, Lo{\"\i}c and {Coulombe}, Louis-Philippe and {Cowan}, Nicolas B. and {Lafreni{\`e}re}, David and {L'Heureux}, Alexandrine and {Piaulet-Ghorayeb}, Caroline and {Benneke}, Bj{\"o}rn and {Cloutier}, Ryan and {Charnay}, Benjamin and {Cook}, Neil J. and {Fournier-Tondreau}, Marylou and {Plotnykov}, Mykhaylo and {Valencia}, Diana},
        title = "{Transmission Spectroscopy of the Habitable Zone Exoplanet LHS 1140 b with JWST/NIRISS}",
      journal = {\apjl},
         year = 2024,
        month = jul,
       volume = {970},
       number = {1},
          eid = {L2},
        pages = {L2},
          doi = {10.3847/2041-8213/ad5afa},
archivePrefix = {arXiv},
       eprint = {2406.15136},
 primaryClass = {astro-ph.EP},
       adsurl = {https://ui.adsabs.harvard.edu/abs/2024ApJ...970L...2C}
}

@ARTICLE{Radica2025,
       author = {{Radica}, Michael and {Piaulet-Ghorayeb}, Caroline and {Taylor}, Jake and {Coulombe}, Louis-Philippe and {Benneke}, Bj{\"o}rn and {Albert}, Loic and {Artigau}, {\'E}tienne and {Cowan}, Nicolas B. and {Doyon}, Ren{\'e} and {Lafreni{\`e}re}, David and {L'Heureux}, Alexandrine and {Lim}, Olivia},
        title = "{Promise and Peril: Stellar Contamination and Strict Limits on the Atmosphere Composition of TRAPPIST-1 c from JWST NIRISS Transmission Spectra}",
      journal = {\apjl},
         year = 2025,
        month = jan,
       volume = {979},
       number = {1},
          eid = {L5},
        pages = {L5},
          doi = {10.3847/2041-8213/ada381},
archivePrefix = {arXiv},
       eprint = {2409.19333},
 primaryClass = {astro-ph.EP},
       adsurl = {https://ui.adsabs.harvard.edu/abs/2025ApJ...979L...5R}
}

@ARTICLE{Benneke2024,
       author = {{Benneke}, Bj{\"o}rn and {Roy}, Pierre-Alexis and {Coulombe}, Louis-Philippe and {Radica}, Michael and {Piaulet}, Caroline and {Ahrer}, Eva-Maria and {Pierrehumbert}, Raymond and {Krissansen-Totton}, Joshua and {Schlichting}, Hilke E. and {Hu}, Renyu and {Yang}, Jeehyun and {Christie}, Duncan and {Thorngren}, Daniel and {Young}, Edward D. and {Pelletier}, Stefan and {Knutson}, Heather A. and {Miguel}, Yamila and {Evans-Soma}, Thomas M. and {Dorn}, Caroline and {Gagnebin}, Anna and {Fortney}, Jonathan J. and {Komacek}, Thaddeus and {MacDonald}, Ryan and {Raul}, Eshan and {Cloutier}, Ryan and {Acuna}, Lorena and {Lafreni{\`e}re}, David and {Cadieux}, Charles and {Doyon}, Ren{\'e} and {Welbanks}, Luis and {Allart}, Romain},
        title = "{JWST Reveals CH$_4$, CO$_2$, and H$_2$O in a Metal-rich Miscible Atmosphere on a Two-Earth-Radius Exoplanet}",
      journal = {arXiv e-prints},
         year = 2024,
        month = mar,
          eid = {arXiv:2403.03325},
        pages = {arXiv:2403.03325},
          doi = {10.48550/arXiv.2403.03325},
archivePrefix = {arXiv},
       eprint = {2403.03325},
 primaryClass = {astro-ph.EP},
       adsurl = {https://ui.adsabs.harvard.edu/abs/2024arXiv240303325B}
}

@ARTICLE{Madhusudhan2023K2-18b,
       author = {{Madhusudhan}, Nikku and {Sarkar}, Subhajit and {Constantinou}, Savvas and {Holmberg}, M{\r{a}}ns and {Piette}, Anjali A.~A. and {Moses}, Julianne I.},
        title = "{Carbon-bearing Molecules in a Possible Hycean Atmosphere}",
      journal = {\apjl},
         year = 2023,
        month = oct,
       volume = {956},
       number = {1},
          eid = {L13},
        pages = {L13},
          doi = {10.3847/2041-8213/acf577},
archivePrefix = {arXiv},
       eprint = {2309.05566},
 primaryClass = {astro-ph.EP},
       adsurl = {https://ui.adsabs.harvard.edu/abs/2023ApJ...956L..13M}
}

@ARTICLE{Madhusudhan2025K2-18b,
       author = {{Madhusudhan}, Nikku and {Constantinou}, Savvas and {Holmberg}, M{\r{a}}ns and {Sarkar}, Subhajit and {Piette}, Anjali A.~A. and {Moses}, Julianne I.},
        title = "{New Constraints on DMS and DMDS in the Atmosphere of K2-18 b from JWST MIRI}",
      journal = {\apjl},
         year = 2025,
        month = apr,
       volume = {983},
       number = {2},
          eid = {L40},
        pages = {L40},
          doi = {10.3847/2041-8213/adc1c8},
archivePrefix = {arXiv},
       eprint = {2504.12267},
 primaryClass = {astro-ph.EP},
       adsurl = {https://ui.adsabs.harvard.edu/abs/2025ApJ...983L..40M}
}

@ARTICLE{Beky2014,
       author = {{B{\'e}ky}, Bence and {Kipping}, David M. and {Holman}, Matthew J.},
        title = "{SPOTROD: a semi-analytic model for transits of spotted stars}",
      journal = {\mnras},
         year = 2014,
        month = aug,
       volume = {442},
       number = {4},
        pages = {3686-3699},
          doi = {10.1093/mnras/stu1061},
archivePrefix = {arXiv},
       eprint = {1407.4465},
 primaryClass = {astro-ph.EP},
       adsurl = {https://ui.adsabs.harvard.edu/abs/2014MNRAS.442.3686B}
}

@ARTICLE{Cabot2024,
       author = {{Cabot}, Samuel H.~C. and {Madhusudhan}, Nikku and {Constantinou}, Savvas and {Valencia}, Diana and {Vos}, Johanna M. and {Masseron}, Thomas and {Cheverall}, Connor J.},
        title = "{High-resolution Spectroscopic Reconnaissance of a Temperate Sub-Neptune}",
      journal = {\apjl},
         year = 2024,
        month = may,
       volume = {966},
       number = {1},
          eid = {L10},
        pages = {L10},
          doi = {10.3847/2041-8213/ad3828},
archivePrefix = {arXiv},
       eprint = {2403.18891},
 primaryClass = {astro-ph.EP},
       adsurl = {https://ui.adsabs.harvard.edu/abs/2024ApJ...966L..10C}
}

@article{yurchenko_exomol_2024,
    title = {{ExoMol} line lists – {LVII}. {High} accuracy ro-vibrational line list for methane ({CH4})},
    volume = {528},
    issn = {0035-8711},
    url = {https://doi.org/10.1093/mnras/stae148},
    doi = {10.1093/mnras/stae148},
    number = {2},
    urldate = {2025-06-24},
    journal = {Monthly Notices of the Royal Astronomical Society},
    author = {Yurchenko, Sergei N and Owens, Alec and Kefala, Kyriaki and Tennyson, Jonathan},
    month = feb,
    year = {2024},
    pages = {3719--3729},
}

@article{yurchenko_exomol_2020,
    title = {{ExoMol} line lists - {XXXIX}. {Ro}-vibrational molecular line list for {CO2}},
    volume = {496},
    issn = {0035-8711},
    url = {https://ui.adsabs.harvard.edu/abs/2020MNRAS.496.5282Y},
    doi = {10.1093/mnras/staa1874},
    urldate = {2025-06-24},
    journal = {Monthly Notices of the Royal Astronomical Society},
    author = {Yurchenko, S. N. and Mellor, Thomas M. and Freedman, Richard S. and Tennyson, J.},
    month = aug,
    year = {2020},
    note = {Publisher: OUP
ADS Bibcode: 2020MNRAS.496.5282Y},
    pages = {5282--5291},
}

@article{karman_update_2019,
    title = {Update of the {HITRAN} collision-induced absorption section},
    volume = {328},
    issn = {0019-1035},
    url = {https://ui.adsabs.harvard.edu/abs/2019Icar..328..160K},
    doi = {10.1016/j.icarus.2019.02.034},
    urldate = {2025-06-24},
    journal = {Icarus},
    author = {Karman, Tijs and Gordon, Iouli E. and van der Avoird, Ad and Baranov, Yury I. and Boulet, Christian and Drouin, Brian J. and Groenenboom, Gerrit C. and Gustafsson, Magnus and Hartmann, Jean-Michel and Kurucz, Robert L. and Rothman, Laurence S. and Sun, Kang and Sung, Keeyoon and Thalman, Ryan and Tran, Ha and Wishnow, Edward H. and Wordsworth, Robin and Vigasin, Andrey A. and Volkamer, Rainer and van der Zande, Wim J.},
    month = aug,
    year = {2019},
    note = {ADS Bibcode: 2019Icar..328..160K},
    pages = {160--175},
}

@misc{welbanks_challenges_2025,
    title = {The {Challenges} of {Detecting} {Gases} in {Exoplanet} {Atmospheres}},
    url = {http://arxiv.org/abs/2504.21788},
    doi = {10.48550/arXiv.2504.21788},
    language = {en},
    urldate = {2025-05-02},
    publisher = {arXiv},
    author = {Welbanks, Luis and Nixon, Matthew C. and McGill, Peter and Tilke, Lana J. and Wiser, Lindsey S. and Rotman, Yoav and Mukherjee, Sagnick and Feinstein, Adina and Line, Michael R. and Seager, Sara and Beatty, Thomas G. and Seligman, Darryl Z. and Parmentier, Vivien and Sing, David},
    month = apr,
    year = {2025},
    note = {arXiv:2504.21788 [astro-ph]},
}

@misc{pica-ciamarra_systematic_2025,
    title = {A {Systematic} {Search} for {Trace} {Molecules} in {Exoplanet} {K2}-18 b},
    url = {http://arxiv.org/abs/2505.10539},
    doi = {10.48550/arXiv.2505.10539},
    urldate = {2025-06-18},
    publisher = {arXiv},
    author = {Pica-Ciamarra, Lorenzo and Madhusudhan, Nikku and Cooke, Gregory J. and Constantinou, Savvas and Binet, Martin},
    month = may,
    year = {2025},
    note = {arXiv:2505.10539 [astro-ph]},
}

@ARTICLE{Schlawin2024,
       author = {{Schlawin}, Everett and {Ohno}, Kazumasa and {Bell}, Taylor J. and {Murphy}, Matthew M. and {Welbanks}, Luis and {Beatty}, Thomas G. and {Greene}, Thomas P. and {Fortney}, Jonathan J. and {Parmentier}, Vivien and {Edelman}, Isaac R. and {Gill}, Samuel and {Anderson}, David R. and {Wheatley}, Peter J. and {Henry}, Gregory W. and {Mehta}, Nishil and {Kreidberg}, Laura and {Rieke}, Marcia J.},
        title = "{Possible Carbon Dioxide above the Thick Aerosols of GJ 1214 b}",
      journal = {\apjl},
         year = 2024,
        month = oct,
       volume = {974},
       number = {2},
          eid = {L33},
        pages = {L33},
          doi = {10.3847/2041-8213/ad7fef},
archivePrefix = {arXiv},
       eprint = {2410.10183},
 primaryClass = {astro-ph.EP},
       adsurl = {https://ui.adsabs.harvard.edu/abs/2024ApJ...974L..33S}
}

@ARTICLE{Bello-Arufe2025,
       author = {{Bello-Arufe}, Aaron and {Damiano}, Mario and {Bennett}, Katherine A. and {Hu}, Renyu and {Welbanks}, Luis and {MacDonald}, Ryan J. and {Seligman}, Darryl Z. and {Sing}, David K. and {Tokadjian}, Armen and {Oza}, Apurva V. and {Yang}, Jeehyun},
        title = "{Evidence for a Volcanic Atmosphere on the Sub-Earth L 98-59 b}",
      journal = {\apjl},
         year = 2025,
        month = feb,
       volume = {980},
       number = {2},
          eid = {L26},
        pages = {L26},
          doi = {10.3847/2041-8213/adaf22},
archivePrefix = {arXiv},
       eprint = {2501.18680},
 primaryClass = {astro-ph.EP},
       adsurl = {https://ui.adsabs.harvard.edu/abs/2025ApJ...980L..26B}
}

@ARTICLE{Banerjee2024,
       author = {{Banerjee}, Agnibha and {Barstow}, Joanna K. and {Gressier}, Am{\'e}lie and {Espinoza}, N{\'e}stor and {Sing}, David K. and {Allen}, Natalie H. and {Birkmann}, Stephan M. and {Challener}, Ryan C. and {Crouzet}, Nicolas and {Haswell}, Carole A. and {Lewis}, Nikole K. and {Lewis}, Stephen R. and {Yang}, Jingxuan},
        title = "{Atmospheric Retrievals Suggest the Presence of a Secondary Atmosphere and Possible Sulfur Species on L98-59 d from JWST Nirspec G395H Transmission Spectroscopy}",
      journal = {\apjl},
         year = 2024,
        month = nov,
       volume = {975},
       number = {1},
          eid = {L11},
        pages = {L11},
          doi = {10.3847/2041-8213/ad73d0},
archivePrefix = {arXiv},
       eprint = {2408.15707},
 primaryClass = {astro-ph.EP},
       adsurl = {https://ui.adsabs.harvard.edu/abs/2024ApJ...975L..11B}
}

@ARTICLE{Ahrer2025,
       author = {{Ahrer}, Eva-Maria and {Radica}, Michael and {Piaulet-Ghorayeb}, Caroline and {Raul}, Eshan and {Wiser}, Lindsey and {Welbanks}, Luis and {Acu{\~n}a}, Lorena and {Allart}, Romain and {Coulombe}, Louis-Philippe and {Louca}, Amy and {MacDonald}, Ryan and {Saidel}, Morgan and {Evans-Soma}, Thomas M. and {Benneke}, Bj{\"o}rn and {Christie}, Duncan and {Beatty}, Thomas G. and {Cadieux}, Charles and {Cloutier}, Ryan and {Doyon}, Ren{\'e} and {Fortney}, Jonathan J. and {Gagnebin}, Anna and {Gapp}, Cyril and {Innes}, Hamish and {Knutson}, Heather A. and {Komacek}, Thaddeus and {Krissansen-Totton}, Joshua and {Miguel}, Yamila and {Pierrehumbert}, Raymond and {Roy}, Pierre-Alexis and {Schlichting}, Hilke E.},
        title = "{Escaping Helium and a Highly Muted Spectrum Suggest a Metal-enriched Atmosphere on Sub-Neptune GJ 3090 b from JWST Transit Spectroscopy}",
      journal = {\apjl},
         year = 2025,
        month = may,
       volume = {985},
       number = {1},
          eid = {L10},
        pages = {L10},
          doi = {10.3847/2041-8213/add010},
archivePrefix = {arXiv},
       eprint = {2504.20428},
 primaryClass = {astro-ph.EP},
       adsurl = {https://ui.adsabs.harvard.edu/abs/2025ApJ...985L..10A}
}

@ARTICLE{Piaulet-Ghorayeb2024,
       author = {{Piaulet-Ghorayeb}, Caroline and {Benneke}, Bj{\"o}rn and {Radica}, Michael and {Raul}, Eshan and {Coulombe}, Louis-Philippe and {Ahrer}, Eva-Maria and {Kubyshkina}, Daria and {Howard}, Ward S. and {Krissansen-Totton}, Joshua and {MacDonald}, Ryan J. and {Roy}, Pierre-Alexis and {Louca}, Amy and {Christie}, Duncan and {Fournier-Tondreau}, Marylou and {Allart}, Romain and {Miguel}, Yamila and {Schlichting}, Hilke E. and {Welbanks}, Luis and {Cadieux}, Charles and {Dorn}, Caroline and {Evans-Soma}, Thomas M. and {Fortney}, Jonathan J. and {Pierrehumbert}, Raymond and {Lafreni{\`e}re}, David and {Acu{\~n}a}, Lorena and {Komacek}, Thaddeus and {Innes}, Hamish and {Beatty}, Thomas G. and {Cloutier}, Ryan and {Doyon}, Ren{\'e} and {Gagnebin}, Anna and {Gapp}, Cyril and {Knutson}, Heather A.},
        title = "{JWST/NIRISS Reveals the Water-rich ``Steam World'' Atmosphere of GJ 9827 d}",
      journal = {\apjl},
         year = 2024,
        month = oct,
       volume = {974},
       number = {1},
          eid = {L10},
        pages = {L10},
          doi = {10.3847/2041-8213/ad6f00},
archivePrefix = {arXiv},
       eprint = {2410.03527},
 primaryClass = {astro-ph.EP},
       adsurl = {https://ui.adsabs.harvard.edu/abs/2024ApJ...974L..10P}
}

@ARTICLE{Alderson2024,
       author = {{Alderson}, Lili and {Batalha}, Natasha E. and {Wakeford}, Hannah R. and {Wallack}, Nicole L. and {Aguichine}, Artyom and {Teske}, Johanna and {Adams Redai}, Jea and {Alam}, Munazza K. and {Batalha}, Natalie M. and {Gao}, Peter and {Kirk}, James and {L{\'o}pez-Morales}, Mercedes and {Moran}, Sarah E. and {Scarsdale}, Nicholas and {Wogan}, Nicholas F. and {Wolfgang}, Angie},
        title = "{JWST COMPASS: NIRSpec/G395H Transmission Observations of the Super-Earth TOI-836b}",
      journal = {\aj},
         year = 2024,
        month = may,
       volume = {167},
       number = {5},
          eid = {216},
        pages = {216},
          doi = {10.3847/1538-3881/ad32c9},
archivePrefix = {arXiv},
       eprint = {2404.00093},
 primaryClass = {astro-ph.EP},
       adsurl = {https://ui.adsabs.harvard.edu/abs/2024AJ....167..216A}
}

@ARTICLE{Davenport2025,
       author = {{Davenport}, Brian and {Kempton}, Eliza M. -R. and {Nixon}, Matthew C. and {Ih}, Jegug and {Deming}, Drake and {Fu}, Guangwei and {May}, E.~M. and {Bean}, Jacob L. and {Gao}, Peter and {Rogers}, Leslie and {Malik}, Matej},
        title = "{TOI-421 b: A Hot Sub-Neptune with a Haze-free, Low Mean Molecular Weight Atmosphere}",
      journal = {\apjl},
         year = 2025,
        month = may,
       volume = {984},
       number = {2},
          eid = {L44},
        pages = {L44},
          doi = {10.3847/2041-8213/adcd76},
archivePrefix = {arXiv},
       eprint = {2501.01498},
 primaryClass = {astro-ph.EP},
       adsurl = {https://ui.adsabs.harvard.edu/abs/2025ApJ...984L..44D}
}

@ARTICLE{Morely2017,
       author = {{Morley}, Caroline V. and {Kreidberg}, Laura and {Rustamkulov}, Zafar and {Robinson}, Tyler and {Fortney}, Jonathan J.},
        title = "{Observing the Atmospheres of Known Temperate Earth-sized Planets with JWST}",
      journal = {\apj},
         year = 2017,
        month = dec,
       volume = {850},
       number = {2},
          eid = {121},
        pages = {121},
          doi = {10.3847/1538-4357/aa927b},
archivePrefix = {arXiv},
       eprint = {1708.04239},
 primaryClass = {astro-ph.EP},
       adsurl = {https://ui.adsabs.harvard.edu/abs/2017ApJ...850..121M}
}

@ARTICLE{Lustig-Yaeger2019,
       author = {{Lustig-Yaeger}, Jacob and {Meadows}, Victoria S. and {Lincowski}, Andrew P.},
        title = "{The Detectability and Characterization of the TRAPPIST-1 Exoplanet Atmospheres with JWST}",
      journal = {\aj},
         year = 2019,
        month = jul,
       volume = {158},
       number = {1},
          eid = {27},
        pages = {27},
          doi = {10.3847/1538-3881/ab21e0},
archivePrefix = {arXiv},
       eprint = {1905.07070},
 primaryClass = {astro-ph.EP},
       adsurl = {https://ui.adsabs.harvard.edu/abs/2019AJ....158...27L}
}

@ARTICLE{Mukherjee2025_CS2,
       author = {{Mukherjee}, Sagnick and {Fortney}, Jonathan J. and {Wogan}, Nicholas F. and {Sing}, David K. and {Ohno}, Kazumasa},
        title = "{Effects of Planetary Parameters on Disequilibrium Chemistry in Irradiated Planetary Atmospheres: From Gas Giants to Sub-Neptunes}",
      journal = {\apj},
         year = 2025,
        month = jun,
       volume = {985},
       number = {2},
          eid = {209},
        pages = {209},
          doi = {10.3847/1538-4357/adc7b3},
archivePrefix = {arXiv},
       eprint = {2410.17169},
 primaryClass = {astro-ph.EP},
       adsurl = {https://ui.adsabs.harvard.edu/abs/2025ApJ...985..209M}
}

@ARTICLE{Visscher2006,
       author = {{Visscher}, Channon and {Lodders}, Katharina and {Fegley}, Jr., Bruce},
        title = "{Atmospheric Chemistry in Giant Planets, Brown Dwarfs, and Low-Mass Dwarf Stars. II. Sulfur and Phosphorus}",
      journal = {\apj},
         year = 2006,
        month = sep,
       volume = {648},
       number = {2},
        pages = {1181-1195},
          doi = {10.1086/506245},
archivePrefix = {arXiv},
       eprint = {astro-ph/0511136},
 primaryClass = {astro-ph},
       adsurl = {https://ui.adsabs.harvard.edu/abs/2006ApJ...648.1181V}
}

@ARTICLE{Zahnle2009soot,
       author = {{Zahnle}, K. and {Marley}, M.~S. and {Fortney}, J.~J.},
        title = "{Thermometric Soots on Warm Jupiters?}",
      journal = {arXiv e-prints},
         year = 2009,
        month = nov,
          eid = {arXiv:0911.0728},
        pages = {arXiv:0911.0728},
          doi = {10.48550/arXiv.0911.0728},
archivePrefix = {arXiv},
       eprint = {0911.0728},
 primaryClass = {astro-ph.EP},
       adsurl = {https://ui.adsabs.harvard.edu/abs/2009arXiv0911.0728Z}
}

@INPROCEEDINGS{Moses2024,
       author = {{Moses}, Julianne and {Tsai}, Shang-Min and {Fortney}, Jonathan and {Constantinou}, Savvas and {Madhusudhan}, Nikku and {Visscher}, Channon and {Yu}, Xinting and {Plane}, John and {Yang}, Jeehyun and {Zahnle}, Kevin and {Lee}, Elspeth},
        title = "{Sulfur photochemistry on warm sub-Neptune and Neptune-class exoplanets}",
    booktitle = {56th Annual Meeting of the Division for Planetary Sciences},
         year = 2024,
       series = {AAS/Division for Planetary Sciences Meeting Abstracts},
       volume = {56},
        month = oct,
          eid = {308.06},
        pages = {308.06},
       adsurl = {https://ui.adsabs.harvard.edu/abs/2024DPS....5630806M}
}

@ARTICLE{Veillet2025,
       author = {{Veillet}, R. and {Venot}, O. and {Sirjean}, B. and {Citrangolo Destro}, F. and {Fournet}, R. and {Al-Refaie}, A. and {H{\'e}brard}, E. and {Glaude}, P-A. and {Bounaceur}, R.},
        title = "{Inclusion of sulfur chemistry in a validated C/H/O/N chemical network: identification of key C/S coupling pathways}",
      journal = {arXiv e-prints},
         year = 2025,
        month = may,
          eid = {arXiv:2505.12152},
        pages = {arXiv:2505.12152},
          doi = {10.48550/arXiv.2505.12152},
archivePrefix = {arXiv},
       eprint = {2505.12152},
 primaryClass = {astro-ph.EP},
       adsurl = {https://ui.adsabs.harvard.edu/abs/2025arXiv250512152V}
}

@ARTICLE{Luque2022,
       author = {{Luque}, Rafael and {Pall{\'e}}, Enric},
        title = "{Density, not radius, separates rocky and water-rich small planets orbiting M dwarf stars}",
      journal = {Science},
         year = 2022,
        month = sep,
       volume = {377},
       number = {6611},
        pages = {1211-1214},
          doi = {10.1126/science.abl7164},
archivePrefix = {arXiv},
       eprint = {2209.03871},
 primaryClass = {astro-ph.EP},
       adsurl = {https://ui.adsabs.harvard.edu/abs/2022Sci...377.1211L}
}

@article{owens_exomol_2024,
    title = {{ExoMol} line lists - {LVIII}. {High}-temperature molecular line list of carbonyl sulphide ({OCS})},
    volume = {530},
    issn = {0035-8711},
    url = {https://ui.adsabs.harvard.edu/abs/2024MNRAS.530.4004O},
    doi = {10.1093/mnras/stae1110},
    urldate = {2025-07-08},
    journal = {Monthly Notices of the Royal Astronomical Society},
    author = {Owens, Alec and Yurchenko, Sergei N. and Tennyson, Jonathan},
    month = jun,
    year = {2024},
    note = {Publisher: OUP
ADS Bibcode: 2024MNRAS.530.4004O},
    pages = {4004--4015},
}

@article{hargreaves_spectroscopic_2019,
    title = {Spectroscopic line parameters of {NO}, {NO2}, and {N2O} for the {HITEMP} database},
    volume = {232},
    issn = {0022-4073},
    url = {https://ui.adsabs.harvard.edu/abs/2019JQSRT.232...35H},
    doi = {10.1016/j.jqsrt.2019.04.040},
    urldate = {2025-07-08},
    journal = {Journal of Quantitative Spectroscopy and Radiative Transfer},
    author = {Hargreaves, Robert J. and Gordon, Iouli E. and Rothman, Laurence S. and Tashkun, Sergey A. and Perevalov, Valery I. and Lukashevskaya, Anastasiya A. and Yurchenko, Sergey N. and Tennyson, Jonathan and Müller, Holger S. P.},
    month = jul,
    year = {2019},
    note = {Publisher: Elsevier
ADS Bibcode: 2019JQSRT.232...35H},
    pages = {35--53},
}

@ARTICLE{Beatty2024,
       author = {{Beatty}, Thomas G. and {Welbanks}, Luis and {Schlawin}, Everett and {Bell}, Taylor J. and {Line}, Michael R. and {Murphy}, Matthew and {Edelman}, Isaac and {Greene}, Thomas P. and {Fortney}, Jonathan J. and {Henry}, Gregory W. and {Mukherjee}, Sagnick and {Ohno}, Kazumasa and {Parmentier}, Vivien and {Rauscher}, Emily and {Wiser}, Lindsey S. and {Arnold}, Kenneth E.},
        title = "{Sulfur Dioxide and Other Molecular Species in the Atmosphere of the Sub-Neptune GJ 3470 b}",
      journal = {\apjl},
         year = 2024,
        month = jul,
       volume = {970},
       number = {1},
          eid = {L10},
        pages = {L10},
          doi = {10.3847/2041-8213/ad55e9},
archivePrefix = {arXiv},
       eprint = {2406.04450},
 primaryClass = {astro-ph.EP},
       adsurl = {https://ui.adsabs.harvard.edu/abs/2024ApJ...970L..10B}
}

@ARTICLE{Radica2024_LTT9779,
       author = {{Radica}, Michael and {Coulombe}, Louis-Philippe and {Taylor}, Jake and {Albert}, Loic and {Allart}, Romain and {Benneke}, Bj{\"o}rn and {Cowan}, Nicolas B. and {Dang}, Lisa and {Lafreni{\`e}re}, David and {Thorngren}, Daniel and {Artigau}, {\'E}tienne and {Doyon}, Ren{\'e} and {Flagg}, Laura and {Johnstone}, Doug and {Pelletier}, Stefan and {Roy}, Pierre-Alexis},
        title = "{Muted Features in the JWST NIRISS Transmission Spectrum of Hot Neptune LTT 9779b}",
      journal = {\apjl},
         year = 2024,
        month = feb,
       volume = {962},
       number = {1},
          eid = {L20},
        pages = {L20},
          doi = {10.3847/2041-8213/ad20e4},
archivePrefix = {arXiv},
       eprint = {2401.15548},
 primaryClass = {astro-ph.EP},
       adsurl = {https://ui.adsabs.harvard.edu/abs/2024ApJ...962L..20R}
}

@ARTICLE{Coulombe2025,
       author = {{Coulombe}, Louis-Philippe and {Radica}, Michael and {Benneke}, Bj{\"o}rn and {D'Aoust}, {\'E}lyse and {Dang}, Lisa and {Cowan}, Nicolas B. and {Parmentier}, Vivien and {Albert}, Lo{\"\i}c and {Lafreni{\`e}re}, David and {Taylor}, Jake and {Roy}, Pierre-Alexis and {Pelletier}, Stefan and {Allart}, Romain and {Artigau}, {\'E}tienne and {Doyon}, Ren{\'e} and {Jayawardhana}, Ray and {Johnstone}, Doug and {Kaltenegger}, Lisa and {Langeveld}, Adam B. and {MacDonald}, Ryan J. and {Rowe}, Jason F. and {Turner}, Jake D.},
        title = "{Highly reflective white clouds on the western dayside of an exo-Neptune}",
      journal = {Nature Astronomy},
         year = 2025,
        month = apr,
       volume = {9},
        pages = {512-525},
          doi = {10.1038/s41550-025-02488-9},
archivePrefix = {arXiv},
       eprint = {2501.14016},
 primaryClass = {astro-ph.EP},
       adsurl = {https://ui.adsabs.harvard.edu/abs/2025NatAs...9..512C}
}

@INPROCEEDINGS{Ferruit2014,
       author = {{Ferruit}, P. and {Birkmann}, S. and {B{\"o}ker}, T. and {Sirianni}, M. and {Giardino}, G. and {de Marchi}, G. and {Alves de Oliveira}, C. and {Dorner}, B.},
        title = "{Observing transiting exoplanets with JWST/NIRSpec}",
    booktitle = {Space Telescopes and Instrumentation 2014: Optical, Infrared, and Millimeter Wave},
         year = 2014,
       editor = {{Oschmann}, Jr., Jacobus M. and {Clampin}, Mark and {Fazio}, Giovanni G. and {MacEwen}, Howard A.},
       series = {Society of Photo-Optical Instrumentation Engineers (SPIE) Conference Series},
       volume = {9143},
        month = aug,
          eid = {91430A},
        pages = {91430A},
          doi = {10.1117/12.2054756},
       adsurl = {https://ui.adsabs.harvard.edu/abs/2014SPIE.9143E..0AF}
}

@INPROCEEDINGS{Doyon2012,
       author = {{Doyon}, Ren{\'e} and {Hutchings}, John B. and {Beaulieu}, Mathilde and {Albert}, Loic and {Lafreni{\`e}re}, David and {Willott}, Chris and {Touahri}, Driss and {Rowlands}, Neil and {Maszkiewicz}, Micheal and {Fullerton}, Alex W. and {Volk}, Kevin and {Martel}, Andr{\'e} R. and {Chayer}, Pierre and {Sivaramakrishnan}, Anand and {Abraham}, Roberto and {Ferrarese}, Laura and {Jayawardhana}, Ray and {Johnstone}, Doug and {Meyer}, Michael and {Pipher}, Judith L. and {Sawicki}, Marcin},
        title = "{The JWST Fine Guidance Sensor (FGS) and Near-Infrared Imager and Slitless Spectrograph (NIRISS)}",
    booktitle = {Space Telescopes and Instrumentation 2012: Optical, Infrared, and Millimeter Wave},
         year = 2012,
       editor = {{Clampin}, Mark C. and {Fazio}, Giovanni G. and {MacEwen}, Howard A. and {Oschmann}, Jr., Jacobus M.},
       series = {Society of Photo-Optical Instrumentation Engineers (SPIE) Conference Series},
       volume = {8442},
        month = sep,
          eid = {84422R},
        pages = {84422R},
          doi = {10.1117/12.926578},
       adsurl = {https://ui.adsabs.harvard.edu/abs/2012SPIE.8442E..2RD}
}

@ARTICLE{Cheverall2024,
       author = {{Cheverall}, Connor J. and {Madhusudhan}, Nikku},
        title = "{Feasibility of High-resolution Transmission Spectroscopy for Low-velocity Exoplanets}",
      journal = {\aj},
         year = 2024,
        month = jun,
       volume = {167},
       number = {6},
          eid = {272},
        pages = {272},
          doi = {10.3847/1538-3881/ad380c},
archivePrefix = {arXiv},
       eprint = {2403.18894},
 primaryClass = {astro-ph.EP},
       adsurl = {https://ui.adsabs.harvard.edu/abs/2024AJ....167..272C}
}

@ARTICLE{Awiphan2016,
       author = {{Awiphan}, S. and {Kerins}, E. and {Pichadee}, S. and {Komonjinda}, S. and {Dhillon}, V.~S. and {Rujopakarn}, W. and {Poshyachinda}, S. and {Marsh}, T.~R. and {Reichart}, D.~E. and {Ivarsen}, K.~M. and {Haislip}, J.~B.},
        title = "{Transit timing variation and transmission spectroscopy analyses of the hot Neptune GJ3470b}",
      journal = {\mnras},
         year = 2016,
        month = dec,
       volume = {463},
       number = {3},
        pages = {2574-2582},
          doi = {10.1093/mnras/stw2148},
archivePrefix = {arXiv},
       eprint = {1606.02962},
 primaryClass = {astro-ph.EP},
       adsurl = {https://ui.adsabs.harvard.edu/abs/2016MNRAS.463.2574A}
}

@ARTICLE{Jenkins2020,
       author = {{Jenkins}, James S. and {D{\'\i}az}, Mat{\'\i}as R. and {Kurtovic}, Nicol{\'a}s T. and {Espinoza}, N{\'e}stor and {Vines}, Jose I. and {Rojas}, Pablo A. Pe{\~n}a and {Brahm}, Rafael and {Torres}, Pascal and {Cort{\'e}s-Zuleta}, P{\'\i}a and {Soto}, Maritza G. and {Lopez}, Eric D. and {King}, George W. and {Wheatley}, Peter J. and {Winn}, Joshua N. and {Ciardi}, David R. and {Ricker}, George and {Vanderspek}, Roland and {Latham}, David W. and {Seager}, Sara and {Jenkins}, Jon M. and {Beichman}, Charles A. and {Bieryla}, Allyson and {Burke}, Christopher J. and {Christiansen}, Jessie L. and {Henze}, Christopher E. and {Klaus}, Todd C. and {McCauliff}, Sean and {Mori}, Mayuko and {Narita}, Norio and {Nishiumi}, Taku and {Tamura}, Motohide and {de Leon}, Jerome Pitogo and {Quinn}, Samuel N. and {Villase{\~n}or}, Jesus Noel and {Vezie}, Michael and {Lissauer}, Jack J. and {Collins}, Karen A. and {Collins}, Kevin I. and {Isopi}, Giovanni and {Mallia}, Franco and {Ercolino}, Andrea and {Petrovich}, Cristobal and {Jord{\'a}n}, Andr{\'e}s and {Acton}, Jack S. and {Armstrong}, David J. and {Bayliss}, Daniel and {Bouchy}, Fran{\c{c}}ois and {Belardi}, Claudia and {Bryant}, Edward M. and {Burleigh}, Matthew R. and {Cabrera}, Juan and {Casewell}, Sarah L. and {Chaushev}, Alexander and {Cooke}, Benjamin F. and {Eigm{\"u}ller}, Philipp and {Erikson}, Anders and {Foxell}, Emma and {G{\"a}nsicke}, Boris T. and {Gill}, Samuel and {Gillen}, Edward and {G{\"u}nther}, Maximilian N. and {Goad}, Michael R. and {Hooton}, Matthew J. and {Jackman}, James A.~G. and {Louden}, Tom and {McCormac}, James and {Moyano}, Maximiliano and {Nielsen}, Louise D. and {Pollacco}, Don and {Queloz}, Didier and {Rauer}, Heike and {Raynard}, Liam and {Smith}, Alexis M.~S. and {Tilbrook}, Rosanna H. and {Titz-Weider}, Ruth and {Turner}, Oliver and {Udry}, St{\'e}phane and {Walker}, Simon. R. and {Watson}, Christopher A. and {West}, Richard G. and {Palle}, Enric and {Ziegler}, Carl and {Law}, Nicholas and {Mann}, Andrew W.},
        title = "{An ultrahot Neptune in the Neptune desert}",
      journal = {Nature Astronomy},
         year = 2020,
        month = jan,
       volume = {4},
        pages = {1148-1157},
          doi = {10.1038/s41550-020-1142-z},
archivePrefix = {arXiv},
       eprint = {2009.12832},
 primaryClass = {astro-ph.EP},
       adsurl = {https://ui.adsabs.harvard.edu/abs/2020NatAs...4.1148J}
}

@ARTICLE{Kosiarek2019,
       author = {{Kosiarek}, Molly R. and {Crossfield}, Ian J.~M. and {Hardegree-Ullman}, Kevin K. and {Livingston}, John H. and {Benneke}, Bj{\"o}rn and {Henry}, Gregory W. and {Howard}, Ward S. and {Berardo}, David and {Blunt}, Sarah and {Fulton}, Benjamin J. and {Hirsch}, Lea A. and {Howard}, Andrew W. and {Isaacson}, Howard and {Petigura}, Erik A. and {Sinukoff}, Evan and {Weiss}, Lauren and {Bonfils}, X. and {Dressing}, Courtney D. and {Knutson}, Heather A. and {Schlieder}, Joshua E. and {Werner}, Michael and {Gorjian}, Varoujan and {Krick}, Jessica and {Morales}, Farisa Y. and {Astudillo-Defru}, Nicola and {Almenara}, J. -M. and {Delfosse}, X. and {Forveille}, T. and {Lovis}, C. and {Mayor}, M. and {Murgas}, F. and {Pepe}, F. and {Santos}, N.~C. and {Udry}, S. and {Corbett}, H.~T. and {Fors}, Octavi and {Law}, Nicholas M. and {Ratzloff}, Jeffrey K. and {del Ser}, Daniel},
        title = "{Bright Opportunities for Atmospheric Characterization of Small Planets: Masses and Radii of K2-3 b, c, and d and GJ3470 b from Radial Velocity Measurements and Spitzer Transits}",
      journal = {\aj},
         year = 2019,
        month = mar,
       volume = {157},
       number = {3},
          eid = {97},
        pages = {97},
          doi = {10.3847/1538-3881/aaf79c},
archivePrefix = {arXiv},
       eprint = {1812.08241},
 primaryClass = {astro-ph.EP},
       adsurl = {https://ui.adsabs.harvard.edu/abs/2019AJ....157...97K}
}

@article{daviau_experimental_2021,
    title = {Experimental {Constraints} on {Solid} {Nitride} {Phases} in {Rocky} {Mantles} of {Reduced} {Planets} and {Implications} for {Observable} {Atmosphere} {Compositions}},
    volume = {126},
    copyright = {© 2021. The Authors.},
    issn = {2169-9100},
    url = {https://onlinelibrary.wiley.com/doi/abs/10.1029/2020JE006687},
    doi = {10.1029/2020JE006687},
    language = {en},
    number = {9},
    urldate = {2024-02-21},
    journal = {Journal of Geophysical Research: Planets},
    author = {Daviau, Kierstin and Lee, Kanani K. M.},
    year = {2021},
    note = {\_eprint: https://onlinelibrary.wiley.com/doi/pdf/10.1029/2020JE006687},
    pages = {e2020JE006687},
}

@article{macdonald_poseidon_2024,
    title = {{POSEIDON}: {Multidimensional} atmospheric retrieval of exoplanet spectra},
    shorttitle = {{POSEIDON}},
    url = {https://ui.adsabs.harvard.edu/abs/2024ascl.soft12028M},
    urldate = {2025-07-10},
    journal = {Astrophysics Source Code Library},
    author = {MacDonald, Ryan J. and Madhusudhan, Nikku},
    month = dec,
    year = {2024},
    note = {ADS Bibcode: 2024ascl.soft12028M},
    pages = {ascl:2412.028},
}

@ARTICLE{Speagle2020,
       author = {{Speagle}, Joshua S.},
        title = "{DYNESTY: a dynamic nested sampling package for estimating Bayesian posteriors and evidences}",
      journal = {\mnras},
         year = 2020,
        month = apr,
       volume = {493},
       number = {3},
        pages = {3132-3158},
          doi = {10.1093/mnras/staa278},
archivePrefix = {arXiv},
       eprint = {1904.02180},
 primaryClass = {astro-ph.IM},
       adsurl = {https://ui.adsabs.harvard.edu/abs/2020MNRAS.493.3132S}
}

@article{owens_exomol_2021,
    title = {{ExoMol} line lists – {XLI}. {High}-temperature molecular line lists for the alkali metal hydroxides {KOH} and {NaOH}},
    volume = {502},
    issn = {0035-8711},
    url = {https://doi.org/10.1093/mnras/staa4041},
    doi = {10.1093/mnras/staa4041},
    number = {1},
    urldate = {2025-11-07},
    journal = {Monthly Notices of the Royal Astronomical Society},
    author = {Owens, A and Tennyson, J and Yurchenko, S N},
    month = mar,
    year = {2021},
    pages = {1128--1135},
}

@article{li_direct_2011,
    title = {Direct fit of experimental ro-vibrational intensities to the dipole moment function: {Application} to {HCl}},
    volume = {112},
    issn = {0022-4073},
    shorttitle = {Direct fit of experimental ro-vibrational intensities to the dipole moment function},
    url = {https://www.sciencedirect.com/science/article/pii/S0022407311001336},
    doi = {10.1016/j.jqsrt.2011.03.014},
    number = {10},
    urldate = {2025-11-07},
    journal = {Journal of Quantitative Spectroscopy and Radiative Transfer},
    author = {Li, G. and Gordon, I. E. and Bernath, P. F. and Rothman, L. S},
    month = jul,
    year = {2011},
    pages = {1543--1550},
}

@article{barton_exomol_2014,
    title = {{ExoMol} molecular line lists {V}: the ro-vibrational spectra of {NaCl} and {KCl}},
    volume = {442},
    issn = {0035-8711},
    shorttitle = {{ExoMol} molecular line lists {V}},
    url = {https://doi.org/10.1093/mnras/stu944},
    doi = {10.1093/mnras/stu944},
    number = {2},
    urldate = {2025-11-07},
    journal = {Monthly Notices of the Royal Astronomical Society},
    author = {Barton, Emma J. and Chiu, Christopher and Golpayegani, Shirin and Yurchenko, Sergei N. and Tennyson, Jonathan and Frohman, Daniel J. and Bernath, Peter F.},
    month = aug,
    year = {2014},
    pages = {1821--1829},
}

@article{hill_temperature-dependent_2013,
    title = {Temperature-dependent molecular absorption cross sections for exoplanets and other atmospheres},
    volume = {226},
    issn = {0019-1035},
    url = {https://www.sciencedirect.com/science/article/pii/S0019103512003041},
    doi = {10.1016/j.icarus.2012.07.028},
    number = {2},
    urldate = {2025-11-07},
    journal = {Icarus},
    author = {Hill, Christian and Yurchenko, Sergei N. and Tennyson, Jonathan},
    month = nov,
    year = {2013},
    pages = {1673--1677},
}

@article{li_reference_2013,
    title = {Reference spectroscopic data for hydrogen halides. {Part} {I}: {Construction} and validation of the ro-vibrational dipole moment functions},
    volume = {121},
    issn = {0022-4073},
    shorttitle = {Reference spectroscopic data for hydrogen halides. {Part} {I}},
    url = {https://www.sciencedirect.com/science/article/pii/S0022407313000599},
    doi = {10.1016/j.jqsrt.2013.02.005},
    urldate = {2025-11-07},
    journal = {Journal of Quantitative Spectroscopy and Radiative Transfer},
    author = {Li, Gang and Gordon, Iouli E. and Le Roy, Robert J. and Hajigeorgiou, Photos G. and Coxon, John A. and Bernath, Peter F. and Rothman, Laurence S.},
    month = may,
    year = {2013},
    pages = {78--90},
}

@article{coxon_improved_2015,
    title = {Improved direct potential fit analyses for the ground electronic states of the hydrogen halides: {HF}/{DF}/{TF}, {HCl}/{DCl}/{TCl}, {HBr}/{DBr}/{TBr} and {HI}/{DI}/{TI}},
    volume = {151},
    issn = {0022-4073},
    shorttitle = {Improved direct potential fit analyses for the ground electronic states of the hydrogen halides},
    url = {https://www.sciencedirect.com/science/article/pii/S0022407314003781},
    doi = {10.1016/j.jqsrt.2014.08.028},
    urldate = {2025-11-07},
    journal = {Journal of Quantitative Spectroscopy and Radiative Transfer},
    author = {Coxon, John A. and Hajigeorgiou, Photos G.},
    month = jan,
    year = {2015},
    pages = {133--154},
}

@article{mellor_exomol_2023,
    title = {{ExoMol} line lists – {XLVIII}. {High}-temperature line list of thioformaldehyde ({H2CS})},
    volume = {520},
    issn = {0035-8711},
    url = {https://doi.org/10.1093/mnras/stad111},
    doi = {10.1093/mnras/stad111},
    number = {2},
    urldate = {2025-11-07},
    journal = {Monthly Notices of the Royal Astronomical Society},
    author = {Mellor, Thomas and Owens, Alec and Tennyson, Jonathan and Yurchenko, Sergei N},
    month = apr,
    year = {2023},
    pages = {1997--2008},
}

@article{madhusudhan_exoplanetary_2016,
    title = {Exoplanetary {Atmospheres} - {Chemistry}, {Formation} {Conditions}, and {Habitability}},
    volume = {205},
    issn = {0038-6308, 1572-9672},
    url = {http://arxiv.org/abs/1604.06092},
    doi = {10.1007/s11214-016-0254-3},
    language = {en},
    number = {1-4},
    urldate = {2025-11-25},
    journal = {Space Science Reviews},
    author = {Madhusudhan, Nikku and Agúndez, Marcelino and Moses, Julianne I. and Hu, Yongyun},
    month = dec,
    year = {2016},
    note = {arXiv:1604.06092 [astro-ph]},
    pages = {285--348},
}

@article{madhusudhan_exploring_2025,
    title = {Exploring the sub-{Neptune} frontier with {JWST}},
    volume = {122},
    url = {https://www.pnas.org/doi/10.1073/pnas.2416194122},
    doi = {10.1073/pnas.2416194122},
    number = {39},
    urldate = {2025-11-28},
    journal = {Proceedings of the National Academy of Sciences},
    author = {Madhusudhan, Nikku and Holmberg, Måns and Constantinou, Savvas and Cooke, Gregory J.},
    month = sep,
    year = {2025},
    note = {Publisher: Proceedings of the National Academy of Sciences},
    pages = {e2416194122},
}

@article{rigby_jwst_2025,
    title = {A {JWST} {Transmission} {Spectrum} of the {Temperate} {Sub}-{Neptune} {TOI}-732 c},
    volume = {995},
    issn = {2041-8205, 2041-8213},
    url = {https://iopscience.iop.org/article/10.3847/2041-8213/ae247d},
    doi = {10.3847/2041-8213/ae247d},
    language = {en},
    number = {2},
    urldate = {2026-02-20},
    journal = {The Astrophysical Journal Letters},
    author = {Rigby, Frances E. and Madhusudhan, Nikku and Sarkar, Subhajit and Pica-Ciamarra, Lorenzo and Holmberg, Måns and Moses, Julianne I.},
    month = dec,
    year = {2025},
    pages = {L70},
}

@article{constantinou_atmospheric_2026,
    title = {The atmospheric composition of {TOI}-270 d},
    volume = {705},
    copyright = {https://creativecommons.org/licenses/by/4.0},
    issn = {0004-6361, 1432-0746},
    url = {https://www.aanda.org/10.1051/0004-6361/202452192},
    doi = {10.1051/0004-6361/202452192},
    language = {en},
    urldate = {2026-01-13},
    journal = {Astronomy \& Astrophysics},
    author = {Constantinou, Savvas and Madhusudhan, Nikku and Holmberg, Måns},
    month = jan,
    year = {2026},
    pages = {A25},
}

@ARTICLE{Alderson2025,
       author = {{Alderson}, Lili and {Moran}, Sarah E. and {Wallack}, Nicole L. and {Batalha}, Natasha E. and {Wogan}, Nicholas F. and {Dattilo}, Anne and {Wakeford}, Hannah R. and {Redai}, Jea Adams and {Alam}, Munazza K. and {Aguichine}, Artyom and {Batalha}, Natalie M. and {Gagnebin}, Anna and {Gao}, Peter and {Kirk}, James and {L{\'o}pez-Morales}, Mercedes and {Meech}, Annabella and {Teske}, Johanna and {Wolfgang}, Angie},
        title = "{JWST COMPASS: NIRSpec/G395H Transmission Observations of the Super-Earth TOI-776 b}",
      journal = {\aj},
         year = 2025,
        month = mar,
       volume = {169},
       number = {3},
          eid = {142},
        pages = {142},
          doi = {10.3847/1538-3881/adad64},
archivePrefix = {arXiv},
       eprint = {2501.14596},
 primaryClass = {astro-ph.EP},
       adsurl = {https://ui.adsabs.harvard.edu/abs/2025AJ....169..142A}
}

@ARTICLE{Wallack2026,
       author = {{Wallack}, Nicole L. and {Gao}, Peter and {Greklek-McKeon}, Michael and {Meech}, Annabella and {Aguichine}, Artyom and {Alam}, Munazza K. and {Alderson}, Lili and {Batalha}, Natasha E. and {Batalha}, Natalie M. and {Gagnebin}, Anna and {Gordon}, Tyler A. and {Kirk}, James and {L{\'o}pez-Morales}, Mercedes and {Moran}, Sarah E. and {Redai}, Jea Iyanla and {Scarsdale}, Nicholas and {Teske}, Johanna and {Wakeford}, Hannah R. and {Wogan}, Nicholas F. and {Wolfgang}, Angie},
        title = "{JWST COMPASS: NIRSpec/G395H Transmission Observations of the Sub-Neptune HD 15337 c}",
      journal = {\aj},
         year = 2026,
        month = mar,
       volume = {171},
       number = {3},
          eid = {180},
        pages = {180},
          doi = {10.3847/1538-3881/ae2d12},
archivePrefix = {arXiv},
       eprint = {2602.22327},
 primaryClass = {astro-ph.EP},
       adsurl = {https://ui.adsabs.harvard.edu/abs/2026AJ....171..180W}
}

@ARTICLE{Fisher2026,
       author = {{Fisher}, Chloe E. and {Hooton}, Matthew J. and {Gressier}, Am{\'e}lie and {Zgraggen}, Merlin and {Tian}, Meng and {Heng}, Kevin and {Allen}, Natalie H. and {Chatterjee}, Richard D. and {Morris}, Brett M. and {Borsato}, Nicholas W. and {Espinoza}, N{\'e}stor and {Kitzmann}, Daniel and {Meier}, Tobias G. and {Buchhave}, Lars A. and {Burgasser}, Adam J. and {Demory}, Brice-Olivier and {Fortune}, Mark and {Hoeijmakers}, H. Jens and {Luque}, Raphael and {Meier Vald{\'e}s}, Erik A. and {Mendon{\c{c}}a}, Jo{\~a}o M. and {Prinoth}, Bibiana and {Rathcke}, Alexander D. and {Taylor}, Jake},
        title = "{JWST NIRSpec finds no clear signs of an atmosphere on TOI-1685 b}",
      journal = {\mnras},
         year = 2026,
        month = feb,
       volume = {545},
       number = {4},
          eid = {staf2187},
        pages = {staf2187},
          doi = {10.1093/mnras/staf2187},
archivePrefix = {arXiv},
       eprint = {2512.15338},
 primaryClass = {astro-ph.EP},
       adsurl = {https://ui.adsabs.harvard.edu/abs/2026MNRAS.545f2187F}
}

@ARTICLE{Kipping2013,
       author = {{Kipping}, David M.},
        title = "{Efficient, uninformative sampling of limb darkening coefficients for two-parameter laws}",
      journal = {\mnras},
         year = 2013,
        month = nov,
       volume = {435},
       number = {3},
        pages = {2152-2160},
          doi = {10.1093/mnras/stt1435},
archivePrefix = {arXiv},
       eprint = {1308.0009},
 primaryClass = {astro-ph.SR},
       adsurl = {https://ui.adsabs.harvard.edu/abs/2013MNRAS.435.2152K}
}

@ARTICLE{holmberg2023,
       author = {{Holmberg}, M{\r{a}}ns and {Madhusudhan}, Nikku},
        title = "{Exoplanet spectroscopy with JWST NIRISS: diagnostics and case studies}",
      journal = {\mnras},
         year = 2023,
        month = sep,
       volume = {524},
       number = {1},
        pages = {377-402},
          doi = {10.1093/mnras/stad1580},
archivePrefix = {arXiv},
       eprint = {2306.04676},
 primaryClass = {astro-ph.EP},
       adsurl = {https://ui.adsabs.harvard.edu/abs/2023MNRAS.524..377H}
}

@article{kreidberg_batman_2015,
	title = {batman : {BAsic} {Transit} {Model} {cAlculatioN} in {Python}},
	volume = {127},
	issn = {00046280, 15383873},
	shorttitle = {batman},
	url = {http://iopscience.iop.org/article/10.1086/683602},
	doi = {10.1086/683602},
	language = {en},
	number = {957},
	urldate = {2022-09-27},
	journal = {Publications of the Astronomical Society of the Pacific},
	author = {Kreidberg, Laura},
	month = nov,
	year = {2015},
	pages = {1161--1165},
}

@article{horne_optimal_1986,
	title = {An optimal extraction algorithm for {CCD} spectroscopy},
	volume = {98},
	issn = {0004-6280, 1538-3873},
	url = {http://iopscience.iop.org/article/10.1086/131801},
	doi = {10.1086/131801},
	language = {en},
	urldate = {2021-01-16},
	journal = {Publications of the Astronomical Society of the Pacific},
	author = {Horne, K.},
	month = jun,
	year = {1986},
	pages = {609},
}

@misc{moran_high_2023,
	title = {High {Tide} or {Riptide} on the {Cosmic} {Shoreline}? {A} {Water}-{Rich} {Atmosphere} or {Stellar} {Contamination} for the {Warm} {Super}-{Earth} {GJ}{\textasciitilde}486b from {JWST} {Observations}},
	shorttitle = {High {Tide} or {Riptide} on the {Cosmic} {Shoreline}?},
	url = {http://arxiv.org/abs/2305.00868},
	urldate = {2023-05-02},
	publisher = {arXiv},
	author = {Moran, Sarah E. and Stevenson, Kevin B. and Sing, David K. and MacDonald, Ryan J. and Kirk, James and Lustig-Yaeger, Jacob and Peacock, Sarah and Mayorga, L. C. and Bennett, Katherine A. and López-Morales, Mercedes and May, E. M. and Rustamkulov, Zafar and Valenti, Jeff A. and Redai, Jéa I. Adams and Alam, Munazza K. and Batalha, Natasha E. and Fu, Guangwei and Gonzalez-Quiles, Junellie and Highland, Alicia N. and Kruse, Ethan and Lothringer, Joshua D. and Ceballos, Kevin N. Ortiz and Sotzen, Kristin S. and Wakeford, Hannah R.},
	month = may,
	year = {2023},
	note = {arXiv:2305.00868 [astro-ph]},
}

@article{foreman-mackey_emcee_2013,
	title = {emcee : {The} {MCMC} {Hammer}},
	volume = {125},
	issn = {00046280, 15383873},
	shorttitle = {emcee},
	url = {http://iopscience.iop.org/article/10.1086/670067},
	doi = {10.1086/670067},
	language = {en},
	number = {925},
	urldate = {2021-11-28},
	journal = {Publications of the Astronomical Society of the Pacific},
	author = {Foreman-Mackey, Daniel and Hogg, David W. and Lang, Dustin and Goodman, Jonathan},
	month = mar,
	year = {2013},
	pages = {306--312},
}

@ARTICLE{greene2016,
       author = {{Greene}, Thomas P. and {Line}, Michael R. and {Montero}, Cezar and
         {Fortney}, Jonathan J. and {Lustig-Yaeger}, Jacob and {Luther}, Kyle},
        title = "{Characterizing Transiting Exoplanet Atmospheres with JWST}",
      journal = {\apj},
         year = 2016,
        month = jan,
       volume = {817},
       number = {1},
          eid = {17},
        pages = {17},
          doi = {10.3847/0004-637X/817/1/17},
archivePrefix = {arXiv},
       eprint = {1511.05528},
 primaryClass = {astro-ph.EP},
       adsurl = {https://ui.adsabs.harvard.edu/abs/2016ApJ...817...17G}
}

@ARTICLE{rogers2015,
       author = {{Rogers}, Leslie A.},
        title = "{Most 1.6 Earth-radius Planets are Not Rocky}",
      journal = {\apj},
         year = 2015,
        month = mar,
       volume = {801},
       number = {1},
          eid = {41},
        pages = {41},
          doi = {10.1088/0004-637X/801/1/41},
archivePrefix = {arXiv},
       eprint = {1407.4457},
 primaryClass = {astro-ph.EP},
       adsurl = {https://ui.adsabs.harvard.edu/abs/2015ApJ...801...41R}
}

@ARTICLE{Benneke2019,
       author = {{Benneke}, Bj{\"o}rn and {Wong}, Ian and {Piaulet}, Caroline and {Knutson}, Heather A. and {Lothringer}, Joshua and {Morley}, Caroline V. and {Crossfield}, Ian J.~M. and {Gao}, Peter and {Greene}, Thomas P. and {Dressing}, Courtney and {Dragomir}, Diana and {Howard}, Andrew W. and {McCullough}, Peter R. and {Kempton}, Eliza M. -R. and {Fortney}, Jonathan J. and {Fraine}, Jonathan},
        title = "{Water Vapor and Clouds on the Habitable-zone Sub-Neptune Exoplanet K2-18b}",
      journal = {\apjl},
         year = 2019,
        month = dec,
       volume = {887},
       number = {1},
          eid = {L14},
        pages = {L14},
          doi = {10.3847/2041-8213/ab59dc},
archivePrefix = {arXiv},
       eprint = {1909.04642},
 primaryClass = {astro-ph.EP},
       adsurl = {https://ui.adsabs.harvard.edu/abs/2019ApJ...887L..14B}
}

@ARTICLE{macdonald2017,
       author = {{MacDonald}, Ryan J. and {Madhusudhan}, Nikku},
        title = "{HD 209458b in new light: evidence of nitrogen chemistry, patchy clouds and sub-solar water}",
      journal = {\mnras},
         year = "2017",
        month = "Aug",
       volume = {469},
       number = {2},
        pages = {1979-1996},
          doi = {10.1093/mnras/stx804},
archivePrefix = {arXiv},
       eprint = {1701.01113},
 primaryClass = {astro-ph.EP},
       adsurl = {https://ui.adsabs.harvard.edu/abs/2017MNRAS.469.1979M}
}

@ARTICLE{pinhas2019,
       author = {{Pinhas}, Arazi and {Madhusudhan}, Nikku and {Gandhi}, Siddharth and
         {MacDonald}, Ryan},
        title = "{H$_{2}$O abundances and cloud properties in ten hot giant exoplanets}",
      journal = {\mnras},
         year = "2019",
        month = "Jan",
       volume = {482},
       number = {2},
        pages = {1485-1498},
          doi = {10.1093/mnras/sty2544},
archivePrefix = {arXiv},
       eprint = {1811.00011},
 primaryClass = {astro-ph.EP},
       adsurl = {https://ui.adsabs.harvard.edu/abs/2019MNRAS.482.1485P}
}

@ARTICLE{rogers2010,
       author = {{Rogers}, L.~A. and {Seager}, S.},
        title = "{Three Possible Origins for the Gas Layer on GJ 1214b}",
      journal = {\apj},
         year = "2010",
        month = "Jun",
       volume = {716},
       number = {2},
        pages = {1208-1216},
          doi = {10.1088/0004-637X/716/2/1208},
archivePrefix = {arXiv},
       eprint = {0912.3243},
 primaryClass = {astro-ph.EP},
       adsurl = {https://ui.adsabs.harvard.edu/abs/2010ApJ...716.1208R}
}

@article{madhu2020,
	doi = {10.3847/2041-8213/ab7229},
	url = {https://doi.org/10.3847%2F2041-8213%2Fab7229},
	year = 2020,
	month = {feb},
	publisher = {American Astronomical Society},
	volume = {891},
	number = {1},
	pages = {L7},
	author = {Nikku Madhusudhan and Matthew C. Nixon and Luis Welbanks and Anjali A. A. Piette and Richard A. Booth},
	title = {The Interior and Atmosphere of the Habitable-zone Exoplanet K2-18b},
	journal = {The Astrophysical Journal}}

@ARTICLE{VanEylen2021,
       author = {{Van Eylen}, Vincent and {Astudillo-Defru}, N. and {Bonfils}, X. and {Livingston}, J. and {Hirano}, T. and {Luque}, R. and {Lam}, K.~W.~F. and {Justesen}, A.~B. and {Winn}, J.~N. and {Gandolfi}, D. and {Nowak}, G. and {Palle}, E. and {Albrecht}, S. and {Dai}, F. and {Campos Estrada}, B. and {Owen}, J.~E. and {Foreman-Mackey}, D. and {Fridlund}, M. and {Korth}, J. and {Mathur}, S. and {Forveille}, T. and {Mikal-Evans}, T. and {Osborne}, H.~L.~M. and {Ho}, C.~S.~K. and {Almenara}, J.~M. and {Artigau}, E. and {Barrag{\'a}n}, O. and {Bouchy}, F. and {Cabrera}, J. and {Caldwell}, D.~A. and {Charbonneau}, D. and {Chaturvedi}, P. and {Cochran}, W.~D. and {Csizmadia}, S. and {Damasso}, M. and {Delfosse}, X. and {De Medeiros}, J.~R. and {D{\'\i}az}, R.~F. and {Doyon}, R. and {Esposito}, M. and {F{\H{u}}r{\'e}sz}, G. and {Figueira}, P. and {Georgieva}, I. and {Goffo}, E. and {Grziwa}, S. and {Guenther}, E. and {Hatzes}, A.~P. and {Jenkins}, J.~M. and {Kabath}, P. and {Knudstrup}, E. and {Latham}, D.~W. and {Lavie}, B. and {Lovis}, C. and {Mennickent}, R.~E. and {Mullally}, S.~E. and {Murgas}, F. and {Narita}, N. and {Pepe}, F.~A. and {Persson}, C.~M. and {Redfield}, S. and {Ricker}, G.~R. and {Santos}, N.~C. and {Seager}, S. and {Serrano}, L.~M. and {Smith}, A.~M.~S. and {Su{\'a}rez Mascare{\~n}o}, A. and {Subjak}, J. and {Twicken}, J.~D. and {Udry}, S. and {Vanderspek}, R. and {Zapatero Osorio}, M.~R.},
        title = "{Masses and compositions of three small planets orbiting the nearby M dwarf L231-32 (TOI-270) and the M dwarf radius valley}",
      journal = {arXiv e-prints},
         year = 2021,
        month = jan,
          eid = {arXiv:2101.01593},
        pages = {arXiv:2101.01593},
archivePrefix = {arXiv},
       eprint = {2101.01593},
 primaryClass = {astro-ph.EP},
       adsurl = {https://ui.adsabs.harvard.edu/abs/2021arXiv210101593V}
}

@ARTICLE{Knutson2014,
       author = {{Knutson}, Heather A. and {Benneke}, Bj{\"o}rn and {Deming}, Drake and {Homeier}, Derek},
        title = "{A featureless transmission spectrum for the Neptune-mass exoplanet GJ436b}",
      journal = {\nat},
         year = 2014,
        month = jan,
       volume = {505},
       number = {7481},
        pages = {66-68},
          doi = {10.1038/nature12887},
archivePrefix = {arXiv},
       eprint = {1401.3350},
 primaryClass = {astro-ph.EP},
       adsurl = {https://ui.adsabs.harvard.edu/abs/2014Natur.505...66K}
}

@ARTICLE{Sarkar2024,
       author = {{Sarkar}, Subhajit and {Madhusudhan}, Nikku and {Constantinou}, Savvas and {Holmberg}, M{\r{a}}ns},
        title = "{Exoplanet transit spectroscopy with JWST NIRSpec: diagnostics and homogeneous case study of WASP-39 b}",
      journal = {\mnras},
         year = 2024,
        month = jun,
       volume = {531},
       number = {2},
        pages = {2731-2756},
          doi = {10.1093/mnras/stae1230},
archivePrefix = {arXiv},
       eprint = {2405.06737},
 primaryClass = {astro-ph.EP},
       adsurl = {https://ui.adsabs.harvard.edu/abs/2024MNRAS.531.2731S}
}

@software{tsiaras_2016_pylightcurve,
       author = {{Tsiaras}, A. and {Waldmann}, I.~P. and {Rocchetto}, M. and {Varley}, R. and {Morello}, G. and {Damiano}, M. and {Tinetti}, G.},
        title = "{pylightcurve: Exoplanet lightcurve model}",
 howpublished = {Astrophysics Source Code Library, record ascl:1612.018},
         year = 2016,
        month = dec,
          eid = {ascl:1612.018},
       adsurl = {https://ui.adsabs.harvard.edu/abs/2016ascl.soft12018T}
}

@ARTICLE{Hu2025,
       author = {{Hu}, Renyu and {Bello-Arufe}, Aaron and {Tokadjian}, Armen and {Yang}, Jeehyun and {Damiano}, Mario and {Roy}, Pierre-Alexis and {Coulombe}, Louis-Philippe and {Madhusudhan}, Nikku and {Constantinou}, Savvas and {Benneke}, Bj{\"o}rn},
        title = "{A water-rich interior in the temperate sub-Neptune K2-18 b revealed by JWST}",
      journal = {arXiv e-prints},
         year = 2025,
        month = jul,
          eid = {arXiv:2507.12622},
        pages = {arXiv:2507.12622},
          doi = {10.48550/arXiv.2507.12622},
archivePrefix = {arXiv},
       eprint = {2507.12622},
 primaryClass = {astro-ph.EP},
       adsurl = {https://ui.adsabs.harvard.edu/abs/2025arXiv250712622H}
}

@ARTICLE{Mukherjee2025,
       author = {{Mukherjee}, Sagnick and {Schlawin}, Everett and {Bell}, Taylor J. and {Fortney}, Jonathan J. and {Beatty}, Thomas G. and {Greene}, Thomas P. and {Ohno}, Kazumasa and {Murphy}, Matthew M. and {Parmentier}, Vivien and {Line}, Michael R. and {Welbanks}, Luis and {Wiser}, Lindsey S. and {Rieke}, Marcia J.},
        title = "{A JWST Panchromatic Thermal Emission Spectrum of the Warm Neptune Archetype GJ 436b}",
      journal = {\apjl},
         year = 2025,
        month = apr,
       volume = {982},
       number = {2},
          eid = {L39},
        pages = {L39},
          doi = {10.3847/2041-8213/adba46},
archivePrefix = {arXiv},
       eprint = {2502.17418},
 primaryClass = {astro-ph.EP},
       adsurl = {https://ui.adsabs.harvard.edu/abs/2025ApJ...982L..39M}
}

@preamble{ " \newcommand{\noop}[1]{} " }

@article{HITRAN2020,
author = {I.~E. {Gordon} AND L.~S. {Rothman} AND R.~J. {Hargreaves} AND R. {Hashemi} AND E.~V. {Karlovets} AND F.~M. {Skinner} AND E.~K. {Conway} AND C. {Hill} AND R.~V. {Kochanov} AND Y. {Tan} AND P. {Wcis{\l}o} AND A.~A. {Finenko} AND K. {Nelson} AND P.~F. {Bernath} AND M. {Birk} AND V. {Boudon} AND A. {Campargue} AND K.~V. {Chance} AND A. {Coustenis} AND B.~J. {Drouin} AND J.-M. {Flaud} AND R.~R. {Gamache} AND J.~T. {Hodges} AND D. {Jacquemart} AND E.~J. {Mlawer} AND A.~V. {Nikitin} AND V.~I. {Perevalov} AND M. {Rotger} AND J. {Tennyson} AND G.~C. {Toon} AND H. {Tran} AND V.~G. {Tyuterev} AND E.~M. {Adkins} AND A. {Baker} AND A. {Barbe} AND E. {Can{\`{e}}} AND A.~G. {Cs{'{a}}sz{'{a}}r} AND A. {Dudaryonok} AND O. {Egorov} AND A.~J. {Fleisher} AND H. {Fleurbaey} AND A. {Foltynowicz} AND T. {Furtenbacher} AND J.~J. {Harrison} AND J.-M. {Hartmann} AND V.-M. {Horneman} AND X. {Huang} AND T. {Karman} AND J. {Karns} AND S. {Kassi} AND I. {Kleiner} AND V. {Kofman} AND F. {Kwabia-Tchana} AND N.~N. {Lavrentieva} AND T.~J. {Lee} AND D.~A. {Long} AND A.~A. {Lukashevskaya} AND O.~M. {Lyulin} AND V.~Yu. {Makhnev} AND W. {Matt} AND S.~T. {Massie} AND M. {Melosso} AND S.~N. {Mikhailenko} AND D. {Mondelain} AND H.~S.~P. {M{"{u}}ller} AND O.~V. {Naumenko} AND A. {Perrin} AND O.~L. {Polyansky} AND E. {Raddaoui} AND P.~L. {Raston} AND Z.~D. {Reed} AND M. {Rey} AND C. {Richard} AND R. {T{'{o}}bi{'{a}}s} AND I. {Sadiek} AND D.~W. {Schwenke} AND E. {Starikova} AND K. {Sung} AND F. {Tamassia} AND S.~A. {Tashkun} AND J. {Vander Auwera} AND I.~A. {Vasilenko} AND A.~A. {Vigasin} AND G.~L. {Villanueva} AND B. {Vispoel} AND G. {Wagner} AND A. {Yachmenev} AND S.~N. {Yurchenko}},
title = {The {HITRAN2020} Molecular Spectroscopic Database},
journal = {Journal of Quantitative Spectroscopy and Radiative Transfer},
year = {2022},
volume = {277},
pages = {107949},
doi = {10.1016/j.jqsrt.2021.107949},
}

@ARTICLE{Cooke2024,
       author = {{Cooke}, Gregory J. and {Madhusudhan}, Nikku},
        title = "{Considerations for Photochemical Modeling of Possible Hycean Worlds}",
      journal = {arXiv e-prints},
         year = 2024,
        month = oct,
          eid = {arXiv:2410.07313},
        pages = {arXiv:2410.07313},
          doi = {10.48550/arXiv.2410.07313},
archivePrefix = {arXiv},
       eprint = {2410.07313},
 primaryClass = {astro-ph.EP},
       adsurl = {https://ui.adsabs.harvard.edu/abs/2024arXiv241007313C}
}

@ARTICLE{Batalha2017,
       author = {{Batalha}, Natasha E. and {Mandell}, Avi and {Pontoppidan}, Klaus and {Stevenson}, Kevin B. and {Lewis}, Nikole K. and {Kalirai}, Jason and {Earl}, Nick and {Greene}, Thomas and {Albert}, Lo{\"\i}c and {Nielsen}, Louise D.},
        title = "{PandExo: A Community Tool for Transiting Exoplanet Science with JWST \& HST}",
      journal = {\pasp},
         year = 2017,
        month = jun,
       volume = {129},
       number = {976},
        pages = {064501},
          doi = {10.1088/1538-3873/aa65b0},
archivePrefix = {arXiv},
       eprint = {1702.01820},
 primaryClass = {astro-ph.IM},
       adsurl = {https://ui.adsabs.harvard.edu/abs/2017PASP..129f4501B}
}

@ARTICLE{Lodders2002,
       author = {{Lodders}, Katharina and {Fegley}, Bruce},
        title = "{Atmospheric Chemistry in Giant Planets, Brown Dwarfs, and Low-Mass Dwarf Stars. I. Carbon, Nitrogen, and Oxygen}",
      journal = {\icarus},
         year = 2002,
        month = feb,
       volume = {155},
       number = {2},
        pages = {393-424},
          doi = {10.1006/icar.2001.6740},
       adsurl = {https://ui.adsabs.harvard.edu/abs/2002Icar..155..393L}
}

@ARTICLE{Yu2021,
       author = {{Yu}, Xinting and {Moses}, Julianne I. and {Fortney}, Jonathan J. and {Zhang}, Xi},
        title = "{How to Identify Exoplanet Surfaces Using Atmospheric Trace Species in Hydrogen-dominated Atmospheres}",
      journal = {\apj},
         year = 2021,
        month = jun,
       volume = {914},
       number = {1},
          eid = {38},
        pages = {38},
          doi = {10.3847/1538-4357/abfdc7},
archivePrefix = {arXiv},
       eprint = {2104.09843},
 primaryClass = {astro-ph.EP},
       adsurl = {https://ui.adsabs.harvard.edu/abs/2021ApJ...914...38Y}
}

@ARTICLE{Hu2021,
       author = {{Hu}, Renyu and {Damiano}, Mario and {Scheucher}, Markus and {Kite}, Edwin and {Seager}, Sara and {Rauer}, Heike},
        title = "{Unveiling Shrouded Oceans on Temperate sub-Neptunes via Transit Signatures of Solubility Equilibria versus Gas Thermochemistry}",
      journal = {\apjl},
         year = 2021,
        month = nov,
       volume = {921},
       number = {1},
          eid = {L8},
        pages = {L8},
        doi = {10.3847/2041-8213/ac1f92},
archivePrefix = {arXiv},
       eprint = {2108.04745},
 primaryClass = {astro-ph.EP},
       adsurl = {https://ui.adsabs.harvard.edu/abs/2021ApJ...921L...8H}
}

@ARTICLE{Tsai2021,
       author = {{Tsai}, Shang-Min and {Innes}, Hamish and {Lichtenberg}, Tim and {Taylor}, Jake and {Malik}, Matej and {Chubb}, Katy and {Pierrehumbert}, Raymond},
        title = "{Inferring Shallow Surfaces on Sub-Neptune Exoplanets with JWST}",
      journal = {\apjl},
         year = 2021,
        month = dec,
       volume = {922},
       number = {2},
          eid = {L27},
        pages = {L27},
          doi = {10.3847/2041-8213/ac399a},
archivePrefix = {arXiv},
       eprint = {2111.06429},
 primaryClass = {astro-ph.EP},
       adsurl = {https://ui.adsabs.harvard.edu/abs/2021ApJ...922L..27T}
}

@ARTICLE{Hu2021b,
       author = {{Hu}, Renyu},
        title = "{Photochemistry and Spectral Characterization of Temperate and Gas-rich Exoplanets}",
      journal = {\apj},
         year = 2021,
        month = nov,
       volume = {921},
       number = {1},
          eid = {27},
        pages = {27},
          doi = {10.3847/1538-4357/ac1789},
archivePrefix = {arXiv},
       eprint = {2108.04419},
 primaryClass = {astro-ph.EP},
       adsurl = {https://ui.adsabs.harvard.edu/abs/2021ApJ...921...27H}
}

@ARTICLE{Lopez2014,
       author = {{Lopez}, Eric D. and {Fortney}, Jonathan J.},
        title = "{Understanding the Mass-Radius Relation for Sub-neptunes: Radius as a Proxy for Composition}",
      journal = {ApJ},
         year = 2014,
        month = sep,
       volume = {792},
       number = {1},
          eid = {1},
        pages = {1},
          doi = {10.1088/0004-637X/792/1/1},
 archivePrefix = {arXiv},
       eprint = {1311.0329},
 primaryClass = {astro-ph.EP},
       adsurl = {https://ui.adsabs.harvard.edu/abs/2014ApJ...792....1L}
}

@ARTICLE{Moses2013,
       author = {{Moses}, J.~I. and {Line}, M.~R. and {Visscher}, C. and {Richardson}, M.~R. and {Nettelmann}, N. and {Fortney}, J.~J. and {Barman}, T.~S. and {Stevenson}, K.~B. and {Madhusudhan}, N.},
        title = "{Compositional Diversity in the Atmospheres of Hot Neptunes, with Application to GJ 436b}",
      journal = {\apj},
         year = 2013,
        month = nov,
       volume = {777},
       number = {1},
          eid = {34},
        pages = {34},
          doi = {10.1088/0004-637X/777/1/34},
archivePrefix = {arXiv},
       eprint = {1306.5178},
 primaryClass = {astro-ph.EP},
       adsurl = {https://ui.adsabs.harvard.edu/abs/2013ApJ...777...34M}
}

@ARTICLE{Ohno2022,
       author = {{Ohno}, Kazumasa and {Fortney}, Jonathan J.},
        title = "{Nitrogen as a Tracer of Giant Planet Formation. II.: Comprehensive Study of Nitrogen Photochemistry and Implications for Observing NH3 and HCN in Transmission and Emission Spectra}",
      journal = {arXiv e-prints},
         year = 2022,
        month = nov,
          eid = {arXiv:2211.16877},
        pages = {arXiv:2211.16877},
          doi = {10.48550/arXiv.2211.16877},
archivePrefix = {arXiv},
       eprint = {2211.16877},
 primaryClass = {astro-ph.EP},
       adsurl = {https://ui.adsabs.harvard.edu/abs/2022arXiv221116877O}
}

@ARTICLE{Seager2016,
       author = {{Seager}, S. and {Bains}, W. and {Petkowski}, J.~J.},
        title = "{Toward a List of Molecules as Potential Biosignature Gases for the Search for Life on Exoplanets and Applications to Terrestrial Biochemistry}",
      journal = {Astrobiology},
         year = 2016,
        month = jun,
       volume = {16},
       number = {6},
        pages = {465-485},
          doi = {10.1089/ast.2015.1404},
       adsurl = {https://ui.adsabs.harvard.edu/abs/2016AsBio..16..465S}
}

@ARTICLE{Barstow2016,
       author = {{Barstow}, J.~K. and {Irwin}, P.~G.~J.},
        title = "{Habitable worlds with JWST: transit spectroscopy of the TRAPPIST-1 system?}",
      journal = {\mnras},
         year = 2016,
        month = sep,
       volume = {461},
       number = {1},
        pages = {L92-L96},
          doi = {10.1093/mnrasl/slw109},
archivePrefix = {arXiv},
       eprint = {1605.07352},
 primaryClass = {astro-ph.EP},
       adsurl = {https://ui.adsabs.harvard.edu/abs/2016MNRAS.461L..92B}
}

@ARTICLE{Seager2021,
       author = {{Seager}, Sara and {Petkowski}, Janusz J. and {G{\"u}nther}, Maximilian N. and {Bains}, William and {Mikal-Evans}, Thomas and {Deming}, Drake},
        title = "{Possibilities for an Aerial Biosphere in Temperate Sub Neptune-Sized Exoplanet Atmospheres}",
      journal = {Universe},
         year = 2021,
        month = may,
       volume = {7},
       number = {6},
        pages = {172},
          doi = {10.3390/universe7060172},
archivePrefix = {arXiv},
       eprint = {2106.07729},
 primaryClass = {astro-ph.EP},
       adsurl = {https://ui.adsabs.harvard.edu/abs/2021Univ....7..172S}
}

@article{madhusudhan_habitability_2021,
	title = {Habitability and {Biosignatures} of {Hycean} {Worlds}},
	volume = {918},
	issn = {0004-637X},
	url = {https://dx.doi.org/10.3847/1538-4357/abfd9c},
	doi = {10.3847/1538-4357/abfd9c},
	language = {en},
	number = {1},
	urldate = {2023-11-29},
	journal = {The Astrophysical Journal},
	author = {Madhusudhan, Nikku and Piette, Anjali A. A. and Constantinou, Savvas},
	month = aug,
	year = {2021},
	note = {Publisher: The American Astronomical Society},
	pages = {1},
}

@article{madhusudhan_carbon-bearing_2023,
	title = {Carbon-bearing {Molecules} in a {Possible} {Hycean} {Atmosphere}},
	volume = {956},
	issn = {2041-8205, 2041-8213},
	url = {https://iopscience.iop.org/article/10.3847/2041-8213/acf577},
	doi = {10.3847/2041-8213/acf577},
	language = {en},
	number = {1},
	urldate = {2024-02-05},
	journal = {The Astrophysical Journal Letters},
	author = {Madhusudhan, Nikku and Sarkar, Subhajit and Constantinou, Savvas and Holmberg, Måns and Piette, Anjali A. A. and Moses, Julianne I.},
	month = oct,
	year = {2023},
	pages = {L13},
}

@ARTICLE{Nixon21_ocean,
       author = {{Nixon}, Matthew C. and {Madhusudhan}, Nikku},
        title = "{How deep is the ocean? Exploring the phase structure of water-rich sub-Neptunes}",
      journal = {\mnras},
         year = 2021,
        month = aug,
       volume = {505},
       number = {3},
        pages = {3414-3432},
          doi = {10.1093/mnras/stab1500},
archivePrefix = {arXiv},
       eprint = {2106.02061},
 primaryClass = {astro-ph.EP},
       adsurl = {https://ui.adsabs.harvard.edu/abs/2021MNRAS.505.3414N}
}

@misc{benneke_jwst_2024,
	title = {{JWST} {Reveals} {CH}\$\_4\$, {CO}\$\_2\$, and {H}\$\_2\${O} in a {Metal}-rich {Miscible} {Atmosphere} on a {Two}-{Earth}-{Radius} {Exoplanet}},
	url = {http://arxiv.org/abs/2403.03325},
	urldate = {2024-03-12},
	publisher = {arXiv},
	author = {Benneke, Björn and Roy, Pierre-Alexis and Coulombe, Louis-Philippe and Radica, Michael and Piaulet, Caroline and Ahrer, Eva-Maria and Pierrehumbert, Raymond and Krissansen-Totton, Joshua and Schlichting, Hilke E. and Hu, Renyu and Yang, Jeehyun and Christie, Duncan and Thorngren, Daniel and Young, Edward D. and Pelletier, Stefan and Knutson, Heather A. and Miguel, Yamila and Evans-Soma, Thomas M. and Dorn, Caroline and Gagnebin, Anna and Fortney, Jonathan J. and Komacek, Thaddeus and MacDonald, Ryan and Raul, Eshan and Cloutier, Ryan and Acuna, Lorena and Lafrenière, David and Cadieux, Charles and Doyon, René and Welbanks, Luis and Allart, Romain},
	month = mar,
	year = {2024},
	note = {arXiv:2403.03325 [astro-ph]},
}

@misc{wogan_jwst_2024,
       author = {{Wogan}, Nicholas F. and {Batalha}, Natasha E. and {Zahnle}, Kevin J. and {Krissansen-Totton}, Joshua and {Tsai}, Shang-Min and {Hu}, Renyu},
        title = "{JWST Observations of K2-18b Can Be Explained by a Gas-rich Mini-Neptune with No Habitable Surface}",
      journal = {\apjl},
         year = 2024,
        month = mar,
       volume = {963},
       number = {1},
          eid = {L7},
        pages = {L7},
          doi = {10.3847/2041-8213/ad2616},
archivePrefix = {arXiv},
       eprint = {2401.11082},
 primaryClass = {astro-ph.EP},
       adsurl = {https://ui.adsabs.harvard.edu/abs/2024ApJ...963L...7W}
}

@article{rigby_ocean_2024,
	title = {On the ocean conditions of {Hycean} worlds},
	volume = {529},
	issn = {0035-8711},
	url = {https://doi.org/10.1093/mnras/stae413},
	doi = {10.1093/mnras/stae413},
	number = {1},
	urldate = {2024-04-13},
	journal = {Monthly Notices of the Royal Astronomical Society},
	author = {Rigby, Frances E and Madhusudhan, Nikku},
	month = mar,
	year = {2024},
	pages = {409--424},
}

@ARTICLE{Rigby2024_gasdwarf,
       author = {{Rigby}, Frances E. and {Pica-Ciamarra}, Lorenzo and {Holmberg}, M{\r{a}}ns and {Madhusudhan}, Nikku and {Constantinou}, Savvas and {Schaefer}, Laura and {Deng}, Jie and {Lee}, Kanani K.~M. and {Moses}, Julianne I.},
        title = "{Toward a Self-consistent Evaluation of Gas Dwarf Scenarios for Temperate Sub-Neptunes}",
      journal = {\apj},
         year = 2024,
        month = nov,
       volume = {975},
       number = {1},
          eid = {101},
        pages = {101},
          doi = {10.3847/1538-4357/ad6c38},
archivePrefix = {arXiv},
       eprint = {2409.03683},
 primaryClass = {astro-ph.EP},
       adsurl = {https://ui.adsabs.harvard.edu/abs/2024ApJ...975..101R}
}

@article{shorttle_distinguishing_2024,
	title = {Distinguishing {Oceans} of {Water} from {Magma} on {Mini}-{Neptune} {K2}-18b},
	volume = {962},
	issn = {2041-8205},
	url = {https://dx.doi.org/10.3847/2041-8213/ad206e},
	doi = {10.3847/2041-8213/ad206e},
	language = {en},
	number = {1},
	urldate = {2024-10-09},
	journal = {The Astrophysical Journal Letters},
	author = {Shorttle, Oliver and Jordan, Sean and Nicholls, Harrison and Lichtenberg, Tim and Bower, Dan J.},
	month = feb,
	year = {2024},
	note = {Publisher: The American Astronomical Society},
	pages = {L8},
}

@article{holmberg_possible_2024,
	title = {Possible {Hycean} conditions in the sub-{Neptune} {TOI}-270 d},
	volume = {683},
	issn = {0004-6361},
	url = {https://ui.adsabs.harvard.edu/abs/2024A&A...683L...2H/abstract},
	doi = {10.1051/0004-6361/202348238},
	language = {en},
	urldate = {2024-05-10},
	journal = {Astronomy and Astrophysics},
	author = {Holmberg, Måns and Madhusudhan, Nikku},
	month = mar,
	year = {2024},
	pages = {L2},
}

@article{Madhusudhan_chem_2023,
       author = {{Madhusudhan}, Nikku and {Moses}, Julianne I. and {Rigby}, Frances and {Barrier}, Edouard},
        title = "{Chemical conditions on Hycean worlds}",
      journal = {Faraday Discussions},
         year = 2023,
        month = sep,
       volume = {245},
        pages = {80-111},
          doi = {10.1039/D3FD00075C},
archivePrefix = {arXiv},
       eprint = {2306.13706},
 primaryClass = {astro-ph.EP},
       adsurl = {https://ui.adsabs.harvard.edu/abs/2023FaDi..245...80M}
}

@ARTICLE{Li2015,
       author = {{Li}, Gang and {Gordon}, Iouli E. and {Rothman}, Laurence S. and {Tan}, Yan and {Hu}, Shui-Ming and {Kassi}, Samir and {Campargue}, Alain and {Medvedev}, Emile S.},
        title = "{Rovibrational Line Lists for Nine Isotopologues of the CO Molecule in the X $^{1}${\ensuremath{\Sigma}}$^{+}$ Ground Electronic State}",
      journal = {\apjs},
         year = 2015,
        month = jan,
       volume = {216},
       number = {1},
          eid = {15},
        pages = {15},
          doi = {10.1088/0067-0049/216/1/15},
       adsurl = {https://ui.adsabs.harvard.edu/abs/2015ApJS..216...15L}
}

@ARTICLE{Wallack2024,
       author = {{Wallack}, Nicole L. and {Batalha}, Natasha E. and {Alderson}, Lili and {Scarsdale}, Nicholas and {Adams Redai}, Jea I. and {Aguichine}, Artyom and {Alam}, Munazza K. and {Gao}, Peter and {Wolfgang}, Angie and {Batalha}, Natalie M. and {Kirk}, James and {L{\'o}pez-Morales}, Mercedes and {Moran}, Sarah E. and {Teske}, Johanna and {Wakeford}, Hannah R. and {Wogan}, Nicholas F.},
        title = "{JWST COMPASS: A NIRSpec/G395H Transmission Spectrum of the Sub-Neptune TOI-836c}",
      journal = {\aj},
         year = 2024,
        month = aug,
       volume = {168},
       number = {2},
          eid = {77},
        pages = {77},
          doi = {10.3847/1538-3881/ad3917},
archivePrefix = {arXiv},
       eprint = {2404.01264},
 primaryClass = {astro-ph.EP},
       adsurl = {https://ui.adsabs.harvard.edu/abs/2024AJ....168...77W}
}

@ARTICLE{Alam2025,
       author = {{Alam}, Munazza K. and {Gao}, Peter and {Adams Redai}, Jea and {Wallack}, Nicole L. and {Wogan}, Nicholas F. and {Aguichine}, Artyom and {Dattilo}, Anne and {Alderson}, Lili and {Batalha}, Natasha E. and {Batalha}, Natalie M. and {Kirk}, James and {L{\'o}pez-Morales}, Mercedes and {Meech}, Annabella and {Moran}, Sarah E. and {Teske}, Johanna and {Wakeford}, Hannah R. and {Wolfgang}, Angie},
        title = "{JWST COMPASS: The First Near- to Mid-infrared Transmission Spectrum of the Hot Super-Earth L 168-9 b}",
      journal = {\aj},
         year = 2025,
        month = jan,
       volume = {169},
       number = {1},
          eid = {15},
        pages = {15},
          doi = {10.3847/1538-3881/ad8eb5},
archivePrefix = {arXiv},
       eprint = {2411.03154},
 primaryClass = {astro-ph.EP},
       adsurl = {https://ui.adsabs.harvard.edu/abs/2025AJ....169...15A}
}

@ARTICLE{Turner2016,
       author = {{Turner}, Jake D. and {Pearson}, Kyle A. and {Biddle}, Lauren I. and {Smart}, Brianna M. and {Zellem}, Robert T. and {Teske}, Johanna K. and {Hardegree-Ullman}, Kevin K. and {Griffith}, Caitlin C. and {Leiter}, Robin M. and {Cates}, Ian T. and {Nieberding}, Megan N. and {Smith}, Carter-Thaxton W. and {Thompson}, Robert M. and {Hofmann}, Ryan and {Berube}, Michael P. and {Nguyen}, Chi H. and {Small}, Lindsay C. and {Guvenen}, Blythe C. and {Richardson}, Logan and {McGraw}, Allison and {Raphael}, Brandon and {Crawford}, Benjamin E. and {Robertson}, Amy N. and {Tombleson}, Ryan and {Carleton}, Timothy M. and {Towner}, Allison P.~M. and {Walker-LaFollette}, Amanda M. and {Hume}, Jeffrey R. and {Watson}, Zachary T. and {Jones}, Christen K. and {Lichtenberger}, Matthew J. and {Hoglund}, Shelby R. and {Cook}, Kendall L. and {Crossen}, Cory A. and {Jorgensen}, Curtis R. and {Romine}, James M. and {Thompson}, Alejandro R. and {Villegas}, Christian F. and {Wilson}, Ashley A. and {Sanford}, Brent and {Taylor}, Joanna M. and {Henz}, Triana N.},
        title = "{Ground-based near-UV observations of 15 transiting exoplanets: constraints on their atmospheres and no evidence for asymmetrical transits}",
      journal = {\mnras},
         year = 2016,
        month = jun,
       volume = {459},
       number = {1},
        pages = {789-819},
          doi = {10.1093/mnras/stw574},
archivePrefix = {arXiv},
       eprint = {1603.02587},
 primaryClass = {astro-ph.EP},
       adsurl = {https://ui.adsabs.harvard.edu/abs/2016MNRAS.459..789T}
}

@software{Newville2025,
       author = {{Newville}, Matthew and {Otten}, Renee and {Nelson}, Andrew and {Stensitzki}, Till and {Ingargiola}, Antonino and {Allan}, Daniel and {Fox}, Austin and {Carter}, Faustin and {Rawlik}, Michal},
        title = "{LMFIT: Non-Linear Least-Squares Minimization and Curve-Fitting for Python}",
         year = 2025,
        month = mar,
          eid = {10.5281/zenodo.15014437},
          doi = {10.5281/zenodo.15014437},
      version = {1.3.3},
    publisher = {Zenodo},
       adsurl = {https://ui.adsabs.harvard.edu/abs/2025zndo..15014437N}
}
\bibliographystyle{aasjournal}

\end{document}